\documentclass[12pt]{article}
\usepackage{epsfig}
\usepackage{graphicx,amssymb,amsmath,amsfonts,hyperref}

\newcommand{\be}{\begin{equation}}
\newcommand{\ee}{\end{equation}}
\newcommand{\bea}{\begin{eqnarray}}
\newcommand{\eea}{\end{eqnarray}}

\def\be{\begin{equation}}
\def\ee{\end{equation}}
\newcommand{\del}{\partial}
\newcommand{\alp}{\ensuremath{\alpha'\:}}

\newcommand{\ssb}{s\bar{s}}

\newcommand{\ccb}{c\bar{c}}
\newcommand{\bbb}{b\bar{b}}
\newcommand{\rchi}[1]{\ensuremath{\chi^2_m/\chi^2_l = #1}}
\newcommand{\ten}[1]{\times10^{#1}}
\newcommand{\mud}{m_{u/d}}

\newcommand{\jph}[2]{\ensuremath{#1/2^{#2}}}

\newcommand{\plm}{\ensuremath{\pm}}
\newcommand{\qqb}{\ensuremath{q\bar{q}}}

\newcommand{\non}{\nonumber \\}
\newcommand{\CR}{\non\cr}
\newcommand{\pa}{\partial}

\def \t {\theta}
\pdfoutput=1
\begin{document}

\title{ \vspace{1cm} Holography Inspired Stringy Hadrons}
 \author{Jacob Sonnenschein \\
 The Raymond and Beverly Sackler School of Physics and Astronomy,\\
 		Tel Aviv University, Ramat Aviv 69978, Israel}
\maketitle
\begin{abstract}
Holography inspired stringy hadrons (HISH) is a set of  models that describe hadrons: mesons, baryons and glueballs as strings in flat four dimensional space time. The models are  based on a ``map" from stringy hadrons of holographic confining  backgrounds.
In this note we review the ``derivation" of the models.  We start with a brief  reminder of the passage from the  $AdS_5\times S^5$ string theory to certain flavored  confining holographic models. We then describe the string configurations in holographic backgrounds  that correspond to
a  Wilson line,a  meson,a  baryon and a glueball. The key ingredients of the four dimensional picture of hadrons are  the ``string endpoint mass"
and the ``baryonic string vertex".  We determine the classical trajectories of the HISH.  We review the current understanding of the quantization of  the hadronic strings. We end with a summary of the
  comparison of  the outcome of the HISH  models with the PDG data about  mesons and baryons. We  extract the values of the tension, masses and intercepts from best fits, write down certain  predictions for higher excited hadrons and  present attempts to identify glueballs.
\end{abstract}
 \newpage	
\tableofcontents
\section {Introduction}
The stringy description of hadrons has been thoroughly investigated since  the  sixties  of the last century\cite{Collins:book}. In this paper I review  the research work  we have been doing in the last few years on a renewed  stringy description of hadrons. This naturally raises the question of what are the reasons   to go back to  ``square one" and revisit this question? Here are several reasons for doing it.
\begin{itemize}
\item
 (i) Up to date, certain  properties, like the hadronic spectrum, the decay width of hadron and  their scattering  cross section,  are hard to derive  from QCD and  relatively easy from a stringy picture.
\item
(ii) Holography, or gauge/string duality, provides a bridge between the underlying theory of QCD (in certain limits) and a bosonic string model of hadrons .
\item
(iii) To  establish a framework that describes the three types of hadrons  mesons, baryons and glueballs  in terms of the same building blocks, namely strings.
\item
(iv)  The passage from the holographic string regime  to  strings in reality is still a tremendous challenge.
\item
(v) Up to date we  lack a full exact procedure of quantizing a rotating string with massive endpoints (which  will see are mandatory for the stringy hadrons).

\item
(vi) There is a wide range of heavy mesonic  and baryonic resonances that have been discovered in recent years.
Thus the challenge is to develop a framework that  can accommodate hadrons with any quark content light, medium and heavy.
\end{itemize}

The holographic duality is an equivalence between certain bulk  string theories and  boundary field theories.
Practically most of the applications of holography are based on relating bulk fields (not strings) and operators on the dual  boundary field theory.
This is based on the usual limit of $\alpha'\rightarrow 0$  with which we go, for instance,  from a closed string theory to a theory of  gravity.
However, to describe realistic  hadrons   it seems that we  need strings since after all in nature  the string tension which is inversely proportional to $\alpha'$ is not  very large. In holography this relates to the fact that one needs to describe nature $\lambda= g^2 N_c$ is  of order one and not very large one.

The main theme of this review paper  is that there is a wide sector of hadronic physical observables  which cannot be faithfully described by bulk fields but rather require  dual  stringy phenomena.
It is well known that this is the case for  a Wilson, a  't Hooft and a  Polyakov lines ( for a review see for instance \cite{Sonnenschein:1999if}).
 We argue here that in fact also  the  spectra, decays and other properties of hadrons:
                    mesons, baryons and glueballs
can be recast only by holographic stringy hadrons and not by fields that reside in the bulk or on flavor branes.
The major  argument against describing the hadron spectra in terms of fluctuations of fields like bulk fields or modes on probe flavor branes is that they generically do not admit properly the Regge  behavior of the spectra.
For  $M^2$      as a function of $J$ we get from  flavor branes only $J=0$ ,$J=1$  mesons and there is  a big  gap of order $\lambda$  ( or certain fractional power of $\lambda$ depending on the model)  in comparison to high $J$  mesons  described  in terms of strings.
Moreover the attempts to get the observed  linearity between   $M^2$     and $n$ the excitation number  is problematic,  whereas for strings it is an obvious property.

The main ideas of this project is (i) To analyze string configurations in holographic string models that correspond to hadrons, (ii)  to bypass the usual transition from the holographic regime of large $N_c$ and large $\lambda$ to the real world via a $\frac{1}{N_c}$ and $\frac{1}{\lambda}$ expansion and state a model of stringy hadrons in flat four dimensions that is inspired by the corresponding holographic strings. (iii) To confront the  outcome of the models   with the experimental data in \cite{Sonnenschein:2014jwa},\cite{Sonnenschein:2014bia} \cite{Sonnenschein:2015zaa}.

Confining holographic models are characterized  by a ``wall" that truncates in one way or another the range of the radial direction (see figure  2). A common feature to all the holographic stringy hadrons is that there is a segment of the string that stretches along a constant radial coordinate in the vicinity of the ``wall". For the stringy glueball it is the whole folded closed string that rotates there and for the open string it is part of the string, the horizontal segment, that connects with vertical segments either to the boundary for a Wilson line or to flavor branes  for the meson and for the baryon. This fact that the classical solutions of the flatly  rotating strings reside at fixed radial direction is behind  the map to rotating strings in flat four dimensional  space-time. A key ingredient of the map is the $m_{sep}$, the `` string endpoint mass" that provides in the four flat space-time description the dual of the vertical string segments. It is important to note (i) This mass parameter is neither the QCD mass parameter nor that of the constituent quark mass. (ii)  As will be seen below the $m_{sep}$ parameter is not an exact map of a vertical segment but rather only an approximation that is more accurate the longer the horizontal string is.

As we have mentioned above the stringy picture of meson has been thoroughly investigated in the past and we will not cite here this huge body of papers. It turns out that also in recent years there were several attempts to describe hadrons in terms of strings. Papers on the subject that have certain overlap with our approach are  for instance \cite{Baker:2002km},\cite{Schreiber:2004ie},\cite{Hellerman:2013kba},\cite{Zahn:2013yma}.
Another approach to the stringy nature of QCD is the approach of low-energy effective theory on long strings. This approach is different but shares certain features with the approach presented in this paper.
 A recent review of the subject can be found in  \cite{Aharony:2013ipa}.

The alternative description of hadrons in terms of fields in the bulk or on flavor branes is not discussed in this paper. For a review paper about this approach and reference therein see \cite{Erdmenger:2007cm}
The paper is organized in the following manner. After this section of an introduction there is a review section that describes the passage from the $AdS_5\times S^5$ string background to that of various confining backgrounds. We describe in some details the prototype model of Sakai and Sugimoto and its generalization.
Section $(\S 3)$ is devoted to hadrons as strings in a holographic background. In this section we separately  describe in $(\S 3.1)$ the holographic Wilson line, in $(\S 3.2)$ the stringy duals of mesons and in $(\S 3.3)$ the glueball as a rotating folded closed string. We then describe in section $(\S 4)$ the HISH model. We  present the map between the holographic strings and strings in flat space-time in $(\S 4.1)$. We classically solve the system of an open string with massive endpoints and we determine its energy and angular momentum $(\S 4.2)$. We then in $(\S 4.3)$ discuss the stringy baryon of HISH and in particular the stability of Y shape string configurations. Next in section $(\S 3.1)$ we present our current understanding of the quantization of the HISH model.
In $\S (
5.1)$ we review the attempt to quantize  the closed string in holographic background. We then describe the derivation of the    Casimir energy for a static case with massive endpoints  in $(\S 5.2)$. The Liouville or Polchinski-Strominger terms for quantizing the string in non-critical dimension is discussed in $(\S 5.3)$.
Phenomenology: Comparison between the stringy models and experimental data is reviewed in section $(\S 6)$. We first describe the fitting models $(\S 6.1)$ and procedure and then present separately the  meson trajectory  fits  $(\S 6.2)$and the baryon trajectory fits$(\S 6.3)$ and finally we describe the search of glueballs $(\S 6.4)$. We summarize and describe several future directions in $(\S  7)$ 
\section{From \texorpdfstring{$AdS_5\times S^5$}{AdS5xS5} to confining string backgrounds}
The original duality equivalence is  between the ${\cal N}=4$ SYM theory and string theory in $AdS_5\times S^5$.
 Obviously the  ${\cal N}=4$ is not the right framework to describe hadrons that should resemble those found in nature. Instead we need  stringy dual of   four dimensional  gauge dynamical system which is  non-supersymmetric, non-conformal and  confining. The main two requirements on  the desired string background is that it  (i) admits confinement and (ii) that it includes matter in the fundamental representation invariant under chiral flavor symmetry  and that the latter is spontaneously broken.
 \subsection{Confining background}
  There are by now  several ways to get a string background which is dual to a confining boundary field theory.
 \begin{itemize}
 \item
Models based on deforming the $AdS_5\times S^5$ by a relevant or marginal operator
which breaks conformal invariance and supersymmetry. This approach was pioneered in  \cite{Polchinski:2000uf}. There were afterwards many followup papers. For a list of references of them  see for instance in the review paper \cite{Aharony:2002up}.
\item
An important class of models is achieved by compactifing  higher dimensional theories
 to four dimensions in a way which (partially) breaks supersymmetry. A prototype model of this approach is that of a compactified D5 or  NS5 brane on $S^2$ . In an appropriate decoupling limit this provides a dual of  the $3 + 1$ dimensional ${\cal N}=1$ SYM theory\cite{Maldacena:2000yy}. Another model of this class is the so called Witten's model \cite{Witten:1998zw} which  is based on the compactification of
  one space coordinate  of a D4 brane on a circle. The sub-manifold spanned by the compactified coordinate and the radial coordinate  has   geometry of a cigar.  Imposing  on the circle  anti-periodic boundary conditions for the fermions, in particular  the gauginos, render them  massive and hence supersymmetry is broken. In the limit of small compactified radius one finds that the dual field theory is a contaminated low energy effective YM theory in four dimensions.  This approach will be described in $(\S 2.3  )$
\item
It was realized early in the game that one can replace the $S^5$ part of the string background with orbifolds of it or by the $T_{11}$ conifold. In this way conformal invariance is maintained, however, parts of the supersymmetry can be broken. In particular the dual of the conifold model has ${\cal N}=1$ instead of ${\cal N}=4$. It was shown in \cite{Klebanov:2000hb}
that one can move from the conformal theory to a confining one by deforming the conifold.  The corresponding gauge theory is a cascading gauge theory.
\item
 In analogy to $Dp$ brane background solutions  of the ten dimensional equations of motion \cite{Itzhaki:1998dd} one can find solutions for the metric dilaton and RR forms in non-critical $d<10$ dimensions.  In particular  there is a six dimensional model of $D4$ branes compactified on a $S^1$ \cite{Kuperstein:2004yf}. This model resembles Witten's model with the difference that the dual field theory is in the large $N_c$ limit but  with 't Hooft coupling $\lambda \sim 1$.

\item
The AdS/QCD is a bottom-up  approach, namely it not a solution of the ten dimensional equations of motion,   based on an $AdS_5$ gravity  background with additional fields residing on it. The idea is to determine the  background  in such a way that the corresponding dual boundary field theory has properties that resemble those of QCD.
 The basic model of this kind is the ``hard wall" model where the  five-dimensional $AdS_5$ space is truncated at a certain value of the radial direction $u=u_\Lambda$. The conformal invariance of the $AdS_5$ space corresponds to the asymptotic free UV region of the gauge theory. Confinement is achieved by the IR hard wall as will be seen $(\S 3.1)$. This construction was first introduced in \cite{Polchinski:2002jw}. Since this  model does not admit a Regge behaviour for the corresponding meson spectra there was a proposal \cite{oai:arXiv.org:hep-ph/0602229}  to improve the model by ``softening" the hard wall into a soft-wall model. It was argued that in this way one finds an excited spectrum of the form $M^2_n\sim n$ but, as will be discussed below also, this model does not describe faithfully the spectra of mesons.
\item
Another   AdS/QCD  model is the improved holographic QCD model (IHQCD)\cite{Gursoy:2007cb} which is essentially a five-dimensional dilaton-gravity system with a non-trivial dilaton potential. By tuning the latter, it was shown that one can build a model that admits certain properties in accordance with lattice gauge theory results both at zero and finite temperature.

\end{itemize}

\subsection{Introducing fundamental quarks}

Since the early days of string theory  it has been understood that fundamental quarks should correspond to open strings. In the modern era of closed string  backgrounds  this obviously calls for D branes.
   It is thus natural
to wonder, whether one can consistently add D brane probes to
supergravity backgrounds duals
of confining gauge theories, which will play the role of fundamental quarks.
In case that the number of D brane probes $N_f << N_c$   one can convincingly argue that the
back reaction of the probe branes on the bulk geometry is negligible.
It was also well known that open strings between parallel $N_f$ D7 branes and $N_c$ D3 branes
  play the role of flavored quarks in the $SU(N_c)$
  gauge theory that resides on  the D3 4d world volume. Karch and Katz \cite{Karch:2002sh}
 proposed to elevate this brane configuration into a supergravity background by introducing
D7 probe  branes  into the $AdS_5\times S^5$ background. Note that these types of  setups are  non-confining ones.
This idea was further explored in \cite{Kruczenski:2003be} and in many other followup works. For a list of them see the review \cite{Erdmenger:2007cm}. There have been certain attempts to go beyond the flavor probe approximation for instance \cite{Burrington:2004id}. The first attempt to incorporate flavor branes in a confining background was in \cite{Sakai:2003wu} where D7 and anti-D7 branes were introduced to the Klebanov-Strassler model \cite{Klebanov:2000hb}. This project was later completed in \cite{Dymarsky:2009cm}. An easier model of incorporating $D8$ and anti $D8$ branes will be discussed in the next subsection.
 Flavor was also introduced in the bottom-up models confining models. For instance the Veneziano limit of large $N_f$ in addition to the large $N_c$ was studied in \cite{Jarvinen:2011qe}.

\subsection{Review of the  Witten-Sakai-Sugimoto model}
Rather than describing a general confining background, or the whole list of such models,  we have chosen  a prototype model, the Sakai Sugimoto model, which will be  described in details in the next subsection.

Witten's model \cite{Witten:1998zw}  describes  the near-horizon
limit of a configuration of large  $N_c$ number of  D4-branes, compactified on a  circle in the
$x_4$ direction ( see figure 1)  with anti-periodic  boundary conditions for the fermions\cite{Witten:1998zw}.
\begin{figure}[ht!] \centering
					\includegraphics[width=.86\textwidth]{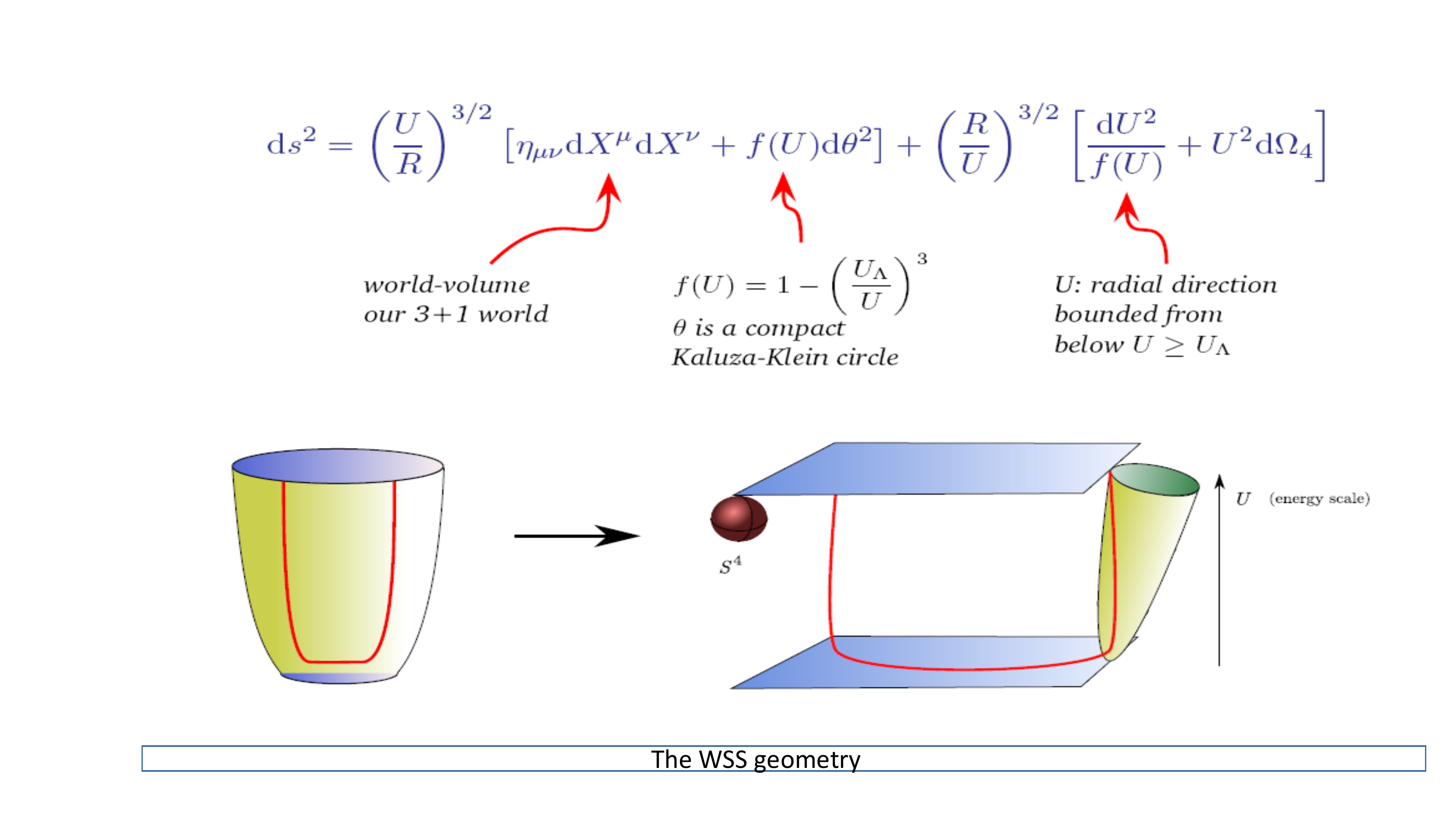}
					\caption{\label{ WSSgeometry}The WSS geometry. The compact direction of the D4 brane denoted in the figure as $\theta$ is referred to as $x_4$ in the text.}
\end{figure}

 This breaks supersymmetry and renders only the gauge field to remain massless whereas the gauginos and scalars become massive. To incorporate flavor  this model is then elevated to the so called   Sakai Sugimoto model \cite{SakSug} to which  a stack of
 $N_f$
D8-branes (located at $x_4=0$) and a stack of  $N_f$ anti-D8-branes (located at the asymptotic location
$x_4=L_{as}$) is added. This is dual to a $4+1$ dimensional maximally
supersymmetric $SU(N_c)$ Yang-Mills theory (with coupling constant $g_5$ and
with a specific UV completion that will not be important for us),
compactified on a circle of radius $R$ with anti-periodic boundary conditions
for the fermions, with $N_f$ left-handed quarks located
at $x_4=0$ and $N_f$ right-handed quarks located at $x_4=L$ (obviously we
can assume $L_{as} \leq \pi R$).

In the limit $N_f \ll N_c$, this background can be described by
$N_f$ probe D8-branes inserted into the
near-horizon limit of a set of $N_c$
D4-branes compactified on a circle with anti-periodic boundary conditions
for the fermions. This background is simply related to the
(near-horizon limit of the) background of near-extremal D4-branes by
exchanging the roles (and signatures) of the time direction and of one
of the spatial directions.
Let us now briefly review this
model, emphasizing the manifestations of confinement and chiral
symmetry breaking.  The background of type IIA string theory
is characterized by the metric, the
RR four-form and a dilaton given by
\bea\label{SSmodel} ds^2&=&\left(
\frac{u}{R_{D4}} \right)^{3/2}\left [- dt^2 +\delta_{ij}dx^i dx^j +
f(u) dx_4^2 \right ] +\left( \frac{R_{D4}}{u} \right)^{3/2} \left [
\frac{du^2}{f(u)} + u^2 d\Omega_4^2 \right ],
\CR
F_{(4)}&=& \frac{2\pi N_c}{V_4}\epsilon_4, \quad
e^\phi = g_s\left( \frac{u}{R_{D4}} \right)^{3/4},
\quad
 R_{D4}^3 \equiv \pi g_s N_c l_s^3,\quad
f(u)\equiv 1-\left( \frac{u_\Lambda}{u} \right)^3,
\eea
where $t$ is the time direction and $x^i$ ($i=1,2,3$) are the
 uncompactified world-volume coordinates of the D4 branes, $x_4$ is a
 compactified direction of the D4-brane world-volume which is transverse to
 the probe D8 branes, the volume of the unit four-sphere $\Omega_4$ is
 denoted by $V_4$ and the corresponding volume form by $\epsilon_4$,
 $l_s$ is the string length and finally $g_s$ is a parameter related
 to the string coupling.  The submanifold of the background spanned by
 $x_4$ and $u$ has the topology of a cigar (as in both  sides of
 figure \ref{D8-barD8SS} below) where the minimum value of $u$ at
 the
 tip of the
 cigar is $u_\Lambda$. The tip of the cigar is non-singular if and
 only if the periodicity of $x_4$ is
\be
\label{relru}
\delta
 x_4 = \frac{4\pi}{3}\left( \frac{R_{D4}^3}{u_\Lambda} \right)^{1/2} = 2\pi R
 \ee
and we identify this with the periodicity of the circle that the
$4+1$-dimensional gauge theory lives on.

The parameters of this gauge theory, the five-dimensional gauge coupling
$g_5$, the low-energy four-dimensional
 gauge coupling $g_4$, the glueball mass scale $M_{gb}$, and the
 string tension $T_{st}$ are determined from the background
 (\ref{SSmodel}) in
 the following form  :
\bea\label{stringauge}
g_5^2&=&(2\pi)^2 g_s l_s,\qquad
g^2_{4}=\frac{g_5^2}{2\pi R}=
3\sqrt{\pi}\left ( \frac{g_s u_\Lambda}{N_c l_s}\right )^{1/2},
 \qquad
M_{gb} = \frac{1}{R},
\CR T_{st} &=& \frac{1}{2\pi
 l_s^2}\sqrt{g_{tt}g_{xx}}|_{u=u_\Lambda}= \frac{1}{2\pi l_s^2}\left(
 \frac{u_\Lambda}{R_{D4}} \right)^{3/2} =\frac{2}{27\pi} \frac{g^2_4 N_c}{R^2}
= \frac{\lambda_5}{27\pi^2 R^3},\eea
where $\lambda_5 \equiv g_5^2 N_c$, $M_{gb}$ is the typical scale of
the glueball masses computed from the spectrum of excitations around
(\ref{SSmodel}), and $T_{st}$ is the confining string tension in this
model (given by the tension of a fundamental string stretched at $u=u_{\Lambda}$
where its energy is minimized). The gravity approximation is valid
whenever $\lambda_5 \gg R$, otherwise the curvature at $u \sim
u_{\Lambda}$ becomes large. Note that as usual in gravity
approximations of confining gauge theories, the string tension is much
larger than the glueball mass scale in this limit. At very large values of $u$ the
dilaton becomes large, but this happens at values which are of order
$N_c^{4/3}$ (in the
large $N_c$ limit with fixed $\lambda_5$), so this will
play no role in the large $N_c$ limit that we will be interested in.
The Wilson line of this gauge theory (before putting in the D8-branes)
admits an area law behavior \cite{Brandhuber:1998er}, which means a confining behavior, as can be easily seen using
the conditions for confinement of \cite{Kinar:1998vq}.

Naively, at energies lower than the Kaluza-Klein scale $1 / R$ the
 dual gauge theory is effectively four dimensional; however, it turns out
that the theory confines and develops a mass gap of order $M_{gb}=1/R$, so (in the
regime where the gravity approximation is valid) there
is no real separation between the confined four-dimensional fields and
the higher Kaluza-Klein modes on the circle. As discussed in \cite{Witten:1998zw},
in the opposite limit of $\lambda_5 \ll R$, the theory approaches the
$3+1$ dimensional pure Yang-Mills theory at energies small compared to
$1/R$, since in this limit the scale of the mass gap is exponentially
small compared to $1/R$. It is believed that there is no phase transition
when varying $\lambda_5/R$ between the gravity regime and the pure
 Yang-Mills regime, but it is not clear how to check this.

Next, we introduce
the probe 8-branes which
span the coordinates $t, x^i, \Omega_4$, and follow some curve $u(x_4)$
in the $(x_4,u)$-plane. Near the boundary at $u\to \infty$ we want to have
$N_f$ D8-branes localized at $x_4=0$ and $N_f$ anti-D8-branes (or D8-branes
with an opposite orientation) localized at $x_4=L_{as}$. Naively one might think
that the D8-branes and anti-D8-branes would go into the interior of the space
and stay disconnected; however, these 8-branes
do not have anywhere to end in the background (\ref{SSmodel}), so the
form of $u(x_4)$ must be such that the D8-branes smoothly connect to the
anti-D8-branes (namely, $u$ must go to infinity at $x_4=0$ and at $x_4=L_{as}$,
and $du/dx_4$ must vanish at some minimal $u$ coordinate $u=u_0$). Such
a configuration spontaneously
breaks the chiral symmetry from the symmetry group which is
visible at large $u$,
$U(N_f)_L\times U(N_f)_R$, to the diagonal $U(N_f)$ symmetry.
Thus, in this configuration the topology forces a breaking of the chiral
symmetry; this is not too surprising since chiral symmetry breaking at
large $N_c$ follows from rather simple considerations.
The most important feature of this solution is the fact that the D8
branes smoothly connect to the anti D8 branes.

In order to find the 8-brane configuration, we need
the induced metric on the D8-branes, which is
\bea
\label{induced}
ds^2_{D8}&=&\left( \frac{u}{R_{D4}} \right)^{3/2}\left [ - dt^2+  \delta_{ij}dx^{i}dx^j \right ]
+\left( \frac{u}{R_{D4}} \right)^{3/2} \left [f(u) + \left( \frac{R_{D4}}{u} \right)^{3} \frac{{u'}^2}{f(u)}\right ]dx_4^2 \CR
 &+&
 \left( \frac{R_{D4}}{u} \right)^{3/2} u^2 d\Omega_4^2
\eea
where $u'=du/dx_4$.
It is easy to check that the Chern-Simons (CS) term in the D8-brane action does not affect the
solution of the equations of motion.  More precisely, the equation
of motion of the gauge field has a classical solution of a vanishing
gauge field, since the CS term includes terms of the form $C_5\wedge
F\wedge F$ and $C_3 \wedge F\wedge F\wedge F$. So, we are left only with
the Dirac-Born-Infeld (DBI) action.  Substituting the determinant of the induced metric
and the dilaton into the DBI action, we obtain
(ignoring the factor of $N_f$ which multiplies all the D8-brane
actions that we will write) :
\bea
\label{eightaction}
S_{DBI} =  T_8 \int dt d^3 x d x_4 d^4\Omega e^{-\phi}
\sqrt{-\det(\hat g)}
        =  \frac{\hat T_8 }{g_s}\int  dx_4 u^4 \sqrt{f(u)+
\left( \frac{R_{D4}}{u} \right)^{3} \frac{{u'}^2}{f(u)}},
\eea
where $\hat g$ is the induced metric (\ref{induced}) and $\hat T_8$ includes the
 outcome of the integration over all the coordinates apart from $d
 x_4$.  From this action it is straightforward to determine the profile of the D8 probe branes.
The form of this profile  which  is drawn in figure
\ref{D8-barD8SS}(a) is given by
\bea
\label{lzerotemp}
L_{as}&=&\int dx_4 = 2\int_{u_0}^{\infty} \frac{du}{u'}=
2 R^{3/2}_{D4}
\int_{u_0}^\infty du \frac{1}{f(u)u^{3/2}\sqrt{ \frac{f(u) u^8}{f(u_0) u_0^8}
-1}} \CR
&=&  \frac{2}{3} \left( \frac{R^3_{D4}}{u_0} \right)^{1/2}
\sqrt{1-y_\Lambda^3}\int_0^1 dz \frac{z^{1/2}}{(1-y_\Lambda^3 z)
\sqrt{1-y_\Lambda^3 z -(1-y_\Lambda^3)z^{8/3}}},
\eea
where $y_\Lambda \equiv u_{\Lambda} / u_0$.
Small values of $L_{as}$ correspond to large values of $u_0$. In this limit
we have $y_{\Lambda}\ll 1$ leading to $L\propto \sqrt{R_{D4}^3/u_0}$.
For general values of $L_{as}$ the dependence of $u_0$ on $L$ is more complicated.

\begin{figure}[t]
\begin{center}
\vspace{3ex}
\includegraphics[width=.65\textwidth]{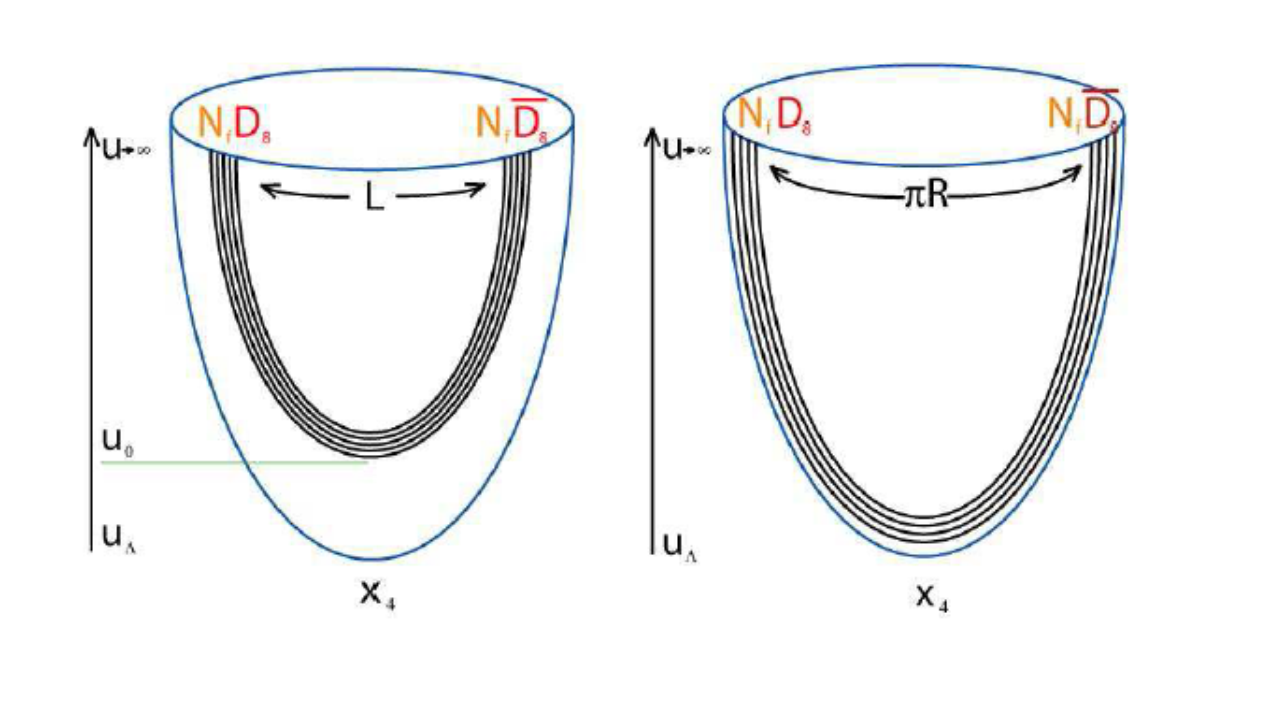}
\end{center}
\caption{The dominant configurations of the D8 and anti-D8 probe
branes in the Sakai-Sugimoto model at zero temperature,
which break the chiral symmetry. The same configurations will turn
out to be relevant also at low temperatures.
On the left a generic configuration with an asymptotic separation of
$L_{as}$, that stretches down to a minimum at $u=u_0$, is drawn. The figure
on the right describes the limiting antipodal case $L_{as}=\pi R$, where
the branes connect at $u_0=u_{\Lambda}$. \label{D8-barD8SS}}
\end{figure}

There is a simple special case of the above solutions, which occurs
when $L_{as} = \pi R$, namely the D8-branes and anti-D8-branes lie at
antipodal points of the circle. In this case the solution for the
branes is simply $x_4(u) = 0$ and $x_4(u)=L_{as}=\pi R$, with the two branches
meeting smoothly at the minimal value $u=u_0=u_{\Lambda}$ to join
the D8-branes and the anti-D8-branes together (see figure 1). This is the solution advocated in \cite{SakSug}. The generalized, not necessarily antipodal configuration that was described above was introduced in \cite{Aharony:2006da}. As will be discussed in section $\S(4,2)$ the difference between the antipodal and the non-antipodal will translate to the difference between stringy meson with massless versus massive endpoints. It was also found out in \cite{Kaplunovsky:2010eh} that, in order to have attraction between flavor instantons that mimic the baryons, the setup has to be non-antipodal.
 In the approach of considering the duals of the mesons as fluctuation modes of the flavor branes\cite{SakSug}, and not as we argue here in this review as string configuration, it turned out that the difference between $u_0$ and $u_\Lambda$ is not the dual of the QCD quark masses. One manifestation of  this is the fact that the Goldstone bosons associated  with the breakdown of the chiral flavor symmetry remain massless even for the non-antipodal case. This led to certain generalizations of the Sakai Sugimoto model  by introducing additional adjoint ``tachyonic"' field into the bulk \cite{Bergman:2007pm},\cite{Casero:2007ae},
\cite{Dhar:2007bz} and by introducing an open Wilson line\cite{Aharony:2008an}.  In view of the HISH these type of generalization are obviously not necessary.

This type of antipodal solution is drawn in figure
\ref{D8-barD8SS}(b).

 It was shown in \cite{Burrington:2007qd} that the classical
configurations both the antipodal and the non-antipodal are   stable.  This was done by a perturbative analysis
 (in $g_sN_f$) of the  backreaction of the  localized D8 branes. The explicit expressions of the backreacted metric, dilaton and RR form were written down and it was found that the backreaction remains small up to a radial value of $u <<  l_sN_f$), and that the background functions are smooth except at the D8 sources. In this perturbative window, the original embedding remains a solution to the equations of motion. Furthermore, the fluctuations around the original embedding, do not become tachyonic due to the backreaction in the perturbative regime. This is is due to a cancellation between the DBI and CS parts of the D8 brane action in the perturbed background.
For further discussion of the pros and cons of the Sakai-Sugimoto model see \cite{Aharony:2006da}.

  The main  results reviewed in this subsection hold  also to an analogous non-critical setup \cite{Kuperstein:2004yf} based on inserting D4 flavor branes into the background of compactified colored D4 branes and in particular the structure of the spontaneous breaking of the flavor chiral symmetry.


\section{Hadrons as strings in holographic background}
 In this section we will analyze various  classical string configurations in confining holographic models. In fact we will define below what is a confining string background according to the  behavior of the classical static string that attaches its boundary.
We will treat four types of strings:

(i) The  string  dual of a Wilson line.

(ii) A rotating  open string that is attached to flavor branes. This will be the dual of a meson.

(iii) A rotating configuration of $N_c$ strings attached to a ``baryonic vertex", the dual of the baryon.

(iv) A rotating folded closed string, the dual of the glueball.

\subsection{The holographic Wilson line\label{sec:The holographic Wilson line}}
One of the most important characteristics of non-abelian  gauge theories, in particular the ones  associated with the group $SU(N_c)$,  is  the Wilson line which is  a non-local gauge invariant expectation value that takes the form
\be
\langle W(C) \rangle = \frac{1}{N_c} Tr\left [ P\ e^{\oint_C A }\right ]= \frac{1}{N_c} Tr\left [ P\ e^{\oint_C A_\mu \dot x^\mu(\tau) d\tau} \right ]
\ee
where $P$ denotes path ordering and $C$ is some given contour.

\begin{figure}[ht!] \centering
					\includegraphics[width=.96\textwidth]{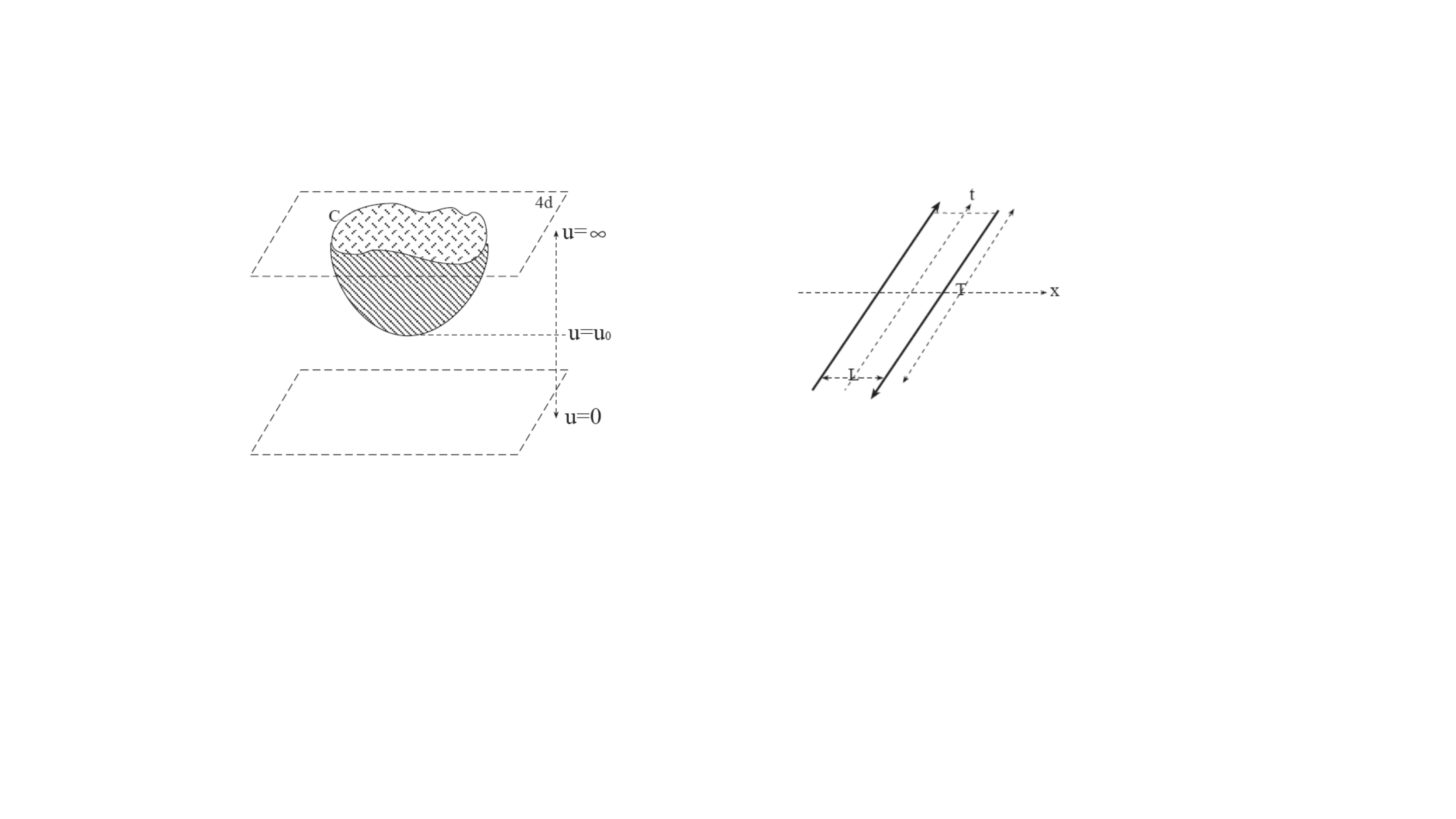}
					\caption{\label{ WSSgeometry}The geometry of the Wilson line. On the right the strip Wilson loop  in real space-time that corresponds to a quark anti-quark potential and on the left the holographic description of a Wilson loop of a general contour $C$}
\end{figure}

 For the special case where $C$ is a strip of length $L$ along one space direction  and $\Delta t$ along  the time direction, one can extract from the corresponding Wilson line, in Euclidean space-time  and for $\Delta t>> L\rightarrow \infty$, the potential $E(L)$ between a quark and an anti-quark as follows
\be
\langle W(strip, L,\Delta t) \rangle = A(L) e^{-\Delta t E(L)}
\ee
The holographic dual of the expectation value of the Wilson line which was determined in \cite{Maldacena:1998im} is given by
\be
\langle W(C) \rangle = e^{ -S^{ren}_{NG}}
\ee
where $S^{ren}_{NG}$ is the renormalized Nambu-Goto action of a string worldsheet whose boundary on the boundary of the bulk space-time is $C$. There are several methods  of renormalizing the result. In particular in the one we discuss below one subtracts the infinite action of the straight strings which  are duals of the masses of quark anti-quark pair.

Next we would like to apply the holographic prescription of computing the Wilson line to a particular  class of
 holographic models. The latter   are characterized by 
 higher than five  dimensional space-time  with a boundary.  The coordinates of these space-times include  the coordinates of the boundary space-time,    a radial coordinate and additional  
 coordinates transverse to the boundary and to  the radial direction. 
We  assume that the corresponding  metric depends only on the radial
coordinate such that its general form is \begin{equation}\label{metric}
ds^{2}=-g_{tt}\left(u\right)dt^{2}+g_{uu}\left(u\right)du^{2}+g_{x_{||}x_{||}}\left(u\right)dx_{||}^{2}+g_{x_{T}x_{T}}\left(u\right)dx_{T}^{2}\end{equation}
 where $t$ is the time direction, $u$ is the radial coordinate,
$x_{||}$ are the space  coordinates on the boundary and $x_{T}$ are the
transverse coordinates. We adopt the notation in which the radial
coordinate is positive defined and the boundary is located at $u=\infty$.
In addition, a ``horizon'' may exist at $u=u_{\Lambda}$, such
that spacetime is defined in the region $u_{\Lambda}<u<\infty$, instead
of $0<u<\infty$ as in the case where no horizon is present.

The construction we  examine is that of the strip discussed above. This takes the form   of an open string living in the bulk of the space with both its ends tied to the boundary.  From the
viewpoint of the field theory living on the boundary the endpoints
of the string are the $q\bar{q}$ pair, so the energy of the string
is related to the energy of the pair. In order to calculate
the energy of this configuration on the classical level we use
the notations and results of \cite{Kinar:1998vq}. First, we define
\begin{equation}\label{fg}
\begin{aligned}f^{2}\left(u\right) & \equiv g_{tt}\left(u\right)g_{x_{||}x_{||}}\left(u\right)\\
g^{2}\left(u\right) & \equiv g_{tt}\left(u\right)g_{uu}\left(u\right)\end{aligned}
\end{equation}
 Upon choosing the worldsheet coordinates $\sigma=x$ ($x$ is a coordinate
on the boundary pointing in the direction from one endpoint of the
string to the other one) and $\tau=t$. This is obviously the most natural gauge choice for the case of the strip Wilson line. If in addition we assume translation invariance
along $t$, the Nambu-Goto action describing the string takes the
form\begin{equation}
\begin{aligned}S_{NG} & =T\int d\sigma d\tau\sqrt{det\left[\partial_{\alpha}X^{\mu}\partial_{\beta}X^{\nu}g_{\mu\nu}\right]}\\
 & =T\int dxdt\sqrt{f^{2}\left(u\left(x\right)\right)+g^{2}\left(u\left(x\right)\right)\left(\partial_{x}u\right)^{2}}\\
 & =T(\Delta t)\int dx\mathcal{L}\end{aligned}
\end{equation}
where we assumes a static configuration,   $\Delta t$ is the length of the strip along the time direction and $T$ is the string tension which from here on we set to be one. Later we will bring it back again. Letting the derivative with respect to $x$, $\pa_x$, playing the role of the time derivative in standard canonical procedure,  then the conjugate momentum and the Hamiltonian are
\begin{equation}
p=\frac{\delta\mathcal{L}}{\delta u'}=\frac{g^{2}\left(u\right)u'}{\sqrt{f^{2}\left(u\right)+g^{2}\left(u\right)u'^{2}}}\label{eq:momentum}\end{equation}
 \begin{equation}
\mathcal{H}=p\cdot u'-\mathcal{L}=-\frac{f^{2}\left(u\right)}{\mathcal{L}}\end{equation}
 As the Hamiltonian does not depend explicitly on $x$, its value
is a constant of motion. We shall deal with the case in which $u\left(x\right)$
is an even function, and therefore there is a minimal value $u_{0}=u\left(0\right)$
for which $u'\left(0\right)=0.$ At that point we see from \eqref{eq:momentum}
that $p=0$. The constant of motion is therefore \begin{equation}
\mathcal{H}=-f\left(u_{0}\right)\end{equation}
 from which we can extract the differential equation of the geodesic
line \begin{equation}
\frac{du}{dx}=\pm\frac{f\left(u\right)}{g\left(u\right)}\frac{\sqrt{f^{2}\left(u\right)-f^{2}\left(u_{0}\right)}}{f\left(u_{0}\right)}\label{eq:geodesic}\end{equation}
 and re-express the on-shell Lagrangian (i.e. the Lagrangian on the
equation of motion) as a function of $f\left(u\right)$ only \begin{equation}
\mathcal{L}=\frac{f^{2}\left(u\right)}{f\left(u_{0}\right)}\end{equation}
 Then the distance between the string's endpoints (or the distance
between the ``quarks'') is
\begin{equation}
\ell\left(u_{0}\right)=\int dx=\int du\left(\frac{du}{dx}\right)^{-1}=2\frac{1}{f\left(u_{0}\right)}\int_{u_{0}}^{u_{b}}du\frac{f^{2}\left(u_{0}\right)}{f^{2}\left(u\right)}\frac{g\left(u\right)}{\sqrt{1-\left(\frac{f\left(u_{0}\right)}{f\left(u\right)}\right)^{2}}}\label{eq:length}\end{equation}
 where $u_{0}$ is the minimal value in the radial direction to which
the string reaches and $u_{b}$ is the value of $u$ on the boundary.
The bare energy of the string is given by\begin{equation}
\begin{aligned}E_{bare}\left(u_{0}\right) & =\int dx\mathcal{L}=\int du\left(\frac{du}{dx}\right)^{-1}\mathcal{L}=2\int_{u_{0}}^{u_{b}}du\frac{g\left(u\right)}{\sqrt{1-\left(\frac{f\left(u_{0}\right)}{f\left(u\right)}\right)^{2}}}\\
 & =f\left(u_{0}\right)\cdot\ell\left(u_{0}\right)+2\int_{u_{0}}^{u_{b}}du\, g\left(u\right)\sqrt{1-\frac{f^{2}\left(u_{0}\right)}{f^{2}\left(u\right)}}\end{aligned}
\label{eq:bareEnergy}\end{equation}
 Generically, the bare energy diverges and hence a renormalization
procedure is needed. There are several renormalization schemes to deal with this infinity. These were summarized in \cite{Kol:2010fq}. Here we follow \cite{Kinar:1998vq}
and use the mass subtraction scheme in which the bare masses of the
quarks are subtracted from the bare energy. The bare external quark  mass is
viewed as a straight string with a constant value of $x$, stretching
from $u=0$ (or $u=u_{\Lambda}$ if there exists an horizon at $u_{\Lambda}$)
to $u=u_{b}$, such that it is given by\footnote{ We refer here to the dual of string that stretches to the boundary as an external quark to distinguish it from a dynamical quark whose dual is a  string that ends  on a flavor brane as will be explained in the next section}
\begin{equation}
m_{q,ext}=\int_{u_{\Lambda}}^{u_{b}}du\, g\left(u\right)
\end{equation}
 Then the renormalized energy would be given by\begin{equation}
E\left(u_{0}\right)=f\left(u_{0}\right)\cdot\ell\left(u_{0}\right)-2\mathcal{K}\left(u_{0}\right)\label{eq:Energy}\end{equation}
 where $\mathcal{K}\left(u_{0}\right)$ is\begin{equation}
\mathcal{K}\left(u_{0}\right)=\int_{u_{0}}^{u_{b}}du\, g\left(u\right)\left(1-\sqrt{1-\frac{f^{2}\left(u_{0}\right)}{f^{2}\left(u\right)}}\right)+\int_{u_{\Lambda}}^{u_{0}}du\, g\left(u\right)\label{eq:Kappa}\end{equation}
It is important to  emphasize
that this result for the energy is only at the classical level and
does not include quantum corrections.

In order to reproduce the QCD heavy quarks potential we first have
to demand the holographic models to reproduce the asymptotic forms
of the potential. This leads to several restrictions on the forms
of the $f$ and $g$ functions. The condition for confining behavior
at large distances was derived in \cite{Kinar:1998vq}:
\begin{enumerate}
\item $f$ has a minimum at $u_{min}$ and $f\left(u_{min}\right)\neq0$
or
\item $g$ diverges at $u_{div}$ and $f\left(u_{div}\right)\neq0$
\end{enumerate}
Then the string tension is given by $f\left(u_{min}\right)$ or $f\left(u_{div}\right)$,
correspondingly. The second asymptotic is perturbative QCD at small
distances.
The conditions on the background to reproduce the leading perturbative
behavior of QCD, which is Coulomb-like, were derived in \cite{Kol:2010fq} .

The physical picture arising from this construction is as follows.
The confining limit is approached as $u_{0}\rightarrow u_{\Lambda}$,
then most of the string lies  at the vicinity of $u_\Lambda$
which  implies in the dual field theory  a string-like interaction between the {}``quarks''
 with a string tension $T_s$ (upon bringing back $T$ that was set above to one)
 \begin{equation}
T_s=  T f\left(u_{\Lambda}\right)\label{eq:string tension}\end{equation}
 and a linear potential\begin{equation}
E=T_s\cdot\ell-2\kappa\end{equation}
 where $\kappa=\mathcal{K}\left(u_{\Lambda}\right)$ is a finite constant.
On the other hand, the Coulomb-like  limit is approached as $u_{0}/u_{\Lambda}\rightarrow\infty$,
then the whole string is far away from $u_\Lambda$ and is ruled
by the geometry near the boundary. The conditions for such a behavior were analyzed in \cite{Kol:2010fq}.

\begin{figure} \centering
\includegraphics [natwidth=1200 bp, natheight=900bp,width= 0.4\textwidth]{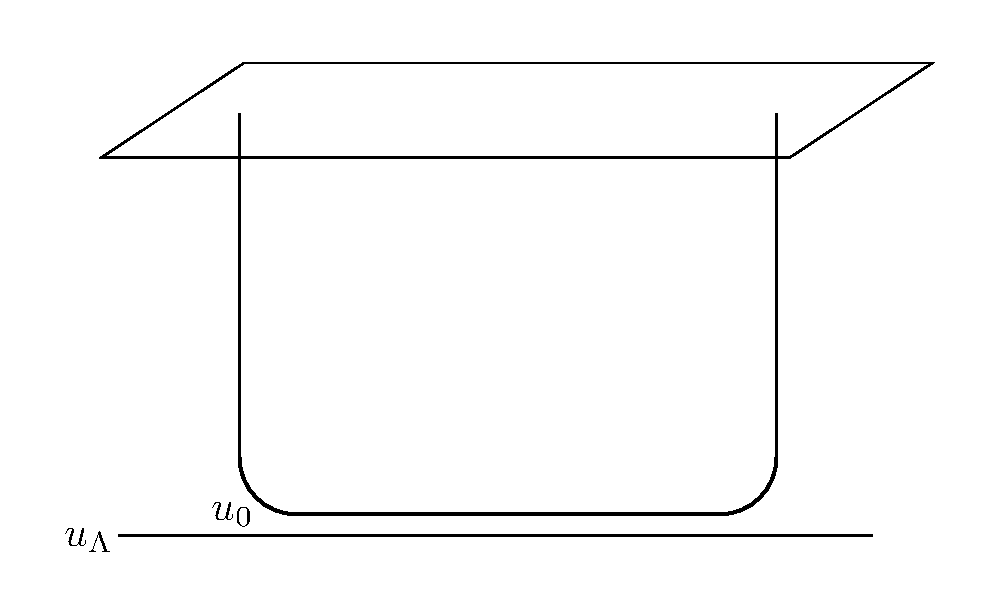}
\includegraphics[natwidth=1200 bp, natheight=900bp,width= 0.4\textwidth]{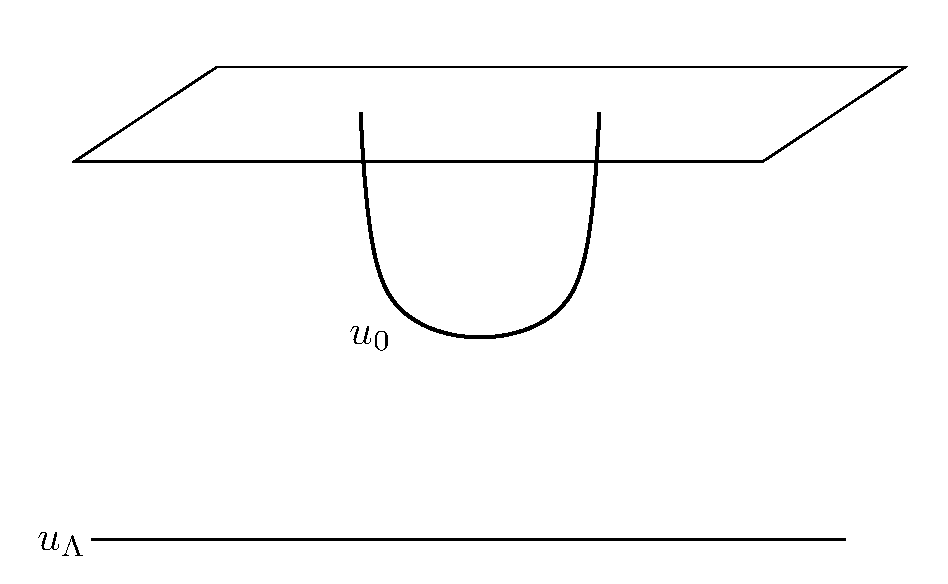}\caption{The physical picture arising from the present construction. Left:
The confining limit is approached as the string is close to the horizon,
then most of the string lies on the horizon and implies a string-like
interaction between the quarks (the endpoints of the string). Right:
The conformal limit is reconstructed when the string is far away from
the horizon such that it is ruled by the geometry near the boundary.}

\end{figure}

\subsubsection{'t Hooft line}
In addition to the Wilson line one discusses also the Polyakov line and the 't Hooft line. The former is a Wilson line for which the contour $C$ is the Euclidean time direction  compactified on a circle. Correspondingly the holographic string realization of the Polyakov loop is a string whose worldsheet wraps the compactified time direction. We will not discuss it here.
In YM theories  the ``electric-magnetic" dual of the Wilson line is
the 't Hooft line.  Just as one extracts the quark anti-quark potential
from the former with the strip contour, the later determines for the same contour the monopole anti-monopole potential.
 
We would like to address now the question of how  to
 construct a string configuration   which associates with the 't Hooft line. The answer for that question is that it should be a configuration of a $D1$ brane starting and ending on the boundary.
The prototype confining background described in $(\S 2.3)$ is that of a compactified D4 branes. This is a type II-A theory where the allowed D branes are only of the form $D2k$ and hence (at least in the supersymmetric case) one cannot embed a D1 brane. Instead we will take
 the stringy realization of  the 't Hooft line as
  a D2-brane ending on the
D4-brane.
 The D2-brane is wrapped  along the  compact space direction $x_4$ so from the point of view
of the four dimensional theory it is a point like object.
 The DBI action of a D2-brane is
\be
S=\frac{1}{(2\pi\alpha')^{3/2}}\int d\tau d\sigma_1 d\sigma_2 e^{-\phi}
\sqrt{det\ h},
\ee
where $h$ is the world-volume  induced metric and $\phi$ is the dilaton.
 For  the  D2-brane we consider,  which is infinite along one
direction ( we denote its length by $Y$ and winds the $x_4$
direction),  we get
\be
S=\frac{Y}{g_{YM}^2}\int dx \sqrt{\partial_x U^2+ \frac{u^3-u_\Lambda^3}{R_4^3}},
\ee
 Note the $1/g_{YM}^2$ factor which is expected for a monopole.
The distance between the monopole and the anti-monopole is
\be
L = 2 \frac{R_4^{3/2}}{u_0^{1/2}} \sqrt{\epsilon}
 \int_1^{\infty} \frac{dy}{\sqrt{(y^{3} - 1)(y^{3}
    -1 + \epsilon)}} ~.
\ee
where $\epsilon = 1 - (U_\Lambda/U_0)^3$.
The energy (after subtracting the energy corresponding to a free
monopole and anti-monopole) is
\be
E=\frac{2 u_0}{(2\pi)^{3/2}
g_{YM}^2}\left[ \int_{1}^{\infty}dy\left( \frac{\sqrt{y^3
        -1+\epsilon}}{{\sqrt{y^3
        -1}}}\right) -1 \right]    +\frac{2(u_\Lambda-u_0)}{(2\pi)^{3/2}}.
\ee
 For $L \Lambda \gg 1$
it is energetically favorable for the system to be in a configuration
of two parallel D2-branes ending on the horizon and wrapping $x_0$.
So in
the "YM region"  we find
 screening of the magnetic
charge which is another indication to confinement of the electric charge. For further details on the holographic determination of the 't Hooft line see for instance \cite{Sonnenschein:1999if}.

\subsection{ The stringy duals of mesons}
Next we want to identify among the strings that can reside in a holographic backgrounds those that correspond to mesons.
Naturally a meson  associates with an  open string and the quark and anti-quark that it is built from with the string endpoints. As we have seen in the previous section, endpoints on the boundary of the holographic bulk correspond to infinitely heavy external quarks.  Holographic backgrounds are associated with type II  closed string theories. In such backgrounds  open strings can start and end only on D-branes (or the boundary). Thus  ``dynamical quarks" should be related to string endpoints which are not on the boundary  but rather they   attach  to   D-branes.
As was discussed in section $\S (2.2)$ flavored confining backgrounds  include  stacks of $N_f$ D branes which are referred  to as ``flavor branes". We consider only the case of $N_f<< N_c$, where $N_c$ is the number of branes that constitute in the near horizon limit the holographic background. Since the endpoints can freely move, then classically the strings do not shrink to zero size only if they rotate and the ``centrifugal force" is balancing the string tension. Thus, to describe the stringy meson we will now discuss
  (i) The conditions on strings rotating at constant radial coordinate.  (ii) Rotating strings attached to flavor branes.

\subsubsection{The conditions on strings rotating at constant radial coordinate}
Consider  the  background metric of (\ref{metric}). Setting aside the transverse coordinates, the metric reads
 \be\label{metricnotrans}  ds^2 = -g_{00}(u)  dt^2
+ g_{ii}(u) ({dx^{i}})^2 + g_{uu} du^2 + ...  \ee where here we denote the metric along the space direction as $g_{ii}$, instead of $g_{||}$ as it was denoted in (\ref{metric}). The metric as well as any other fields of the
background depend only on the radial direction $
u$.
Since we have in mind addressing spinning strings, it is convenient to describe
the space part of the metric as
\be
dx_{i}^2 = dR^2 + R^2 d\t ^2 + dx_3 ^2,
\ee
where $x_3$ is the direction perpendicular to the plane of rotation.
The classical equations of motion of a bosonic string defined on this background can be
formulated on  equal footing in the NG formulation or the  Polyakov action.
Let us use now the latter.  The equations  of motion
associated with the variation of $t, \t, R $ and $u$  respectively are
\bea
&&\pa_\alpha ( g_{00}\pa^\alpha t )  = 0 \CR
&&\pa_\alpha ( g_{ii} R^2\pa^\alpha\t )  = 0 \CR
&&\pa_\alpha ( g_{ii}\pa^\alpha R) - g_{ii} R \pa_\alpha \t \pa^\alpha \t  = 0  \CR
&&2\pa_\alpha ( g_{uu}\pa^\alpha u) + \frac{d g_{00}}{du}  \pa_\alpha t \pa^\alpha t
-\frac{d g_{ii}}{du}  \pa_\alpha x^i \pa^\alpha x^i -  \frac{d g_{uu}}{du}  \pa_\alpha u \pa^\alpha u
= 0, \CR
\eea
where $\alpha$ denotes the worldsheet coordinates $\tau$ and $\sigma$.
In addition in the Polyakov formulation one has to add the Virasoro constraint
\be
-g_{00} (\pa_{\pm} t)^2+g_{ii} (\pa_{\pm} x^i)^2 + g_{uu} (\pa_{\pm} u)^2 + ... =0,
\ee
where $\pa_{\pm}= \pa_{\tau}\pm \pa _ \sigma$, and  $...$ stands for the contribution
to the Virasoro constraint of the rest of the background metric.

Next we would like to find solutions of the equations of motion which describe strings
spinning in space-time. For that purpose we take the following ansatz
\be\label{spinning}
t = \tau \qquad  \t = \omega \tau \qquad R(\sigma,\tau) = R(\sigma)\qquad u=\hat u= {\rm constant}.
\ee
It is obvious that this ansatz solves the first two equations. The third equation
together with the Virasoro constraint is  solved (for the case that $g_{00}=g_{ii}$) by
$R= A cos(\omega\sigma) + B sin(\omega\sigma)$  with $\omega^2 ( A^2 + B^2) =1$.  The
boundary conditions
we want to impose will select
the particular combination of $A$ and $B$.
Let us now investigate the equation of motion associated with $u$ and for the particular ansatz  $u=\hat u$.
This can be  a solution only provided
\be\label{condition}
\frac{d g_{00}}{du}|_{u= \hat u}= 0 \qquad  \frac{d g_{ii}}{du}|_{u= \hat u}=  0.
 \ee
This is just the first condition for having a confining background. The condition
$ g_{00} g_{ii}(\hat u)> 0$ insures that the Virasoro constraint is obeyed in a nontrivial manner.

We would like to check the condition in the class of confining models that was discussed in $\S (2.3)$ namely those of compactified D branes and   in particular the model of   the near
extremal $D4$ brane,  compactified  on an $S^1$. Denoting
the  direction
along the $S^1$ denoted
by $\psi$, the corresponding component of the metric   is  related to that of the $u$ direction as
$g_{\psi\psi }= [g_{uu}]^{-1}$. Thus the  condition for having a
solution of the equation of
motion with $u=\hat u$
includes the condition
\be
\frac{d g_{\psi\psi}}{du}|_{u= \hat u}= -\frac{\frac{d g_{uu}}{du}}{g_{uu}^2}|_{u= \hat u}= 0.
\ee
 This condition is obeyed if at $u=\hat u$   $g_{uu}(\hat u)\rightarrow \infty$, which
 is the second condition for a confining background  with $\hat u= u_\Lambda$, and  with the
the demand of non-vanishing $ g_{00} g_{ii}(\hat u)> 0$ to have a non-trivial Virasoro constraint.

To summarize, we have just realized that there is a close relation between the conditions of having area law
Wilson loop and of having a spinning string configuration at a constant radial coordinate.

\subsubsection{The meson as a rotating string attached to flavor branes}
\begin{figure}[ht!] \centering
					\includegraphics[width=.66\textwidth]{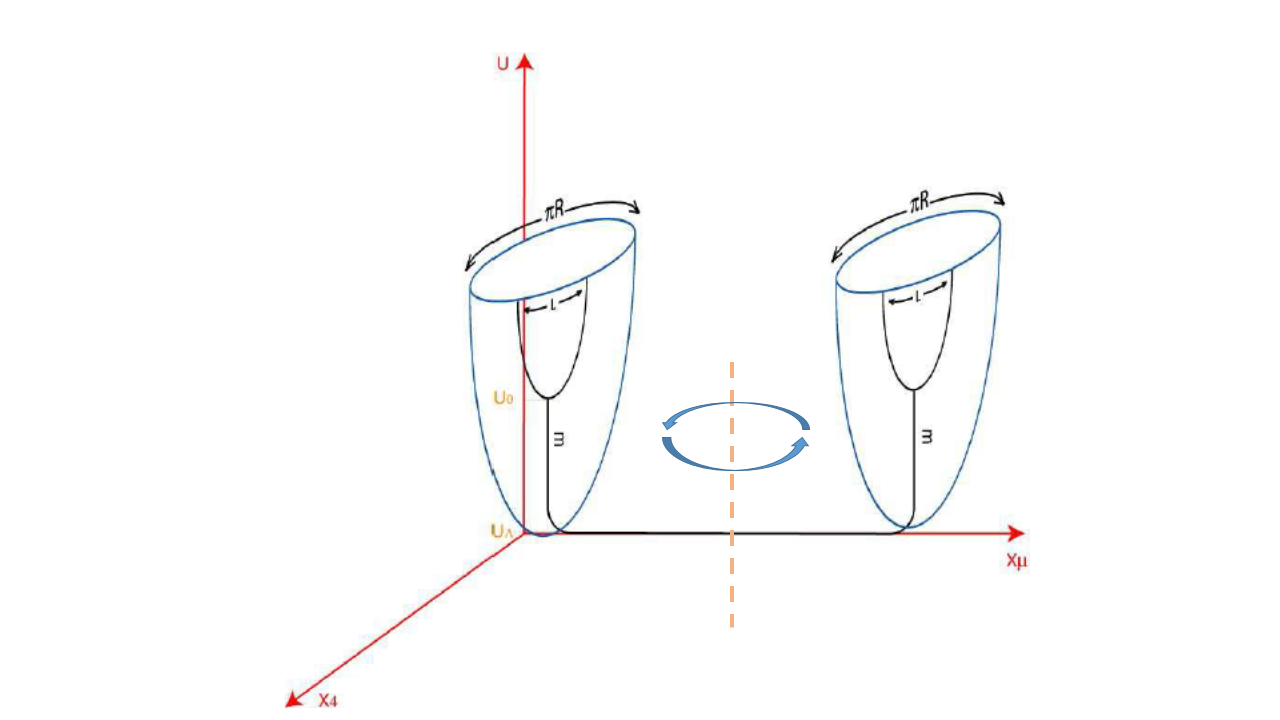}
					\caption{\label{rotstringss} A meson as a rotating open string that stretches from the tip of the flavor brane at two points separated in $X_i$ coordinates.}
\end{figure}

We have seen above the conditions for having a rotating string at constant $u$. We would like to examine  whether rotating strings with endpoints attached to flavor branes can obey these conditions. For the Wilson line in a ``confining background" we found  a string configuration that stretches vertically from  the boundary down to the vicinity of the ``wall"', then flattens along the wall and then ``climbs" up vertically to a flavor brane. In a very similar manner we search for a string of the same type of profile but now not a static one. When the endpoints of the open string are not on the boundary and are not nailed down, the string does not shrink due to the tension only if it rotates and the ``centrifugal force" balances the tension. Therefore we look for a   solution of the string equation of motion of the form of a rotating string in
  a plane spanned by the $x^i$ coordinates. Like the stringy Wilson line, the rotating string in the holographic background will include two vertical segments (region I) which connect the horizontal segment of the string (region II) to the flavor branes.
  Thus, the procedure to determine the holographic dual of the meson    includes a solution of the equation of motion in the vertical segment (region I), a solution in the horizontal segment (region II), matching the two solutions and finally computing the corresponding energy and angular momentum as the Noether charges associated with the  translation symmetry along the time and azimuthal
 directions. Before delving into the equations it should be useful  to have a picture of the `` mesonic rotating string". It appears in figure   (\ref{rotstringss}).

The Nambu-Goto action associated with a rotating string in a background metric of the form (\ref{metricnotrans}) takes the form

\be
S=- T_s\int d\sigma d\tau \sqrt{
((\partial_\tau X^0)^{2}- R^2 (\partial_\tau
\theta)^2)(f^2 (\partial_\sigma R)^2 +
g^2 (\partial_\sigma {u})^2)},
\label{actrot}
\ee
 where we assume here that $g_{00}= g_{xx}$ and where $f^2$ and $g^2$ are defined in  (\ref{fg}).
Since we have in mind analysing rotating strings we do not consider the most general configuration but rather an ansatz similar to the one discussed above (\ref{spinning})
\be\label{spinningf}
X^0 = X^0(\tau) \qquad  \t =  \t(\tau) \qquad R(\sigma,\tau) = R(\sigma)\qquad u= u(\sigma).
\ee
The boundary conditions corresponding to this configuration are Neumann in the directions
parallel to the flavor brane, and Dirichlet in the directions transverse to the brane, namely:
\bea
&&\bigg(
\partial_\sigma X^0=\partial_\sigma\theta=\partial_\sigma R=\partial_\sigma
\lambda\bigg)\bigg|_{\sigma= \pm  \pi/2}=0,\nonumber\\
&& u(\sigma)=u_f\bigg|_{\sigma=-\pi/2,\pi/2}.
\eea
In fact for the rotating string we have $\t= \omega \tau$ and since we can use the static gauge where
$t=\tau$ the action in fact reads
\bea
S &=&- T_s\int d\sigma d\tau \sqrt{
(1- R^2 \omega^2)(f^2 \partial_\sigma R{}^2 +
g^2 \partial_\sigma  u{}^2)}\CR
 &=&-T_s\int d R d\tau \sqrt{
(1- R^2 \omega^2)(f^2  +
g^2  (u')^2)},
\label{actrot2}
\eea
where here $u'= \frac{du}{dR}$

The equation of motion with respect to $\delta u$  is given  by
\be\label{eqnIII}
\frac{d}{dR}\left [\frac{1}{\gamma}\frac{g^2 u'}{\sqrt{(f^2  +
g^2  (u')^2)}}\right] = \frac{1}{\gamma} \frac{d  \sqrt{(f^2 +g^2(u')^2)}}{du}
\ee
 where $\gamma = \frac{1}{\sqrt{(1- R^2 \omega^2)}}$

Let us now examine whether this equation (\ref{eqnIII}) indeed admits a solution of an horizontal part (region II) connected to flavor branes by vertical parts (region I). In regions I that are along
\be
 -\alpha>\sigma\geq -\pi/2     \qquad  \pi/2 \geq \sigma >\alpha
\ee

\be
u'\rightarrow \infty, \
\sqrt{(f^2  +g^2  (u')^2)}\rightarrow g u'\ \ \gamma= \gamma(L)= constant\
\ee

 where along these segments $R=L$ with  $L$ is the length of the string. Thus the l.h.s of the equation  turns into $\frac{d g}{ d R}$ which is  also the value of the r.h.s.
  In region II   which is along
   \be
   \alpha \geq \sigma \geq-\alpha
   \ee
   since we require there that $u=u_0$ is ( approximately)   constant  we expand the equation (\ref{eqnIII})  to leading order in $u'$. For finite $g(u_0)$ the l.h.s of the equation vanishes and thus since in this case
$\sqrt{(f^2  +g^2  (u')^2)}\rightarrow f$ it implies that $f(u)$ has to have an extremum at $u=u_0$. The other option that we want to examine is the case that $lim_{u\rightarrow u_0} g^2(u)\rightarrow \infty$ which for the static case is one of the two sufficient conditions \cite{Kinar:1998vq} to  admit an confining area law behavior.
 
Since we took the string to stretch separately along region of the type I, region II and again region of the type I we have to reexamine the variation of the action with respect to $\delta R$.
\be
\delta S = T\left ( \int_{-\pi/2}^{-\alpha} + \int_{\alpha}^{\pi/2}\right )d\sigma d\tau  \delta R \left [ \gamma R (\dot \theta)^2 g \pa_\sigma u\right] + T\int d\tau \delta R \frac{f^2\pa_\sigma R}{\gamma\sqrt{f^2 + g^2(\pa_\sigma u)^2}}\bigg|_{-\alpha}^{\alpha}
\ee
Upon substituting the solution of the bulk equations of motion and in particular that in the interval $(-\alpha,\alpha)$ $\pa_\sigma u=0$,  we find that the variation of the action can vanish only provided that
\be
\frac{f(u_0)}{\gamma(L)} = \omega^2 L \gamma(L) \int_{u_0}^{u_f}du  g(u)
\ee

Let us define now the notion of a string endpoint mass
\be
m_{sep} = T \int_{u_0}^{u_f} g(u)
\ee
then the condition of the vanishing of the variation of the action takes the form
\be
T_{eff} = m_{sep}\gamma^2 \omega^2 L \qquad T_{eff}= T f(u_0)
\ee
As will be further discussed in section $(\S 4.2)$ this is nothing but a balancing equation between the tension and the centrifugal force acting on the string endpoint, where the tension  is exactly the same one that was found above for the Wilson line.

Next we determine the energy and angular momentum of the rotating string that get contributions from both the horizontal and vertical segments. The energy and angular momentum are the  Noether charges associated with the invariance of the action under shifts of $X^0$ and $\theta$ respectively, given by
\bea
E &=& T\int dR \gamma\sqrt{ f^2 + g^2 (u')^2 } \CR
J &=& T\int dR  R^2 \dot\theta\gamma\sqrt{ f^2 + g^2 (u')^2 }\CR
\eea
The contribution of regions I is therefore given by
\be
E_I=\frac{2m_{sep}}{\sqrt{1-\omega^2 L^2}}, \qquad
J_I=\frac{2m_{sep} \omega L{}^2}{\sqrt{1-\omega^2 L^2}}
\ee
which as will be discussed in section $(\S 4.2)$ maps into the contribution to the energy and angular momentum of the massive endpoints of the string. The contribution of the string that stretches along region II is
\bea
E_{II}&=&T_s\int_{-L}^{L} T\int dR \gamma\sqrt{ f^2 + g^2 (u')^2 } \CR
&=&T_{eff}\frac{2}{\omega}\arcsin(\omega L)
\\
J_{II}&=&T_s\int_{-L}^{L}dR  R^2 \dot\theta\gamma\sqrt{ f^2 + g^2 (u')^2 }\CR
&=&T_{eff}\left[
\frac{1}{\omega^2}(\arcsin(\omega L)-\omega L\sqrt{1-\omega^2
L{}^2}\right]
\eea

Combining together the contributions of regions I and region II we find
\bea
E&=&T_{eff}\frac{2}{\omega}\arcsin(\omega L)+\frac{2m_{sep}}{\sqrt{1-\omega^2 L^2}}
\\
J&=&T_g \frac{1}{\omega^2}(\arcsin(\omega L)-\omega L\sqrt{1-\omega^2
L{}^2})+\frac{2m_{sep} \omega L{}^2}{\sqrt{1-\omega^2 L^2}}.
\eea

\begin{figure}[ht!] \centering
					\includegraphics[width=.90\textwidth, natheight=1077bp,natwidth=2000bp]{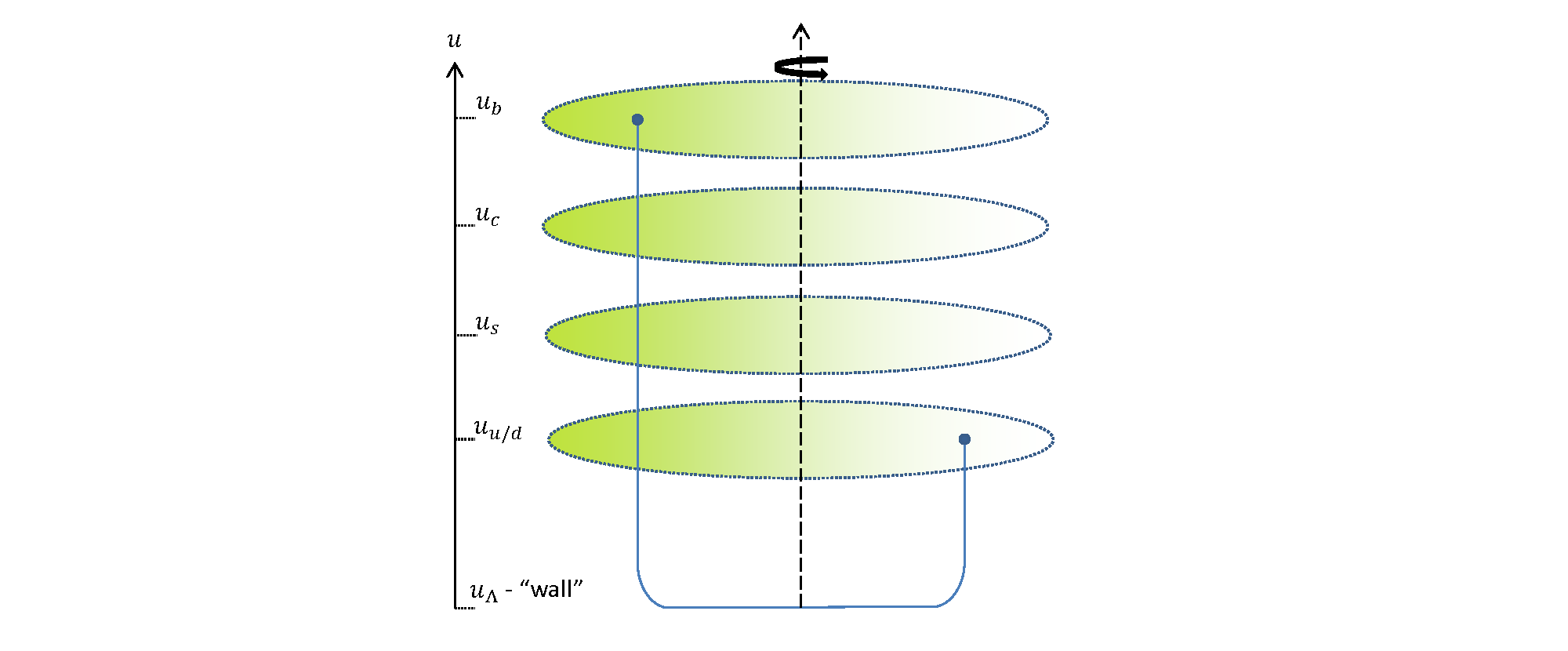}
					\caption{\label{udscbflavor}  Holographic setup with flavor branes associated with the $u,d$ quark, $s$, $c$ and $b$ quarks.}
\end{figure}

So far we have considered stringy holographic baryons that attach to one flavor brane. In holographic backgrounds one can introduce flavor branes at different radial locations thus corresponding to different string endpoint masses, or different quark masses. For instance, a setup that corresponds to $u$ and $d$ quarks of the same $m_{sep}$ mass, a strange $s$ quark, a charm $c$ quark, and a bottom $b$ quark is schematically drawn in figure (\ref{udscbflavor}).  A $B$ meson composed of a bottom quark and a light \(\bar{u}/\bar{d}\) anti-quark was added to the figure.

\subsection{Holographic stringy baryons} \label{sec:holobaryon}

\begin{figure}[t!] \centering
					\includegraphics[width=.75\textwidth, natheight=1029bp,natwidth=2000bp]{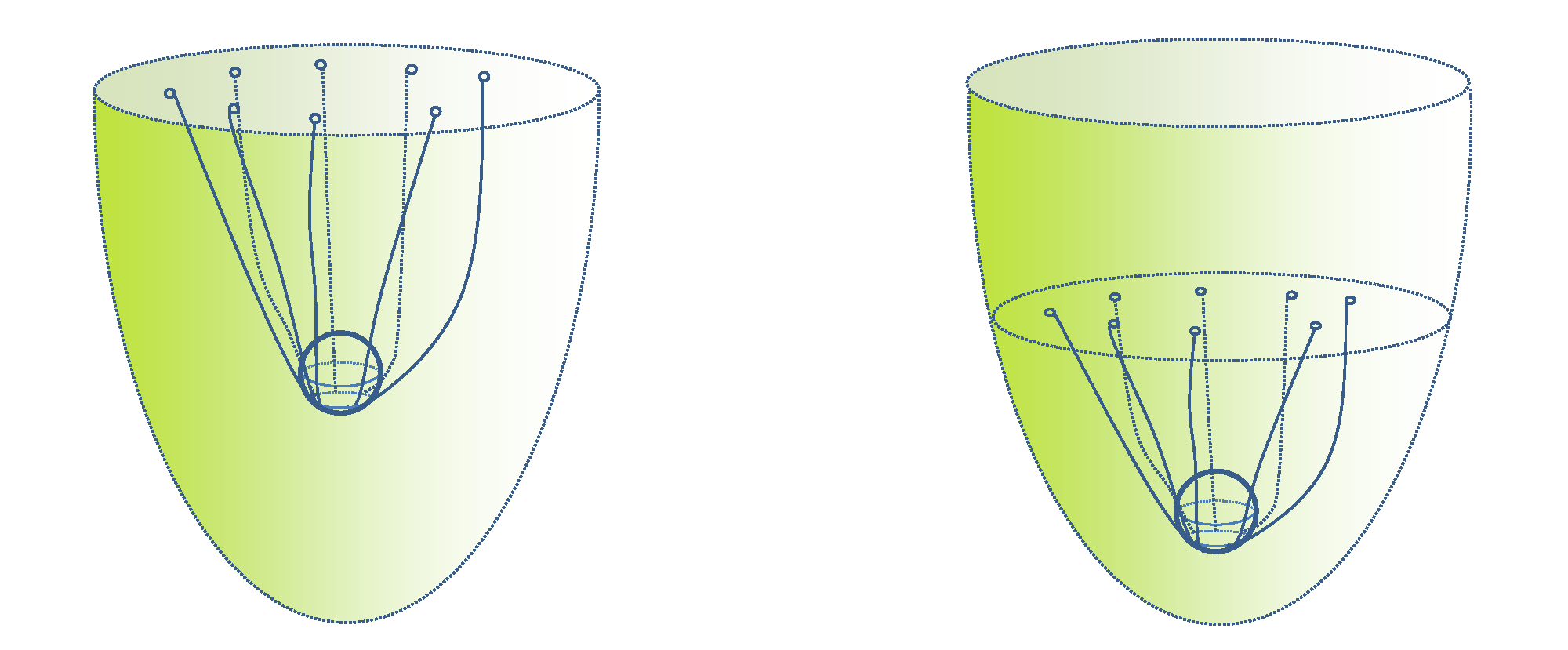}
					\caption{\label{schembaryons}  Schematic picture of holographic baryons. On the left is an external baryon with strings that end on the boundary, while on the right is a dynamical baryon with strings ending on a flavor brane.}
\end{figure}
We have seen in the previous sections that  (i)
 a string with its two ends on the  boundary of a holographic background corresponds in the dual field theory to an external quark, and (ii) a   rotating string with its ends on  flavor probe branes is a dual of a meson.   It is thus clear that a stringy holographic baryon has to include $N_c$ strings that are connected together and end on  flavor branes  (dynamical baryon) or on the boundary (external baryon). The question is what provides the ``baryonic vertex'' that connects together $N_c$ strings.
In \cite{Witten:1998xy} it was shown  that in the $AdS_5\times S^5$ background, which is equipped with  an RR flux of value $N_c$,  a $D5$ brane that wraps the $S^5$ has to have  $N_c$ strings attach to it. This property can be generalized to other  holographic  backgrounds so that   a $Dp$ brane wrapping a non-trivial $p$ cycle with a  flux of an RR field  of value $N_c$  provides a baryonic vertex.   These two possible stringy configurations are
schematically depicted in figure (\ref{schembaryons}). Whereas the dual of  the original proposal \cite{Witten:1998xy} was a conformal field theory, baryons can be constructed also in holographic backgrounds that correspond to confining field theories. A prototype model of this nature is the Sakai Sugimoto  model discussed in section( $\S$ 2.3) of  $N_c$ $D4$ branes background  compactified on a circle with an $N_f$ $D8$--$\overline{D8}$  U-shaped flavor branes \cite{SakSug}. In this model the baryonic vertex is made out of a $D4$ brane that wraps an $S^4$.  Another model for baryons in  a confining background is  the deformed conifold model  with $D7$--$\overline{D7}$ U-shaped flavor branes \cite{Dymarsky:2010ci}. In this model the baryon is a $D3$ brane that wraps the three-cycle of the deformed conifold.

The argument why a  $Dp$ brane wrapping a fluxed  $p$ cycle  is a baryonic vertex is in fact very simple.  The  world-volume action of the wrapped $Dp$ brane has the form of  $S= S_{DBI}+S_{CS}$. The CS term takes the following form
\be
S_{CS}= \int\limits_{S_p\times R_1} \!\! \sum_i c_{p_i} \wedge e^{F}
= \int\limits_{S_p\times R_1} \!\!\! c_{p-1}\wedge F
= -\!\!\!\! \int\limits_{S_p\times R_1} \!\!\! F_p\wedge A
= -N_c\int\limits_{R_1} \!\! A
\ee
where (i) the sum over \(i\) is over the RR $p$-forms that  reside in the background, (ii) from the sum one particular RR form  was chosen,  the $c_{p-1}$-form that couples to the Abelian field-strength F,  (iii) for simplicity we took the $p$ cycle to be $S^p$, (iv) $A$  is the Abelian  connection  that resides on the wrapped brane, (v) $F_p$ is the RR $p$-form field strength, and (vi) in the last step we have made use of the fact that $\int_{S^p} {F_p} = N_c$.
This implies that there is a charge $N_c$  for the Abelian gauge field. Since in a compact space one cannot have non-balanced charges and since the endpoint of a string carries a charge one, there  must be $N_c$  strings attached to it.

 It is interesting to note that a baryonic vertex, rather than being a ``fractional'' $D0$ brane of the form of a $Dp$ brane wrapping a $p$ cycle, can also be a $D0$ brane in an $N_c$ fluxed background. This is the case in the non-critical string backgrounds like \cite{Kuperstein:2004yf} where there is no non-trivial cycle to wrap branes over, but an ``ordinary" $D0$ brane in this background there is  also a CS term of the form $N_c\int_{R_1} A$.

\begin{figure}[t!] \centering
					\includegraphics[width=.80\textwidth, natheight = 549bp, natwidth = 831bp]{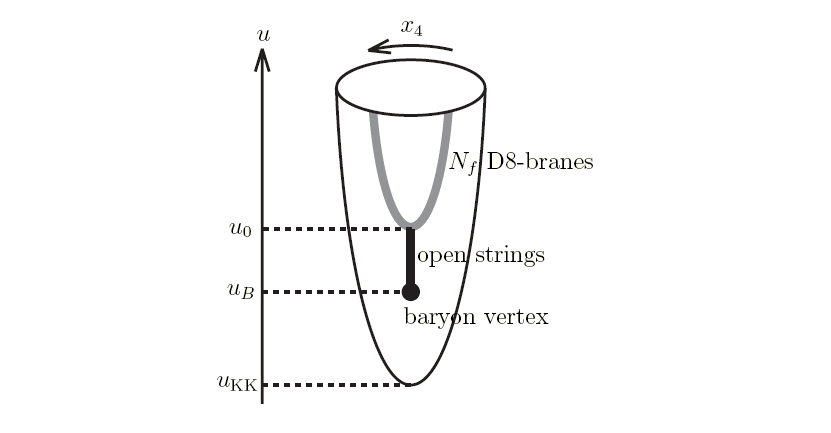}
					\caption{\label{baryonloc} The location of the baryonic vertex  along the radial direction of the Sakai-Sugimoto model. $u_0$ is the tip of the flavor branes and $u_{kk}$ is the ``confining scale" of the model.}
\end{figure}

Next we would like to determine the  location of the baryonic vertex in the radial dimension.  In particular the question is whether it is located on the flavor branes or below them.   This  is schematically depicted in figure (\ref{baryonloc}) for the model of \cite{SakSug}.

 In \cite{Seki:2008mu} it was shown  that in the model of \cite{SakSug} by minimizing the ``mechanical energy" of the $N_c$ strings and the wrapped brane that it is preferable for the baryonic vertex to be located on the flavor branes of the model. It was also found that for the baryonic vertex of the model of \cite{Dymarsky:2009cm}, if the tip of the U-shaped flavor brane is close to the  lowest point of the deformed conifold the baryonic vertex is  on  the flavor  branes,  but there is also a range of parameters where it will end up below the flavor branes. It is interesting to note that  for a background that corresponds to the  deconfining phase of the dual gauge theory the baryonic vertex falls into the ``black hole" and thus the baryon dissolves.

\begin{figure}[t!] \centering
					\includegraphics[width=.75\textwidth, natheight = 784bp, natwidth = 1374bp]{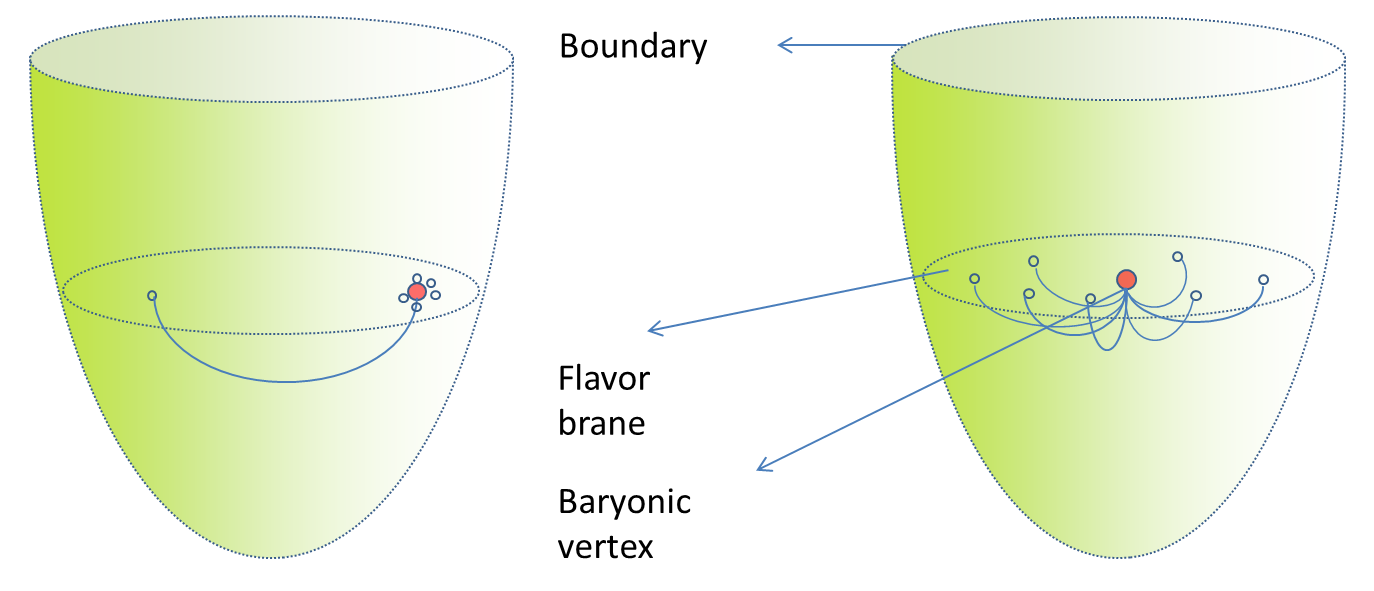}
					\caption{\label{fig:holbaryon} The location of the baryonic vertex on the flavor brane and the corresponding configuration of the baryon for some large \(N_c\). When \(N_c = 3\), the left is the analog of the quark-diquark flat space-time model, and the right the analog of the Y-shape model.}
\end{figure}

The locations of the baryonic vertex and the ends of the $N_c$ strings  on the flavor branes is  a dynamical issue. In figure (\ref{fig:holbaryon}) we draw two possible setups. In one the baryonic vertex is located at the center and the $N_c$ ends of the strings are located around it in a spherically symmetric way. In the ``old" stringy model of baryons for three colors this is the analog of the Y-shape configuration that we will further discuss in the sections describing the flat space-time models. Another possibility is that of a baryonic vertex is  connected by $N_c-1$  very short strings and with one long string  to the flavor branes. Since the product of  $N_c-1$  fundamental representations includes the anti-fundamental one, this configuration can be viewed as a string connecting a quark with an anti-quark. For the case of $N_c=3$ this is the analog of what will be discussed below as the quark-diquark stringy configuration. This latter string configuration (for any $N_c$) is
  similar to the stringy meson, but there is a crucial difference, which is the fact that the stringy baryon includes a baryonic vertex.

 Correspondingly there are many options of holographic stringy baryons that connect to different flavor branes. This obviously relates to baryons that are composed of quarks of different flavor. In fact there are typically more than one option for a given baryon. In addition to the distinction between the central and quark-diquark configurations  there are more than one option just to the latter configuration. We will demonstrate this situation in section \ref{sec:structure}, focusing on the case of the doubly strange $\Xi$ baryon (\(ssd\) or \(ssu\)). The difference between  the two holographic setups is translated to the two options of the diquark being either composed of two $s$ quarks, whereas the other setup features a $ds$ or $us$ diquark. Rather than trying to determine the preferred configuration from holography we will use a comparison with experimental data to investigate this issue.
\subsection{The glueball as a  rotating closed string}
The third type of a hadron is the glueball. The fact that the latter does not include quarks maps under the holographic duality to a string with no endpoints, namely, a  closed string. Before analyzing the closed strings of holographic background we first  review certain general  properties of closed string. Firstly we would like to remind the reader the classical rotating string in flat space-time.
\subsubsection{Classical rotating folded string} \label{sec:classicalFoldedString}
Here we use the Nambu-Goto action for the string
\be S = -\frac{1}{2\pi\alp}\int d\tau d\sigma \sqrt{-h} \,,\ee
with
\be h = \det h_{\alpha\beta}\,, \qquad h_{\alpha\beta} = \eta_{\mu\nu}\partial_\alpha x^\mu \partial_\beta x^\nu \,,\ee
and \be \alp = \frac{1}{2\pi T} \,. \ee
The rotating folded string is the solution
\be x^0 = \tau \qquad x^1 = \frac{1}{\omega}\sin(\omega\sigma)\cos(\omega\tau) \qquad x^2 = \frac{1}{\omega}\sin(\omega\sigma)\sin(\omega\tau) \,.\label{eq:rotsol}\ee
We take \(\sigma \in (-\ell,\ell)\) and correspondingly \(\omega\) takes the value \(\omega = \pi/\ell\). The energy of this configuration is
\be E = T \int_{-\ell}^\ell d\sigma \partial_\tau X^0 = 2T\ell \ee
The angular momentum we can get by going to polar coordinates (\(x^1 = \rho\cos\theta, x^2 = \rho\sin\theta\)), then
\be J = T \int_{-\ell}^\ell d\sigma \rho^2 \partial_\tau \theta =
\frac{T}{\omega} \int_{-\ell}^\ell d\sigma \sin^2(\omega\sigma) = \frac{\pi T}{\omega^2} = \frac{T\ell^2}{\pi}\ee
From the last two equations we can easily see that for the classical rotating folded string
\be J = \frac{1}{4\pi T}E^2 = \frac{1}{2}\alp E^2 \ee

\begin{figure}[ht!] \centering
	\includegraphics[width=0.95\textwidth,natheight=1029bp,natwidth=2000bp]{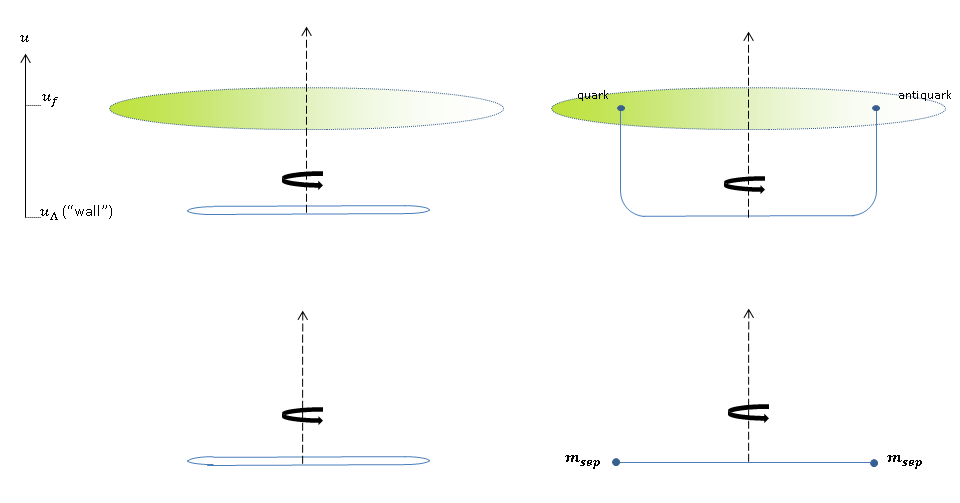} \\
	\caption{\label{fig:closed_open_map} Closed and open strings in holographic backgrounds (above), and their mappings into flat spacetime (below). For the open string, the mapping from curved to flat background adds endpoint masses to the strings, with the vertical segments mapped to the point-like masses in flat space. For the closed string, we look at the simple folded string in both cases. Note that classically the rotating folded string has zero width, and as such would look like an open string with no endpoint masses, and not like in the drawing.}
			\end{figure}

Thus we conclude that the Regge slope for the closed string is
\be \alp_{closed} = \frac12\alp_{open} \,, \ee
A simple explanation for this property is that one can regard  the fully folded string as two strings and hence the tension of the system is twice the tension of  a single open string and correspondingly the slope of the closed string is half of the open one.
This property that  holds also for the rotating folded string in curved holographic background, as will be shown in $\S (6.4)$, will be the main tool for our proposed  phenomenological identification of glueballs (see $\S (3.42)$.

It is well known that this ratio between the slopes ( or tensions)
follows also from gauge theory arguments. This is based on the fact
 that  the potential between two static adjoint SU(N) charges,  is  proportional to the quadratic Casimir operator. For small distances this  group theory factor can be obtained easily from perturbation theory. Calculations in \cite{Bali:2000un} show that what is referred to as the ``Casimir scaling hypothesis'' holds in lattice QCD for large distances as well, and this means that the effective string tension also scales like the Casimir operator (as the potential at large distances is simply \(V(\ell) \approx T_{eff}\ell\)). Therefore, a model of the glueball as two adjoint charges (or constituent gluons) joined by a flux tube predicts the ratio between the glueball and meson (two fundamental charges) slopes to be
\be \frac{\alp\!_{gb}}{\alp\!_{\qqb}} = \frac{C_2(\text{Fundamental})}{C_2(\text{Adjoint})} = \frac{N^2-1}{2N^2} = \frac{4}{9}\,, \ee
where for the last equation we take \(N = 3\). For \(N \rightarrow \infty\) we recover the ratio of \(1/2\). Note, however, this type of determination of the ratio from QCD is not valid at higher orders non-perturbatively.

Other models attempt to tie the closed string to the phenomenological "pomeron". The pomeron slope is measured to be
\be \alp\!_{pom} = 0.25\:\text{GeV}^{-2} \approx 0.28\times\alp\!_{meson} \ee
and the pomeron trajectory commonly associated with both glueballs and closed strings. One string model that predicts a pomeron-like slope was proposed \cite{Isgur:1984bm} and is presented in \cite{Meyer:2004jc}, or in more detail in \cite{Meyer:2004gx}. It is simply the model of a rotating closed string, with a fixed circular shape. This string has two types of trajectories, a phononic trajectory (excitations propagating along the string) which has \(\alp_{phonon} = \frac{1}{4}\alp_{open}\), and an orbital trajectory (the circular string rotating around an axis in the circle's plane), for which \(\alp\!_{orbital} = \frac{3\sqrt{3}}{16}\alp_{open} \approx 0.32\times\alp_{open}\). If the rotating circular loop were allowed to deform, it would have necessarily flowed towards the flattened folded string configuration that we have been discussing, which always maximizes the angular momentum at a given energy.
There are also other possibilities of rigidly rotating closed string of other shapes, as in \cite{Burden:1982zb}, which may give yet another prediction of the ratio between open and closed string Regge slopes.

\subsubsection{The closed string in a curved  Holographic background}
Now that we spelled out the basic properties of the classical rotating folded string in flat space time, we want to explore the same type of configurations but now in holographic confining backgrounds.
The full analysis of rotating closed string in holographic curved backgrounds was performed in \cite{PandoZayas:2003yb}. We present here the key points in short form.

If we look at a curved background metric of the form 
\be ds^2 = h(r)^{-1/2}(-dX^0dX^0+dX^idX^i) + h(r)^{1/2}dr^2 + \ldots \,,\ee, namely with $g_{00}=h(r)^{-1/2}$
with \(i = 1,2,3\) and the ellipsis denoting additional transverse coordinates, the rotating folded string, namely the configuration,
\be X^0 = e\tau \qquad X^1 = e\sin\sigma\cos\tau \qquad X^2 = e\sin\sigma\sin\tau \,,\ee
is still\footnote{We follow a somewhat different normalization here, taking \(\omega = \pi/\ell\) from the previous section to be \(1\), and introducing a common prefactor \(e\), but the solution is essentially the same as the flat space solution of section \ref{sec:classicalFoldedString}.} a solution to the string equations of motion provided we take
\be r(\sigma,\tau) = r_0 = Const. \ee
where \(r_0\) is a point where the metric satisfies the condition
\be \partial_r g_{00}(r)|_{r=r_0} = 0, \qquad g_{00}(r)|_{r=r_0} \neq 0 \ee
This is also one of the sufficient conditions for the dual gauge theory to be confining \cite{Kinar:1998vq}. Compared to the folded string in flat spacetime, the energy and angular momentum take each an additional factor in the form of \(g_{00}(r_0)\):
\be E = \frac{1}{2\pi\alp}\int_{-\pi}^\pi g_{00}(r_0)d\sigma = g_{00}(r_0) \frac{e}{\alp} \ee
\be J = T\int_{-\pi}^\pi g_{00}(r_0)\sin^2\sigma d\sigma = g_{00}(r_0) \frac{e^2}{2\alp} \ee
Defining an effective string tension \(T_{eff} = g_{00}(r_0)T\) and slope \(\alp_{eff} = (2\pi T_{eff})^{-1}\), we can write the relation
\be J = \frac{1}{2}\alp_{eff} E^2 \ee

The same factor of \(g_{00}(r_0)\) multiplies the effective tension in the open string case, and therefore the closed and open string slopes are still related by the factor of one half, although the open string trajectories have the additional modification which can be ascribed to the presence of endpoint masses \cite{Kruczenski:2004me,Sonnenschein:2014jwa}. We draw the two types of strings in figure \ref{fig:closed_open_map}.

\section{ Holography inspired hadronic strings}
In the previous section we have determined the classical string configurations that correspond to  Wilson lines, mesons, baryons and glueballs in holographic background. Analyzing, and in particular quantizing, strings in curved holographic backgrounds is a rather complicated task.   In this section we will argue   that one can approximate the holographic string configurations by  modified strings in flat space time. It is important to emphasize that whereas the stringy holographic  configurations are supposed to be duals of gauge invariant objects of certain confining gauge theory in the large $N$ and large $\lambda$ approximation, the stringy picture  that we discuss from here on  is  {\bf a phenomenological model}. It is by no means a formal passage from the large $N$ large $\lambda$ region to the QCD domain of $N_c=3$ and $\lambda$ of ${\cal O}(1)$.
In this section we first  describe  the ``map" from the holographic models of classical  hadronic strings  to models of classical strings in flat space-time. We then analyze the system of rotating strings with massive endpoint which as will be shown below associate with both mesons and baryons. We discuss the stability of baryonic configurations and certain properties of closed strings
In the next section we will address the problem of quantizing these systems.

 \subsection{ The map from holographic strings to strings in flat space-time}
 The map from holographic strings to strings in flat space-time includes several ingredients:
 \begin{itemize}
 \item
All the  stringy configurations in holographic confining background  that were discussed in section $(\S 3)$ are characterized by having a ``flat string" that stretches close to the ``wall" of the confining background. For the glueball it is a closed string, for the mesons and baryons a string connected to flavor D branes and for the Wilson line connected to the boundary.

In all these cases  the strings  are characterized  by an effective  string tension which is given  by the ``bare" string tension dressed in the following way
\bea
T_{eff}&=& T f(u*)= T\sqrt{ g_{00}(u*) g_{xx}(u*)} \CR
  \mathrm{either}\  f_{min}&=& f(u*) \ \mathrm{or} \  lim_{u\rightarrow u*} g(u) \rightarrow \infty\CR
\eea
Obviously for the closed strings the tension will be $T_{closed}= 2T_{open}$.
 \item
 All the holographic open strings: the Wilson line, the meson and the baryon include, as was shown in section $(\S 3)$, two vertical strings that connect the horizontal string that stretches in the vicinity of the ``wall" and flavor branes or the boundary of the holographic bulk. It was shown there that the action (or energy) of these segments  can  be converted to the action of  massive particles at the ends of the string where the mass is given by 
 \be
m_{sep} = T \int_{u_0}^{u_f} g(u)
\ee
The role of the masses in the map is described in figure (\ref{fig:mapholflat}).
\begin{figure}[t!] \centering
\includegraphics[natwidth=1200bp, natheight=900bp,width= 0.9\textwidth]{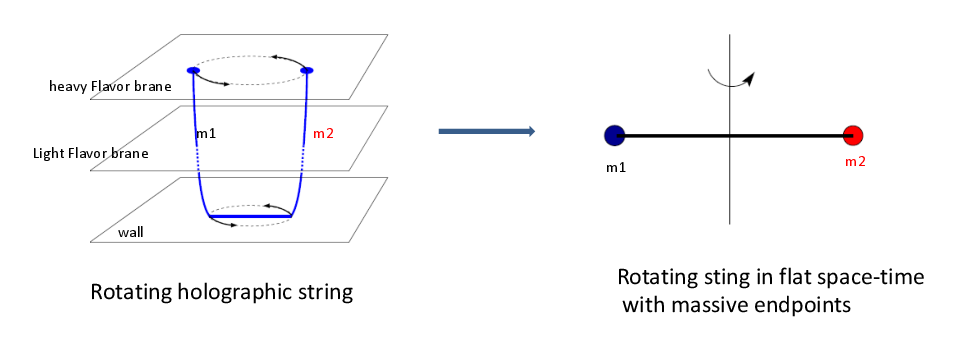}
					\caption{\label{fig:mapholflat} On the left: rotating holographic open string and on the right the corresponding open string with massive endpoints in flat space-time with $m_1=m_2$.}
		\end{figure}
In the next subsection we will classically analyze such a string with massive endpoints and in the next section we will address the issue of its quantization. It is interesting also to note the role of these masses for the stringy Wilson line in flat space time.
   As was discussed in section $(\S 3.1)$ the holographic Wilson line diverges and has to be renormalized. This can be done in several different methods. One of them is exactly the subtraction of the action of the vertical strings, namely the mass at the endpoints. Moreover  the outcome of the subtraction is that in addition to the linear term the energy of the string has also a constant term  (and  terms that depend on negative powers of the length). The interpretation of the constant term  is that of an addition of a  mass (or masses). For a detailed description see \cite{Kinar:1998vq}.
\item
For baryons the transition from the holographic strings to  those of ``Holography inspired models" there is a dramatic difference and that is that whereas in the former the number of colors $N_c$ and hence the number of string arms is large in the latter it is taken to be $N_c=3$. As explained above we do not claim to have a controlled method to pass from large $N_c$ we take as an ansatz that there are three strings connected to the baryonic vertex.
\item
An important ingredient of the transition is the fact that whereas the original holographic models are in critical ten dimensions the model in flat space-time is taken to be non-critical in four space-time dimension. This will have a dramatic effect on the quantum action of the string due to the Liouville term.
\end{itemize}

\subsection{The  rotating string with massive endpoints}\footnote{The theory of a bosonic string with massive endpoints has been already addressed  in the original application of string theory to hadron physics. For instance  see\cite{Chodos:1973gt}. It is worth mentioning thought that up to date the quantization of this system has not been fully solved.}
		We describe the string with massive endpoints (in flat space-time) by adding to the Nambu-Goto action,
		\be S_{NG} = -T\int\!\!{d\tau d\sigma \sqrt{-h}} \ee
		\[ h_{\alpha\beta} \equiv \eta_{\mu\nu} \del_\alpha X^\mu \del_\beta X^\nu \]
		a boundary term - the action of a massive chargeless point particle
		\be S_{pp} = -m\int\!\! d\tau \sqrt{-\dot{X}^2} \ee
		\[ \dot{X}^\mu \equiv \del_\tau X^\mu \]
		at both ends. There can be different masses at the ends, but here we assume, for simplicity's sake, that they are equal. We also define \(\sigma = \pm l\) to be the boundaries, with \(l\) an arbitrary constant with dimensions of length.

		The variation of the action gives the bulk equations of motion
		\be \del_\alpha\left(\sqrt{-h}h^{\alpha\beta}\del_\beta X^\mu\right) = 0 \label{eq:bulk} \ee
		and at the two boundaries the condition
		\be T\sqrt{-h}\del^\sigma X^\mu \pm m\del_\tau\left(\frac{\dot{X}^\mu}{\sqrt{-\dot{X}^2}}\right) = 0 \label{eq:boundary} \ee
		
		It can be shown that the rotating configuration
		\be X^0 = \tau, X^1 = R(\sigma)\cos(\omega\tau), X^2 = R(\sigma)\sin(\omega\tau) \ee
		solves the bulk equations \eqref{eq:bulk} for any choice of \(R(\sigma)\). We will use the simplest choice, \(R(\sigma) = \sigma\), from here on.\footnote{Another common choice is $X^0=\tau, x^1=\sin(\sigma) \cos(\omega\tau), X^2= \sin(\sigma)\sin(\omega\tau)$.} Eq. \eqref{eq:boundary} reduces then to the condition that at the boundary,
		\be \frac{T}{\gamma} = \gamma m \omega^2 l \label{eq:boundaryRot}\ee
		with \(\gamma^{-1} \equiv \sqrt{1-\omega^2 l^2}\).\footnote{Notice that in addition to the usual term $\gamma m$ for the mass, the tension that balances the ``centrifugal force" is $\frac{T}{\gamma}$.}
		We then derive the Noether charges associated with the Poincar\'e invariance of the action, which include contributions both from the string and from the point particles at the boundaries. Calculating them for the rotating solution, we arrive at the expressions for the energy and angular momentum associated with this configuration:
	\be\label{Classtra}	E = -p_0 = 2\gamma m + T \int_{-l}^l\!\frac{d\sigma}{\sqrt{1-\omega^2\sigma^2}} \ee
	\be J = J^{12} = 2\gamma m \omega l^2 + T \omega \int_{-l}^l\! \frac{\sigma^2 d\sigma}{\sqrt{1-\omega^2\sigma^2}} \ee
		Solving the integrals, and defining \(q \equiv \omega l\) - physically, the endpoint velocity - we write the expressions in the form
		\be E = \frac{2m}{\sqrt{1-q^2}} + 2Tl\frac{\arcsin(q)}{q} \ee
		\be J = 2m l \frac{q}{\sqrt{1-q^2}} + Tl^2\left(\frac{\arcsin(q)-q\sqrt{1-q^2}}{q^2}\right) \ee
		The terms proportional to \(m\) are the contributions from the endpoint masses and the term proportional to \(T\) is the string's contribution. These expressions are supplemented by condition \eqref{eq:boundaryRot}, which we rewrite as
		\be Tl = \frac{mq^2}{1-q^2} \label{eq:Tm}\ee
		This last equation can be used to eliminate one of the parameters \(l, m , T,\) and \(q\) from \(J\) and \(E\). Eliminating the string length from the equations we arrive at the final form
		\be E = 2m \left(\frac{q\arcsin(q)+\sqrt{1-q^2}}{1-q^2}\right) \label{eq:massiveE} \ee
		\be J = \frac{m^2}{T}\frac{q^2}{(1-q^2)^2}\left(\arcsin(q)+q\sqrt{1-q^2}\right) \label{eq:massiveJ} \ee
		These two equations are what define the Regge trajectories of the string with massive endpoints. They determine the functional dependence of \(J\) on \(E\), where they are related through the parameter \(0 \leq q < 1\) (\(q = 1\) when \(m = 0\)). Since the expressions are hard to make sense of in their current form, we turn to two opposing limits - the low mass and the high mass approximations.
			In the low mass limit where the endpoints move at a speed close to the speed of light, so \(q \rightarrow 1\), we have an expansion in \((m/E)\):
		\be J = \alp E^2\left(1-\frac{8\sqrt{\pi}}{3}\left(\frac{m}{E}\right)^{3/2} + \frac{2\sqrt{\pi^3}}{5}\left(\frac{m}{E}\right)^{5/2} + \cdots\right) \label{eq:lowMass}\ee
		from which we can easily see that the linear Regge behavior is restored in the limit \(m\rightarrow 0\), and that the first correction is proportional to \(\sqrt{E}\). The Regge slope \(\alp\) is related to the string tension by \(\alp = (2\pi T)^{-1}\).
		The high mass limit, \(q \rightarrow 0\), holds when \((E-2m)/2m \ll 1\). Then the expansion is
		\be J = \frac{4\pi}{3\sqrt{3}}\alp m^{1/2} (E-2m)^{3/2} + \frac{7\pi}{54\sqrt{3}} \alp m^{-1/2} (E-2m)^{5/2}
			+ \cdots \label{eq:highMass} \ee

\subsection{The stringy baryon and its stability }
Next we discuss the stringy baryon. For the case of $N_c=3$, the asymmetric configuration depicted in figure 7. is the analog of what will be discussed below as the quark-diquark stringy configuration. This latter string configuration (for any $N_c$) is similar to the stringy meson, but there is a crucial difference, which is the fact that the stringy baryon includes a baryonic vertex which carries the baryon number.

\begin{figure}[ht!] \centering
					\includegraphics[width=.90\textwidth, natheight=783bp, natwidth=1926bp]{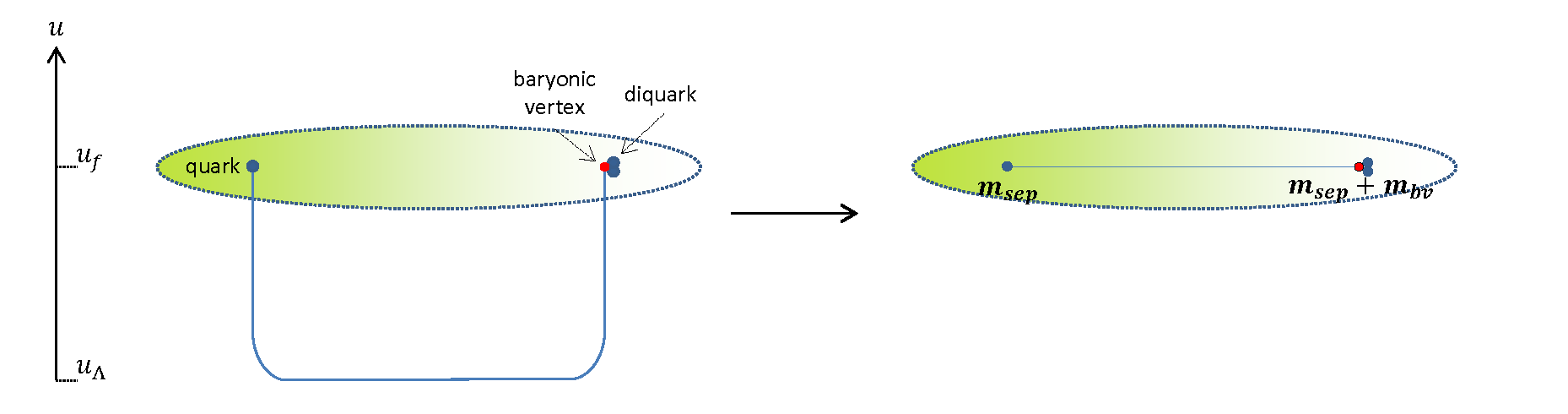}
					\caption{\label{holtoflat}  The holographic setup downlifted to $N_c=3$ of a quark and a diquark is mapped to a similar configuration in flat space-time. The vertical segments of the holographic string are mapped into masses of the endpoints.}
\end{figure}

It was shown in \cite{Kruczenski:2004me} that the classical rotating   holographic stringy configuration of the meson can be mapped into that of a classical rotating bosonic string in flat space-time with massive endpoints. A similar map applies also to  holographic stringy baryons that can be transformed into stringy baryons in flat space time with massive endpoints. We will proceed now to discuss this map for the central and asymmetric layouts of figure (\ref{fig:holbaryon}). The asymmetric holographic configuration of a quark and $N_c-1$ quarks on the two ends of the holographic string depicted in figure (\ref{holtoflat}) is mapped into a similar stringy configuration in flat space-time where the vertical segments of the string are transferred into massive endpoints of the string. On the left hand side of the string in flat space-time there is an endpoint with mass $m_{sep}$ given by
\be
m_{sep} = T\int_{u\Lambda}^{u_f} du \sqrt{g_{00}g_{uu}}
\ee
where $u_\lambda$ is the location of the ``wall'', $u_f$ is the location of the flavor brane, and $g_{00}$ and $g_{uu}$ are the $00$ and $uu$ components of the metric of the background.
On the right hand side the mass of the endpoint is a sum of  $m_{sep}+m_{bv}$. This is the sum of the energy associated with the vertical segment of the string, just like that of the right hand side, and the mass of the baryonic vertex. Note that even though on the right endpoint of the string there are in fact two endpoints or ``two quarks" the mass is that of a single quark since there is a single vertical string. This string setup is obviously
 very similar to that of the meson. The only difference is the baryonic vertex that resides at the diquark endpoint.  Since we do not know how to evaluate the mass of the baryonic vertex, it will be left as a free parameter to determine by the comparison with data. Our basic task in this case will be to distinguish between two options: (i) the mass of the baryonic vertex is much lighter than the endpoint mass, $m_{bv}<<m_{sep}$, in which case the masses at the two endpoints will be roughly the same, and (ii) an asymmetric setup with two different masses if the mass of the baryonic vertex cannot be neglected.

\begin{figure}[ht!] \centering
					\includegraphics[width=.90\textwidth, natheight=882bp, natwidth=1845bp]{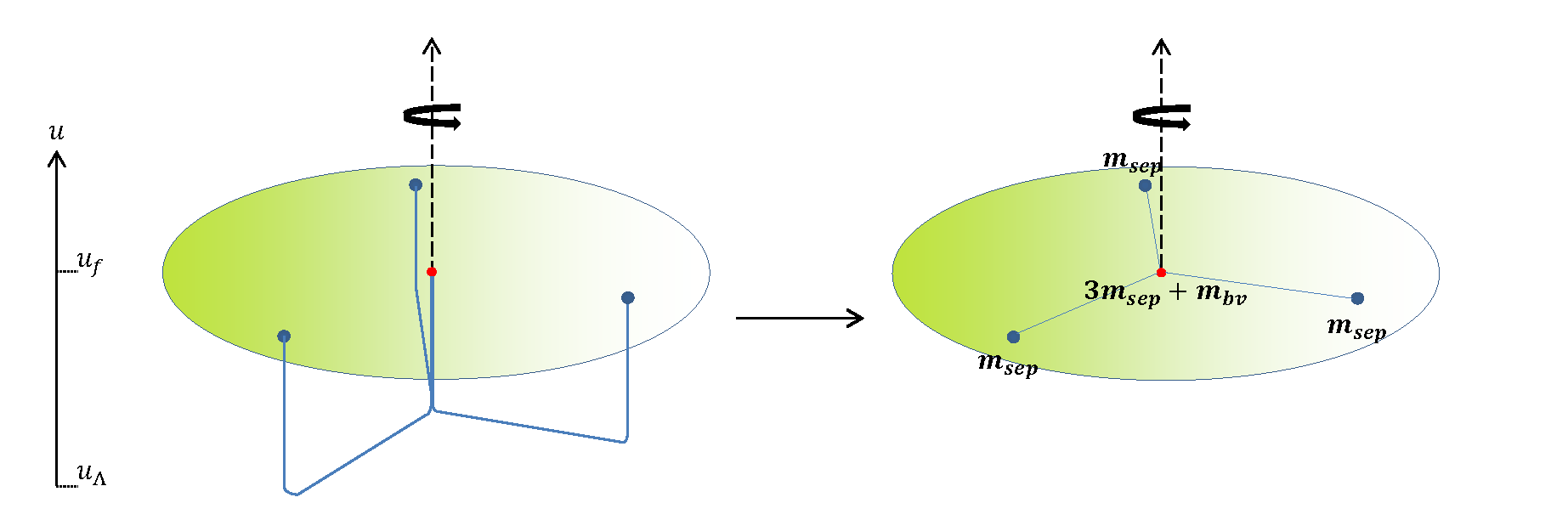}
					\caption{\label{centerholtoflat}   The map between the holographic central configuration (for $N_c=3$) and  the Y-shaped string in flat space-time.}
\end{figure}

The configuration with a central baryonic vertex  can be mapped into the analog of a Y-shaped object with $N_c$ massive endpoints and with a central baryonic vertex of mass $m_c= m_{bv}+N_c m_{sep}$. The factor of $N_c$ is due to the fact that there are $N_c$ strings that stretch from it vertically from the flavor brane to the ``Wall", as can be seen in figure (\ref{centerholtoflat}) for the case of $N_c=3$. In this case, regardless of the ratio between the mass of the baryonic vertex and that of the string endpoint, there is a massive center which is at least as heavy as three sting endpoints.

So far we have considered stringy holographic baryons that attach to one flavor brane. In holographic backgrounds one can introduce flavor branes at different radial locations thus corresponding to different string endpoint masses, or different quark masses. For instance, a setup that corresponds to $u$ and $d$ quarks of the same $m_{sep}$ mass, a strange $s$ quark, a charm $c$ quark, and a bottom $b$ quark is schematically drawn in figure (\ref{udscbflavor}).  A $B$ meson composed of a bottom quark and a light \(\bar{u}/\bar{d}\) antiquark was added to the figure.


 Correspondingly there are many options of holographic stringy baryons that connect to different flavor branes. This obviously relates to baryons that are composed of quarks of different flavor. In fact there are typically more than one option for a given baryon. In addition to the distinction between the central and quark-diquark configurations  there are more than one option just to the latter configuration. We will demonstrate this situation in section \ref{sec:structure}, focussing on the case of the doubly strange $\Xi$ baryon (\(ssd\) or \(ssu\)). The difference between the the two holographic setups is translated to the two options of the diquark being either composed of two $s$ quarks, whereas the other setup it is a $ds/us$ diquark. Rather than trying to determine the preferred configuration from holography we will use the comparison with the observational spectra to investigate this issue.


\begin{figure}[t!] \centering
					\includegraphics[width=.90\textwidth, natheight=592bp,natwidth=1500bp]{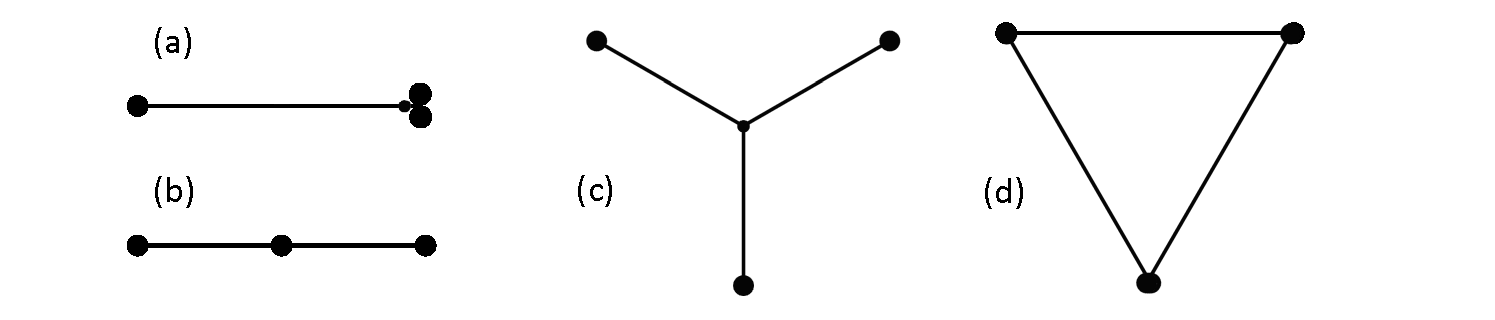}
					\caption{\label{fig:flatbaryons} Four different stringy models for the baryon in flat space-time. (a) is the quark-diquark model, (b) the two-string model, (c) the three-string Y-shape, and (d) the closed \(\Delta\)-shape. (a) and (c) are the models discussed in the preceding discussion of holographic strings and their mappings to flat space-time.}
				\end{figure}

There are some words to be said about the models emerging when mapping the holographic rotating strings to flat space-time.   In this section, we briefly discuss another matter: the stability of the rotating solution. Specifically, we claim that the Y-shape model of the baryon is (classically) unstable. In our analysis of the spectrum we will disregard this potential instability of the model and use the expressions for the energy and angular momentum of the unperturbed rotating solution of the Y-shape model as one of our fitting models, but it is important to remember that there is this theoretical argument against it as a universal setup for baryons before we test it out as a phenomenological model.

Other than the Y-shape and quark-diquark configurations which we analyze, there are two more possible stringy models for the baryon when considering a purely flat space-time point of view. These are drawn, together with the quark-diquark and Y-shape models, in figure (\ref{fig:flatbaryons}). The additional models are the two-string model where one of the quarks is located at the center of the baryon, and the other two are attached to it by a string. The second is the \(\Delta\)-shape model, in which each quark is connected to the other two. It can also be looked at as a closed string with three points along it that carry finite momentum (which can be either massive or massless). While the two-string model may have its analog in holography, with the baryonic vertex lying with the quark at the center of mass, the \(\Delta\)-shaped string cannot be built if we impose the constraint that the three quarks should all be connected to a baryonic vertex.

Two independent analyses of the three-string model\cite{'tHooft:2004he}\cite{Sharov:2000pg} concluded that the rotating solution of the Y-shape configuration is unstable, even before taking quantum effects into account. In another work, it was found that the instability does not show itself in first order perturbation theory\cite{ThesisFederovsky}, but the claim of the unstable nature of the Y-shape model has been verified using numerical methods in \cite{ThesisHarpaz}, where the instability was observed in simulations and its dependence on endpoint masses was examined. To summarize the result, in the three-string Y-shape model, a perturbation to one of the three arms would cause it to shorten until eventually the Y-shape collapses to a form like that of the straight two-string model. From this model in turn a different kind of instability is expected\cite{'tHooft:2004he}\cite{Sharov:2000pg}. The quark in the center of the baryon will move away from the center of mass given a small perturbation and as it approaches one of the quarks at the endpoints quantum effects will induce a collapse to the single string quark-diquark model, it being energetically favorable for two of the quarks to form an diquark bound state in the anti-fundamental color representation. It would seem that all other models have an instability that would cause them to eventually collapse to the quark-diquark configuration as two quarks get close enough to each other.\footnote{More discussion and detailed analyses of the different stringy models of the baryon and their stability are found in the work of G.S. Sharov, most recently in \cite{Sharov:2013tga}.}

From a phenomenological point of view the models differ in their prediction of the slope of the Regge trajectory. Assuming the strings in baryons have the same tension as those in mesons, we can see which of the models offers the best match. We will see that it is the configuration we know to be stable, that of a single string with a quark and a diquark at its endpoints.\footnote{The quark-diquark model was also used to analyze the baryon spectrum in \cite{Selem:2006nd}.}
\subsection{The stringy glueball}
The map of the classical folded rotating closed string in holographic background to a similar string in flat four dimensions is simple. Unlike the case of the open string here there are no vertical segments involved and correspondingly no $m_{sep}$. It is just the string tension dependence on the holographic background that makes it different from an ordinary rotating folded closed string in flat space-time.
However, as will be seen in $(\S 6.1)$, the form of the quantum string yields another significant difference from the ordinary closed string. According to \cite{PandoZayas:2003yb} we find that the relation between the energy and angular momentum is modified from the linear Regge trajectory form and looks like
\be J = \alp_{closed}(E- m_0)^2  +\tilde a  \,.\ee

Thus one may interpret the result as described by an action that includes a closed string plus a massive particle in the center of the closed string thought an interpretation of a  quantum fluctuation as a classical particle may be suspicious.   As will be mention above for the case of Witten's model $m_0$ is positive  and hence it indeed fits the picture of a massive particle in the center whereas for the models discussed in \cite{PandoZayas:2003yb} the mass is negative and thus the interpretation is less clear.  The addition of the massive particle at the center maybe related to the observation made in \cite{Dubovsky:2015zey} about the presence of a massive pseudoscalar axion on the worldsheet of confining flux tubes.

\section {On the quantum  stringy hadrons}
So far we have described the model of  holography inspired stringy hadrons classically. Though the basic property of the spectrum, the relation between the mass and the angular momentum, follows from the classical configuration, the quantum corrections of the classical spectrum are also quite important. It is well known that   the basic classical linear  Regge trajectory in the critical dimension is shifted
 \be
 J=\alpha' M^2  \qquad   \rightarrow \qquad J=\alpha' M^2 + a
 \ee
   where $a$ is the quantum intercept and  furthermore  there are quantum  excitations denoted by $n$ that relate to the mass as  $ n = \alpha' M^2 + a$ so that altogether the linear quantum trajectories is given by
   \be\label{Retra}
   n+J =\alpha' M^2 + a
   \ee
This form of the linear trajectory follows from the quantization of the open string. In \cite{Arvis:1983fp} the quantization of the Nambu Goto action of the  open string with Dirichlet boundary conditions was performed and yielded the following expression for the energy of the string

\be
E_{stat} = \sqrt{ (TL)^2 + 2\pi T \left ( n- \frac{D-2}{24}\right )}
\ee
If instead of the static string we take the rotating one, we are instructed from the classical solution to replace  $TL\rightarrow \sqrt{2\pi TJ}$ so that we find for the rotating case
\be
E_{rot} = \sqrt{ 2\pi T  \left(J+  n- \frac{D-2}{24}\right )}
\ee
which is exactly (\ref{Retra}) when the intercept is taken to be $a=\frac{D-2}{24}$.
It is indeed  well known that the intercept is given in terms of the Casimir energy which is the  sum of the eigenvalues of the worldsheet Hamiltonian $w_n$,  namely
\be
E_{Casimir}\equiv \frac12\sum_{n=1}^{n=\infty} w_n = \frac{\pi (D-2)}{2L}\sum_{n=1}^{n=\infty} n= -\frac{(D-2)}{24}\frac{\pi}{L}= a\frac{\pi}{L}
\ee
where we have taken that $w_n=n$ and we have performed a zeta function regularization.

We have seen above that even at the classical level the stringy meson and baryon differ from the string associated with the linear Regge trajectory.
The question at this point is   to what extent is the quantum stringy hadrons different from that of the linear Regge trajectories on top of the classical difference. As will be shown the quantum picture of the HISH is much more complicated.   Elevating  the classical picture  to the quantum one can be done in principle  in two different  ways: (i) Quantizing the  strings in the holographic curved  background, or (ii) Quantizing  the strings of the  HISH model, namely after mapping to flat four dimensional space-time including the impact of the Liuville mode. In principle the two paths should yield the same result. This was demonstrated in \cite{Aharony:2009gg} where
the effective action up to six-derivative order for the special case of  certain confining
string background  was computed and found out to be the Nambu Goto action in four dimensions.
However, in our prescription
 it is not obvious that the two procedures commute, that is  if we follow (i) and then use a map to flat-space time we get  the same picture as in (ii). In fact this  for the closed string system this is not the case. The map of the holographic folded rotating  closed string  at the classical level yields a folded rotating  closed string in flat space time where the only memory of the holographic background is the  effective tension that depends on the background. Obviously the  quantization of this string is the same as an ordinary closed  string in four dimensions. However, performing the quantization in certain holographic backgrounds as was done in \cite{PandoZayas:2003yb}, yielded a result which is quite different from  the one found following (ii).

We started this section with basic properties of the quantization of the closed and open bosonic string in flat 26 dimensions. We then review the quantization to one loop  of closed string in a holographic background. Next we discuss the quantization of the string in non-critical dimensions.
 The case of a static  string with massive endpoints in the critical dimensions is then described  and finally  we breifly review  certain issues about the quantization   of the rotating string with massive endpoints in four dimensions  and in particular in the limits of small and large string endpoint masses.
\subsection{ The  quantum  rotating closed string in holographic background}\label{sec:holo_string}

In section $(\S 3.4.2 )$ we have described the classical configuration of the folded rotating string in a holographic  confining background. The result is summarize in
\be J = \frac{1}{2}\alp_{eff} E^2 \ee
where $\alp_{eff} = (2\pi T_{eff})^{-1}$  and $T_{eff} = g_{00}(r_0)T$.
Now  we will review the semi-classical quantization of this string.
The calculations of the quantum corrected trajectory of the folded closed string in a curved background in different holographic backgrounds were performed in \cite{PandoZayas:2003yb} and \cite{Bigazzi:2004ze} using semiclassical methods. This was done in the analog of the static gauge, namely with $\tau= x^0$ and $\sigma = \rho=\sqrt{x_1^2+x_2^2}$  by computing the spectrum of quadratic fluctuations, bosonic and fermionic, around the classical configuration of the folded string.
It was shown in \cite{PandoZayas:2003yb} that the Noether charges of the energy  $E$ and angular momentum  $J$ that incorporate the quantum fluctuations, are related to the expectation value of the world-sheet Hamiltonian in the following manner
\cite{Frolov:2002av}
\be
LE -J = \int d\sigma \langle{\cal H}_{ws}\rangle \,.
\ee
The contributions to the expectation value of the world-sheet  Hamiltonian are from several massless bosonic modes, ``massive" bosonic modes and massive fermionic modes.
 For the ``massive" bosonic fluctuations around the rotating solution one gets a \(\sigma\)-dependent mass term, with equations of motion of the form
\be (\partial_\tau^2-\partial_\sigma^2+2m_0^2L^2\cos^2\sigma)\delta x^i = 0 \ee
appearing in both analyses, \(m_0\) being a mass parameter which depends on the particular geometry. A similar mass term, also with \(\cos\sigma\), appears in the equations of motion for some fermionic fluctuations as well, the factor of \(\cos^2\sigma\) in the mass squared coming in both cases from the induced metric calculated for the rotating string, which is \(h_{\alpha\beta} \sim \eta_{\alpha\beta}\cos^2\sigma\). Finally summing up all the quadratic quantum fluctuation it was found in both \cite{PandoZayas:2003yb} and \cite{Bigazzi:2004ze}
 that the Regge trajectories are of the form
\be J = \alp_{closed}(E^2- 2m_0 E) +a  \,.\ee
where $m_0$ is a mass parameter that characterizes the holographic model and $a$ is the intercept which generically takes the form $a= \frac{\pi}{24}(\#\text{bosonic massless  modes} - \#\text{fermionic massless modes})$. The two papers \cite{PandoZayas:2003yb} and \cite{Bigazzi:2004ze} use different holographic models (Klebanov-Strassler and Maldacena-N\'{u}\~{n}ez backgrounds in the former and Witten background in the latter) and predict different signs for \(m_0\), which is given as a combination of the parameters specific to the background. In \cite{PandoZayas:2003yb} \(m_0\) is positive, while in \cite{Bigazzi:2004ze} it is negative. According to \cite{PandoZayas:2003yb} the slope of the closed string trajectory is left unchanged from the classical case
\be \alp\!_{closed} = \frac{1}{2}\alp\!_{open} \,,\ee
\subsection{ On the quantization of the string in non-critical dimensions}
The HISH models are by definition in four dimensional, namely in non-critical dimension.
The question of quantizing  the classical string not in the critical dimensions was addressed
in  the seminal work of Polyakov. It was later addressed again by Polchinski and Strominger in the context of an  effective string theory  theory  \cite{Polchinski:1991ax}.
 Recently, it was discussed  again in \cite{Hellerman:2013kba} in the context  Nambu- Goto formulation and in
 \cite{Hellerman:2014cba} in terms of a Liuville mode. The basic observation is that   for the  quantum  effective string action in D dimensions, to be  2d conformal invariant one has to add a Liouville term of the form
\be
S_L= \frac{26-D}{24\pi} \int d^2 \sigma\sqrt{|g|}[ g^{ab}\pa_a \varphi\pa_b \varphi - {\cal R}_2\varphi]
\ee
where the Liouville field can be  taken to be a composite field of the form
\be
\varphi =-\frac12 Log( g^{ab}\pa_a x^\mu\pa_bx_\mu)
\ee
For a classical rotating string of the form
\be\label{classol}
X\equiv ( x^0,x^1,x^2) = l ( \tau, \cos(\t)\sin(\sigma), \sin(\t)\sin(\sigma))
\ee
one find that
\be
\varphi = - Log(\sqrt 2 l\cos(\sigma)).
\ee
Thus the Liouville term reads
\be\label{Liuv}
S_L= \frac{26-D}{24\pi}\int d\tau \int_{-\delta}^\delta d\sigma tan^2(\sigma) =  \frac{26-D}{12\pi} \left [ \tan(\delta) -\delta\right ]\ee
where the span of the world sheet coordinate $\sigma$ is taken to be  $-\delta\leq  \sigma\leq \delta$

An alternative formulation for the non-critical term was proposed by  Polchinski-Strominger  in (\cite{Polchinski:1991ax}).
In the orthogonal gauge
\be
\dot X^\mu \dot X_\mu +  {X'}^\mu  X'_\mu=0 \qquad \dot X^\mu  X'_\mu =0
\ee
 which can be expressed also as
\be
h_{++}=\pa_+ X^\mu\pa_+ X_\mu= 0 \qquad h_{--}=\pa_- X^\mu\pa_- X_\mu= 0
\ee
it reads
\be\label{PolStr} {\cal S}_{ps} =
\frac{26-D}{24\pi}\int d\t \int_{-\delta}^\delta d\sigma
\frac{(\pa_+^2X\cdot \pa_-X)(\pa_-^2X\cdot \pa_+X)}{(\pa_+X\cdot
\pa_-X)^2}= \ee
Substituing the classical solution (\ref{classol}) we get obviously the same result as in (\ref{classol}).

This term has to be added to   all the HISH models to the closed string one as well as to the open string both with massless or massive endpoints.
\subsubsection{The folded closed string}
Let us start first with the closed string case. In this case $\delta=\frac{\pi}{2}$ and thus the non-critical term diverges. This follows from the fact that  the denominator in the non-critical PS  term is simply \((\dot{X}^2)^2\), so the problem emerges because the endpoints move at the speed of light. This divergence was discussed also in
 \cite{Hellerman:2013kba} where the quantum correction to the linear Regge trajectory was analyzed   for a general dimension $D$.  In dimensions larger than four the string  rotates in two planes and the angular momentum is characterized by two quantum numbers \(J_1\) and \(J_2\). The result obtained there for the Regge trajectory of the closed string is
\be \frac{\alp}{2}M^2 = (J_1+J_2) - \frac{D-2}{12} + \frac{26-D}{24}
\left((\frac{J_1}{J_2})^\frac{1}{4}-(\frac{J_2}{J_1})^\frac{1}{4}\right)^2 \,. \ee
This expression is singular when \(J_2 = 0\), which is necessarily the case when \(D = 4\), since in four dimensions the rotation is in a single plane. Therefore the expression is not usable precisely in the context in which we would like to use it.
 One potential way to regularize it is  to add two masses at the two endpoints of the folded string. The resulting system looks like two open strings connected at their boundaries by these masses, but not interacting in any other way. In the rotating solution the two strings lie on top of one another. The boundary condition, which is the equation of motion of the massive endpoint is modified: it is the same as for the open string, but with an effective double tension \(T \rightarrow 2T\), in accordance with the ratio of the slopes of the open and closed strings discussed above. If this process of adding masses on the closed string and taking then the limit of zero mass is a legitimate way to regularize, then it is probable that the result is also simply double that of the open string, as it is for the critical dimension.

\subsubsection {The regular open string}

 For the regular open string, namely, with no massive endpoints and for which $\delta=\frac{\pi}{2}$ the PS term diverges and correspondingly the intercept.
  Even for small masses $\delta\sim\frac{\pi}{2}$ where the  contribution to the intercept  due to this term will be very large  is un-physical.  Thus there is a question of how to regularize and renormalize this expression. In \cite{Hellerman:2013kba} a procedure to do it was proposed. Here we would like to mention an approach that seems natural for the case of a string with massive endpoints. One option is to add  a term in the world-line action  analogous to the PS or Liouville term can cancel the $tan(\delta)$ term. One  can try to introduce a Liouville term also on the world line but since $h_{\tau\tau}= l^2 cos^2 (\sigma)$ it is clear that a term of the form $\pa_\tau\varphi$ vanishes.
Another option is to introduce a `` counter-term" of the form
\be
S_{ct}= \int d\tau {\cal L}_{ct} = \delta_m\int d\tau \sqrt{ \dot X^\mu \dot X_\mu} = \delta_m l cos(\delta)
\ee
It is thus clear that if we take $\delta_m= \frac{26-D}{24\pi}\frac{T}{m}$ we get $S_{ct}=\frac{26-D}{24\pi} tan(\delta)$. This counter-term cancels the divergent term  of the Polchinski Strominger or Liouville term. If we combine the finite term with the usual intercept we get \cite{Hellerman:2013kba}
\be
a= \frac{D-2}{24} + \frac{26-D}{24} = 1
\ee
This result that the intercept is $D$ independent may look counter-intuitive

\subsection{ The Casimir energy of  a critical  static string with massive endpoints}\label{CasEnrsta}
We have seen that the stringy hadrons both mesons and baryons are described in the HISH approach as strings with massive endpoints rotating in four dimensions.
Classically they were treated in section $\S( 4.2 )$. Before analyzing the quantum corrections to this system we first start with a static string in the critical dimension, namely without considering the Liouville term. In the next subsection will use discuss  the non-critical rotating string with massive endpoints.

Consider the case of a  string of length $L$ with massive endpoints  which is  static (i.e. non-rotating). The endpoints can move in directions perpendicular to the direction along which the string is stretched and the string can fluctuate along those directions
The solutions for the  fluctuations in the transverse directions are
\be
\delta x^\mu = \frac{1}{\sqrt{2 T}} \sum_{n\neq 0} e^{-iw_nt } \frac{\alpha^\mu_n}{w_n} u_n(\rho)
\ee
This solution is in the static gauge where $\tau= x^0$ and $\sigma= \rho$ where $\rho$ is the coordinate along the string and where  $\mu= 1,2,...D-2$.
The orthogonality of the transverse modes is given by
\be
\int_0^L d\rho  u_n(\rho) u_m(\rho) \epsilon(\rho) = \delta_{nm}\qquad \int_0^L dr  u'_n(\rho) u'_m(\rho) = w_n^2 \delta_{nm}
\ee
where $\epsilon(\rho) = 1+ \frac{m}{T} [\delta(\rho+L/2) + \delta(\rho-L/2)]$.

The eigenfrequencies $w_n$ are the roots of the following equation
\be\label{eigenstatic}
tan( w_nL) = \frac{2m Tw_n}{m^2 w_n^2 -T^2}
\ee
where we have used    the gauge $\sigma=\rho$ and hence  $\delta =L$.
Defining the  dimensionless eigenfrequencies $\hat w_n= w_n L$ and $q=\frac{m}{TL}$ we get the equation
\be\label{eigenstatica}
tan( \hat w_n) = \frac{2 q \hat w_n}{q^2 {\hat w_n}^2 -1}
\ee
On the $w$ axis  the roots are placed symmetrically around $w=0$ and hence it is enough to consider only the positive roots.

Upon quantization the creation and annihilation operators $\alpha_n$ obey the algebra
\be
[\alpha^\mu_n,\alpha^\nu_m] = w_n\delta^{\mu\nu} \delta_{n+m,0}
\ee
The energy  due to the fluctuations is given by
\be
E= \frac{T}{2}\int_{=L/2}^{L/2} d\rho (\delta\dot X^i)^2\epsilon(\rho) + (\delta {X^i}')^2
\ee
Substituting the creation and annihilation operators
\bea
E&=&\frac{1}{L} \sum_{n=1}^\infty \sum_{i=1}^{D-2}(\alpha_n^i{\alpha_n^i}^\dagger +{\alpha_n^i}^\dagger\alpha_n^i) \CR
&=& \sum_{n=1}^\infty \sum_{i=1}^{D-2}{\alpha_n^i}^\dagger\alpha_n^i +\frac{D-2}{2}\frac{1}{L}\sum_{n=1}^\infty w_n
\eea
The Casimir energy of this static string reads
\be\label{CasimirE}
E_C(m) = \frac12\sum_{n=1}^\infty w_n
\ee
For the special cases of $m=0$ and $m=\infty$ one gets using the zeta function regularization
\be
E_C(m=\infty)=E_C(m=0) = \frac{\pi}{2 L}\sum_{n=1}^\infty n = -\frac{\pi}{24L}
\ee
For non-trivial and finite $m$  the eigenfrequencies take the form of $w_n = n + f(L)\frac{1}{n}$.
The second term cannot be regularized using zeta function regularization. Instead following \cite{Lambiase:1995st} we use Cauchy's theorem to regularize the sum and  then we perform a renormalization procedure that yields a finite result which for the massless case coincides with the zeta function regularization.

The main idea \cite{Lambiase:1995st} is to express the sum of the eigenfrequencies using the following relation for an analytic function $f(w)$
\be
\frac{1}{2\pi i} \oint_C dw w \frac{f'(w)}{f(w)}= \frac{1}{2\pi i} \oint_C dw w [Log f(w) ]'= \sum_k n_k w_k -\sum_l p_l \tilde w_l
\ee
where $w_k$ denotes a zero of order $k$, $n_k$ the number of them  and $\tilde w_l$ denotes a pole of order $l$ and $p_l$ the number of them.
The analytic function $f(w)$ for our case is the eigenfrequencies equation (\ref{eigenstatic}) taken to avoid having poles in the form
\be
f(w)= 2mTw cos(w L) -( m^2 w^2- T^2) sin( wL) =0
\ee
Substituting this expression in (\ref{CasimirE})  we get
\be
E_C(m)= \frac{1}{4\pi i} \oint_C dw w [Log f(w) ]'
\ee
where the contour $C$ includes the real positive semiaxis where the roots of $f(w)$ occur. Since $f(w)$ does not have poles we deform the contour to be a semi-circle with radius $\Lambda$ and the segment along the imaginary axis  $(-i\Lambda, i\Lambda)$. The integral along the deformed contour is finite for finite $\Lambda$ and hence this can be taken as $E^{(reg)}_C(m , L)$ the regularized Casimir energy.
\bea\label{Ereg}
E^{(reg)}_C(m , L)&=& \frac{1}{2\pi} \int_0^\Lambda dy  Log \left [2m Ty \cosh( yL) + ( m^2 y^2 + T^2) \sinh ( yL) \right]\CR  &+& \frac{1}{4\pi} \left [  w Log [f(w)]\right ]^\Lambda_{-\Lambda} + I_{sc}(\Lambda)\CR
\eea

where the second term is the surface term that follows from the integration by parts and the third term is the semi-circular integral at radius $\Lambda$. For large $\Lambda$  the surface term becomes $2 \Lambda^2 L + \Lambda Log [ -\frac14 ( m\Lambda + T)^4]$. However, if one integrate by parts also the integral along the semi-circle then the surface terms drop out.

To perform the renormalization we subtract $E^{(reg)}_C(m , L\rightarrow \infty)$. In this method the renormalized Casimir energy reads
\be
 E^{(ren)}_C(m , L)= \lim_{\Lambda\rightarrow \infty}[ E^{(reg)}_C(m , L)-E^{(reg)}_C(m , L\rightarrow \infty)]
\ee
The subtracted Casimir energy is the asymptotic value of the regularized Casimir energy when $L\rightarrow \infty$. It also has three contributions. However, it easy  to see that the surface term takes the same value as  the one in (\ref{Ereg}) for large $\Lambda$. Moreover, the  difference between the integrals along the semi-circle radius vanishes. Thus the subtracted energy takes finally the form
\be
E^{(reg)}_C(m , L\rightarrow \infty )= \frac{1}{2\pi} \int_0^\Lambda dy  Log\left [e^{( yL)}\frac{( m y + T)^2 }{2}\right ]
\ee
and hence the renormalized Casimir energy is  given by
\be
E^{(ren)}_C(m , L)= \frac{1}{2\pi L}\int_0^\infty dx   Log \left [ 1-  e^{-2x} \left (  \frac{(x-a)}{x+a)}\right )^2
\right ] \ee
where $a$ is the dimensionless quantity $a = \frac{TL}{m}$. For the special case of massless endpoints one finds
\be\label{Eren}
E^{(ren)}_C(m=0 , L)=E^{(ren)}_C(m=\infty , L)= \frac{1}{2\pi}\int_0^\infty dx   Log \left [ 1-  e^{-2xL} \right ]= -\frac{\pi}{24 L}
\ee



Note that one  cannot  get rid of the divergences by computing the difference of the integral for two different finite length $L$ or two different tensions $T$ or two different masses. The difference in these cases is still divergent.

\begin{figure}[h!] \centering
                    \includegraphics[width=.65\textwidth, natheight = 784bp, natwidth = 1374bp]{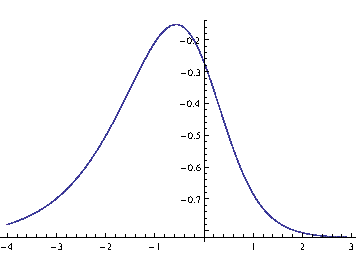}
                    \caption{\label{etaa}  $\eta$ as a function of $Log_{10}(a) $ }
\end{figure}

Defining the ratio of the Casimir energy of a finite and infinite  string endpoint mass one gets
\be
\eta(q) =\frac{E_c^{(ren)}(m,L)}{E_c^{(ren)}(m=\infty,L)}=-\frac{12}{\pi^2}\int_0^\infty dz Log\left[1-e^{-2 z}\left (\frac{1-az}{1+az}\right )^2\right ]
\ee
The dependence of $\eta$ on $Log_{10}(q)$ is drawn in figure(\ref{etaa}).In both limits of $q\rightarrow 0$ and  $q\rightarrow \infty$,
$\eta\rightarrow 1$ and its extremum is around $q=\frac{m}{TL}\sim 2$.

\subsection{ On the quantization of the rotating string with massive endpoints in four dimensions}
We have determined the  classical trajectories of the strings with massive endpoints, we have described the quantizaton of the regular open string in non-critical dimension and the quantization of the static string with massive endpoints. Now we would like to combine all these ingredients to determine the semi-classical stringy meson ( and baryon). This task is under current investigation\cite{ASY}. Here we briefly mention several of its features.
\begin{itemize}
\item
In section $\S(4.2)$ we were using the Nambu Goto action for the string and the length of the endpoints worldline as the particle action. For the string one can use also the Polyakov formulation and for the particle an action with an auxiliary field so that in fact there are four possible formulation of the action of the combined system.
\item
The source of the difficulty of quantizing this system is the fact that in the Polyakov formulation of the string there are only the two diffeomorphism and no scale symmetry. One can gauge fix the local symmetries in various ways in particular the orthogonal gauge  and the analog of the static gauge mentioned above and a gauge where the fluctuations are perpendicular to the classical configuration \cite{oai:arXiv.org:hep-th/9911123}, \cite{Zahn:2013yma}.
\item
The non-linear nature of the system shows up in the boundary equation  that in the formulation without worldsheet and worldline metrics read
\be\label{bequQ}
 T\del_\sigma X^\mu \pm  m\del_\tau\left(\frac{\dot{X}^\mu}{\sqrt{-(\dot X)^2}}\right) = 0
 \ee
\item
The quantum fluctuations transverse to the plane of rotation, transverse to the rotation in the plane of rotation  and those along the direction of the string ( which exists only for the string with massive endpoints), obey different equations of motion and have to be treated separately.
\item
Even without deriving a full semi-classical quantization it seems plausible that the transition from the classical trajectory given in (\ref{Classtra}) to the quantum one will be following

\be
\alpha' E_{cl}^2\rightarrow  \alpha' E_{qm}^2= \alpha' E_{cl}^2 + a = \alpha' E_{cl}^2 + (a_{Cas} + a_{PS})
\ee
and the leftover challenge is to compute or at least estimate the contributions to the intercept $a_{Cas}$ and $a_{PS}$

\item
Needless to say that  eigenvalues of the  excited  string states  with massive endpoints is not anymore just the integers $w_n=n$ but rather the eigenvalues discussed above in the previous section. Some low lying states were determined in \cite{Zahn:2013yma}
\end{itemize}
\section{Phenomenology: Comparison between the stringy models and experimental data}
Now that we have completed the development of the theoretical model starting from stringy configurations in holographic models and then mapping them into strings in  four dimensional flat space-time, we would like to perform a detailed set of fits between the models and the corresponding hadrons. We will first describe the fitting procedure and then present separately the fits of mesons (section \ref{sec:mesons})\cite{Sonnenschein:2014jwa}\cite{Sonnenschein:2014bia}, the fits of baryons (section \ref{sec:baryons}), and the fits of glueballs (section \ref{sec:glueballs}\cite{Sonnenschein:2015zaa}).

\subsection{Fitting models and procedure}
		We define the \emph{linear} fit by
			\be J + n = \alp E^2 + a \label{eq:linear} \ee
		where the fitting parameters are the \emph{slope} \alp and the \emph{intercept}, \(a\).
				
		For the \emph{massive} fit, we use the general expressions for the mass and angular momentum of the rotating string with massive endpoints (of eqs. \ref{eq:massiveE} and \ref{eq:massiveJ}), generalized for the case of two different masses, and we add to them, by hand, an intercept and an extrapolated \(n\) dependence, assuming the same replacement of \(J \rightarrow J + n - a\) when moving from the classical result to the real, quantum world.
			\be E = \sum_{i = 1,2}m_i\left(\frac{q_i\arcsin(q_i)+\sqrt{1-q_i^2}}{1-q_i^2}\right) \label{eq:massFitE} \ee
	\be J + n = a + \sum_{i=1,2}\pi\alp m_i^2\frac{q_i^2}{(1-q_i^2)^2}\left(\arcsin(q_i)+q_i\sqrt{1-q_i^2}\right) \label{eq:massFitJ} \ee
		With the relation between \(q_1\) and \(q_2\) as in eq. \ref{eq:boundaryRot}:
		\be \frac{T}{\omega} = m_1\frac{q_1}{1-q_1^2} = m_2\frac{q_2}{1-q_2^2} \label{eq:boundaryTwo}\ee
		
		With the two additions of \(n\) and \(a\), the two equations reduce to that of the linear fit in \eqref{eq:linear} in the limit where both masses are zero.
		
		Now the fitting parameters are \(a\) and \(\alp\) as before, as well as the the two endpoint masses \(m_1\) and \(m_2\). In many cases we assume \(m_1 = m_2\) and retain only one free mass parameter, \(m\).
		
		\paragraph{Fitting procedure:} For the \emph{meson and baryon fits} of sections \ref{sec:mesons} and \ref{sec:baryons}, the fits are made to minimalize \(\chi^2\), defined here as
						\be \chi^2 = \frac{1}{N-1}\sum_i\left(\frac{M_i^2-E_i^2}{M_i^2}\right)^2 \label{eq:chi_def} \ee
			\(M_i\) and \(E_i\) are, respectively, the measured and calculated value of the mass of the \(i\)-th particle, and \(N\) the number of points in the trajectory. We will also use the subscripts \(l\) (for linear fit) or \(m\) (for massive fit) to denote to which fitting model a given value of \(\chi^2\) belongs.
			We put \(M_i^2\) in the denominator instead of the experimental uncertainty \(\Delta M_i^2\), because our models cannot replicate the high accuracy with which some hadron masses are measured. In the above definition \(\chi^2\) is normalized in such a way that \(\sqrt{\chi^2}\) gives the percentage of deviation in \(M^2\).
			
			For the \emph{glueballs and lattice fits} of section \ref{sec:glueballs}, we work with the definition of \(\chi^2\) with experimental (or computational on the lattice) uncertainty in the denominator
			\be \chi^2 = \frac{1}{N-1}\sum_i\left(\frac{M_i^2-E_i^2}{\Delta (M_i^2)}\right)^2\,.\ee
\subsection{Review of meson trajectory fits} \label{sec:mesons}
We begin with a summary of the main results of the meson fits. Following that we present a ``universal'' fit for the \((J,M^2)\) trajectories of mesons composed of \(u\), \(d\), \(s\), and \(c\) quarks. The latter parts of this section are more detailed discussion of the different meson fits.

\subsubsection{Summary of results for the mesons}

			\begin{table}[ht!] \centering
					\begin{tabular}{|c|c|cc|c|c|c|} \hline
					Traj. & \(N\) & \multicolumn{2}{|c|}{\(m\)} & \alp & \(a\) \\ \hline
					
					\(\pi/b\) & \(4\) & \multicolumn{2}{|c|}{\(\mud = 90-185\)} & \(0.808-0.863\) & \((-0.23)-0.00\) \\
					
					\(\rho/a\) & \(6\) & \multicolumn{2}{|c|}{\(\mud =0-180\)} & \(0.883-0.933\) & \(0.47-0.66\) \\
				
					\(\eta/h\) &  \(5\) & \multicolumn{2}{|c|}{\(\mud = 0-70\)} & \(0.839-0.854\) & \((-0.25)-(-0.21)\) \\
					
					\(\omega\) &  \(6\) & \multicolumn{2}{|c|}{\(\mud = 0-60\)}& \(0.910-0.918\) & \(0.45-0.50\) \\
					
					\(K^*\) &  \(5\) & \(\mud = 0-240\) & \(m_s = 0-390\) & \(0.848-0.927\) & \(0.32-0.62\) \\
					
					\(\phi\) &  \(3\) & \multicolumn{2}{|c|}{\(m_s = 400\)} & \(1.078\) & \(0.82\) \\
					
					\(D\) &  \(3\) & \(\mud = 80\) & \(m_c = 1640\) & \(1.073\) & \(-0.07\) \\
					
					\(D^*_s\) & \(3\) & \(m_s = 400\) & \(m_c = 1580\) & \(1.093\) & \(0.89\) \\
					
					\(\Psi\) &  \(3\) & \multicolumn{2}{|c|}{\(m_c = 1500\)} & \(0.979\) & \(-0.09\) \\
					
					\(\Upsilon\) &  \(3\) & \multicolumn{2}{|c|}{\(m_b = 4730\)} & \(0.635\) & \(1.00\) \\ \hline \end{tabular}
					
					\caption{\label{tab:mes_j} The results of the meson fits in the \((J,M^2)\) plane. For the uneven \(K^*\) fit the higher values of \(m_s\) require \(\mud\) to take a correspondingly low value. \(\mud + m_s\) never exceeds 480 MeV, and the highest masses quoted for the \(s\) are obtained when \(\mud = 0\). The ranges listed are those where \(\chi^2\) is within 10\% of its optimal value. \(N\) is the number of data points in the trajectory. $m$ is given in MeV and $\alp$ in GeV$^{-2}$. }
				\end{table}
				
				\begin{table}[ht!] \centering
					\begin{tabular}{|c|c|c|c|cc|} \hline
					Traj. & \(N\) & \(m\) & \alp & \multicolumn{2}{|c|}{\(a\)} \\ \hline
					
					\(\pi\)/\(\pi_2\) & \(4+3\) & {\(\mud = 110-250\)} & \(0.788-0.852\) & \(a_0 = (-0.22)-0.00\) & \(a_2 = (-0.00)-0.26\) \\
					
					\(a_1\) & \(4\) & \(\mud = 0-390\) & \(0.783-0.849\) & \multicolumn{2}{|c|}{\((-0.18)-0.21\)} \\
					
					\(h_1\) & \(4\) & \(\mud = 0-235\) & \(0.833-0.850\) & \multicolumn{2}{|c|}{\((-0.14)-(-0.02)\)} \\
					
					\(\omega/\omega_3\) & \(5+3\) & \(\mud = 255-390\) & \(0.988-1.18\) & \(a_1 = 0.81-1.00\) & \(a_3 = 0.95-1.15\) \\
					
					\(\phi\) & \(3\) & \(m_s = 510-520\) & \(1.072-1.112\) & \multicolumn{2}{|c|}{\(1.00\)} \\
					
					\(\Psi\) & \(4\) & \(m_c = 1380-1460\) & \(0.494-0.547\) & \multicolumn{2}{|c|}{\(0.71-0.88\)} \\
					
					\(\Upsilon\) & \(6\) & \(m_b = 4725-4740\) & \(0.455-0.471\) & \multicolumn{2}{|c|}{\(1.00\)} \\
					
					\(\chi_b\) & \(3\) & \(m_b = 4800\) & \(0.499\) & \multicolumn{2}{|c|}{\(0.58\)} \\	\hline \end{tabular}
					
					\caption{\label{tab:mes_n} The results of the meson fits in the \((n,M^2)\) plane. The ranges listed are those where \(\chi^2\) is within 10\% of its optimal value. \(N\) is the number of data points in the trajectory.}
\end{table}
Tables \ref{tab:mes_j} and \ref{tab:mes_n} summarize the results of the fits for the mesons in the \((J,M^2)\) plane and \((n,M^2)\) plane respectively.

The higher values of \(\alp\) and \(a\) always correspond to higher values of the endpoint masses, and the ranges listed are those where \(\chi^2\) is within 10\% of its optimal value.

We list in the tables the optimal ranges for each fit done individually, but we want to consider the results as a whole. In the \((J,M^2)\) plane we see that the trajectories fitted, which are all the leading trajectories for mesons of various masses, are quite consistent in terms of the slopes and masses obtained from each trajectory. With the sole exception of the \(b\bar{b}\) trajectory, which gives a much lower slope than the rest, all the trajectories can be fitted well (if not always optimally) using a slope of \(\approx 0.9\) GeV\(^{-2}\). The masses of the light quarks (\(u\) and \(d\)) are not determined. The results are consistent with zero, but do not rule out masses up to 100 MeV. The fits for mesons containing \(s\), \(c\), and \(b\) quarks are always improved when adding masses, in terms of both quality and consistency of the slope obtained. The mass of the \(s\) quark is found around 300--400 MeV, while the \(c\) and \(b\) quarks are at their constituent masses of 1500 MeV and 4730 MeV respectively.

The intercepts appear to be scattered over different values, positive and negative. If we change our \(x\)-axis to the orbital angular momentum \(L\) instead of \(J\) (which is either \(L\) or \(L+1\)), we get only negative values: around \(-0.2\) for the light pseudoscalar trajectories (\(\pi\), \(\eta\)), and \(-0.4\) for the lighter vector trajectories (\(\rho\), \(\omega\), and also \(K^*\)). As the endpoint masses get heavier (\(\phi\), \(D\), \(D_s\), \(\Psi\), \(\Upsilon\)) the intercept moves to zero.

In the \((n,M^2)\) plane we have less consistency between the different trajectories. This might be expected since we base our fits on the conjecture that we can extrapolate the dependence of \(M^2\) on \(n\) directly from its dependence on \(J\). However, we still see that the light mesons can be fitted on linear or close-to-linear trajectories with a slope of some \(0.80-0.85\) GeV\(^{-2}\). The trajectory of the \(s\bar{s}\) has a higher slope. The fits of the \(c\) and \(b\) have much lower slopes than the light mesons, but work surprisingly well when fitted separately.

\subsubsection{Universal slope fits} \label{sec:universal}
		\begin{figure}[t!] \centering
					\includegraphics[natwidth=1200bp, natheight=900bp, width=.90\textwidth]{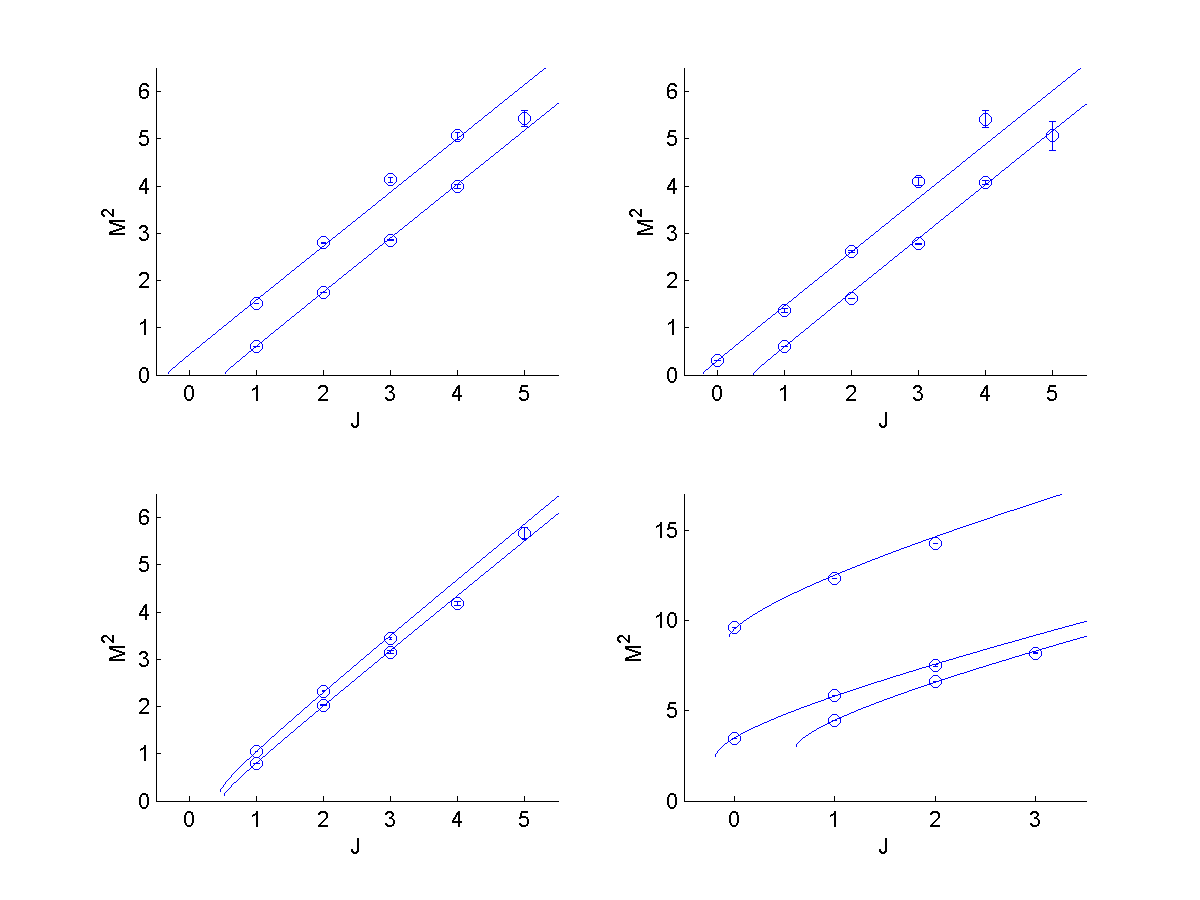}
					\caption{\label{fig:multiFit_j} Nine \((J,M^2)\) trajectories fitted using universal quark masses and slope (\(\mud = 60\), \(m_s = 220\), \(m_c = 1500\), and \(\alp = 0.884\)). Top left: \(\pi\) and \(\rho\), top right: \(\eta\) and \(\omega\), bottom left: \(K^*\) and \(\phi\), bottom right: \(D\), \(D^*_s\), and \(\Psi\).}
				\end{figure}
				
				Based on the combined results of the individual fits for the \((J,M^2)\) trajectories of the \(u\), \(d\), \(s\), and \(c\) quark mesons, we assumed the values
				\be m_{u/d} = 60, m_s = 220, m_c = 1500 \ee
				for the endpoint masses and attempted to find a fit in which the slope is the same for all trajectories. This wish to use a universal slope forces us to exclude the \(\bbb\) trajectory from this fit, but we can include the three trajectories involving a \(c\) quark. For these, with added endpoint masses (and only with added masses), the slope is very similar to that of the light quark trajectories.
								
				The only thing that was allowed to change between different trajectories was the intercept. With the values of the masses fixed, we searched for the value of \(\alp\) and the intercepts that would give the best overall fit to the nine trajectories of the \(\pi/b\), \(\rho/a\), \(\eta/h\), \(\omega/f\), \(K^*\), \(\phi\), \(D\), \(D^*_s\), and \(\Psi\) mesons. The best fit of this sort, with the masses fixed to the above values, was
				\be \alp = 0.884 \ee
				\[ a_\pi = -0.33 \qquad a_\rho = 0.52 \qquad a_\eta = -0.22 \qquad a_\omega = 0.53 \]
				\[ a_{K^*} = 0.50 \qquad a_\phi = 0.46 \qquad a_D = -0.19 \qquad a_{D^*_s} = -0.39 \qquad a_\Psi = -0.06\]
				and it is quite a good fit with \(\chi^2 = 13.13\ten{-4}\). The trajectories and their fits are shown in figure \ref{fig:multiFit_j}.

\subsubsection{Meson trajectories in the \texorpdfstring{$(J,M^2)$}{(J,M2)} plane}
We now present in more detail the fits to the different trajectories, beginning with the orbital trajectories, i.e. trajectories in the \((J,M^2)\) plane.
		\paragraph{Light quark mesons}	
		\begin{figure}[t!] \centering 
						\includegraphics[natwidth=1200bp, natheight=900bp, width=.40\textwidth]{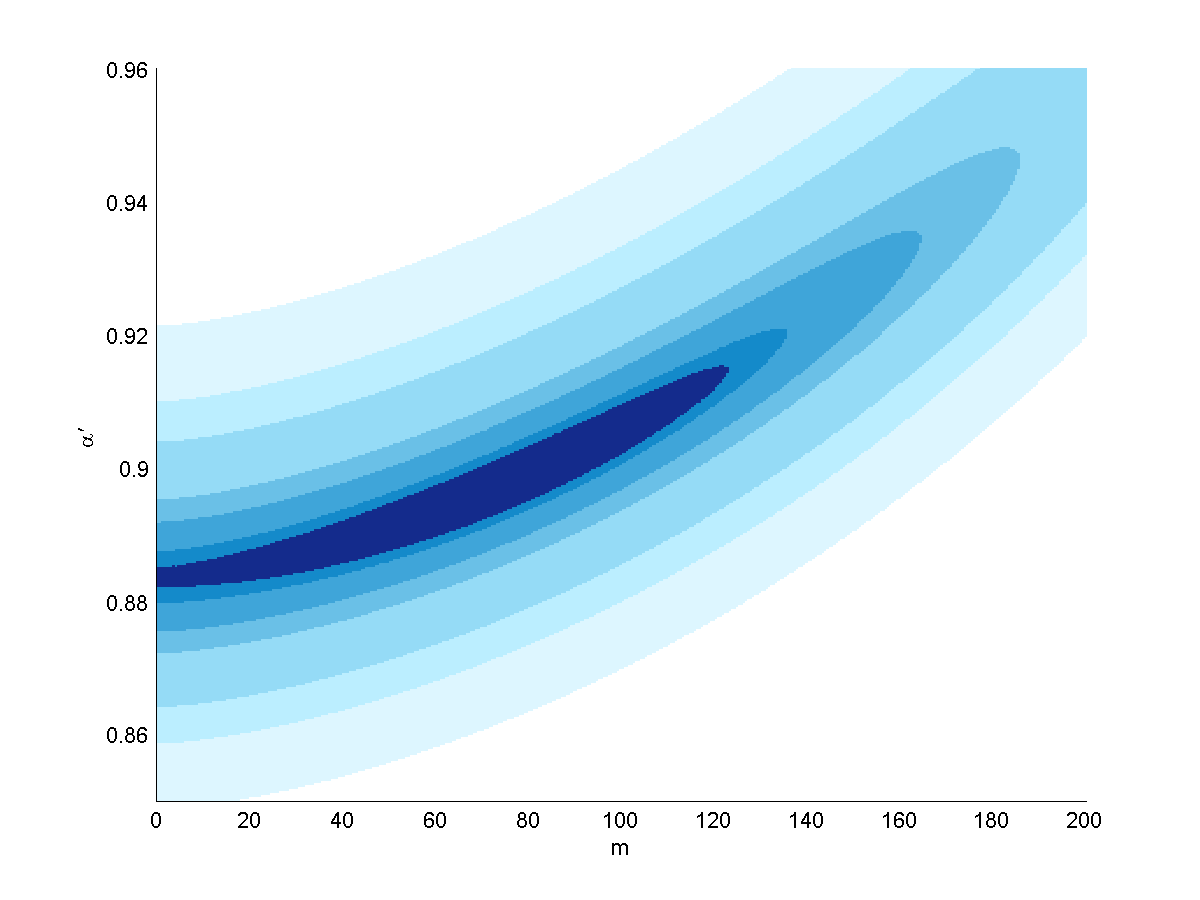}
						\includegraphics[natwidth=392bp, natheight=900bp, width=.13\textwidth]{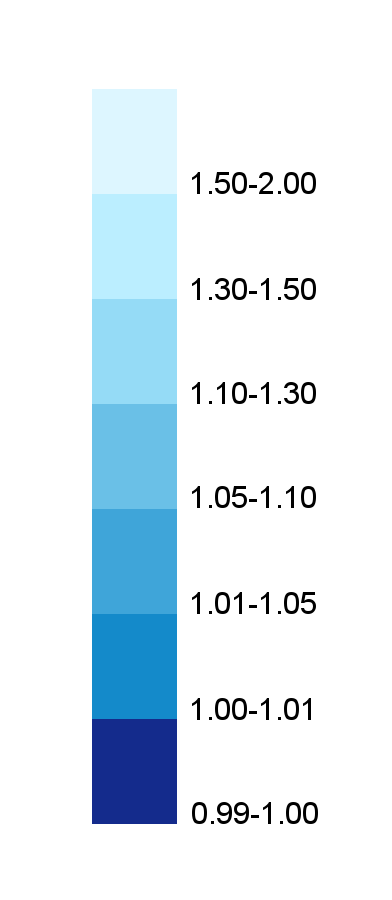}
						\includegraphics[natwidth=1200bp, natheight=900bp, width=.40\textwidth]{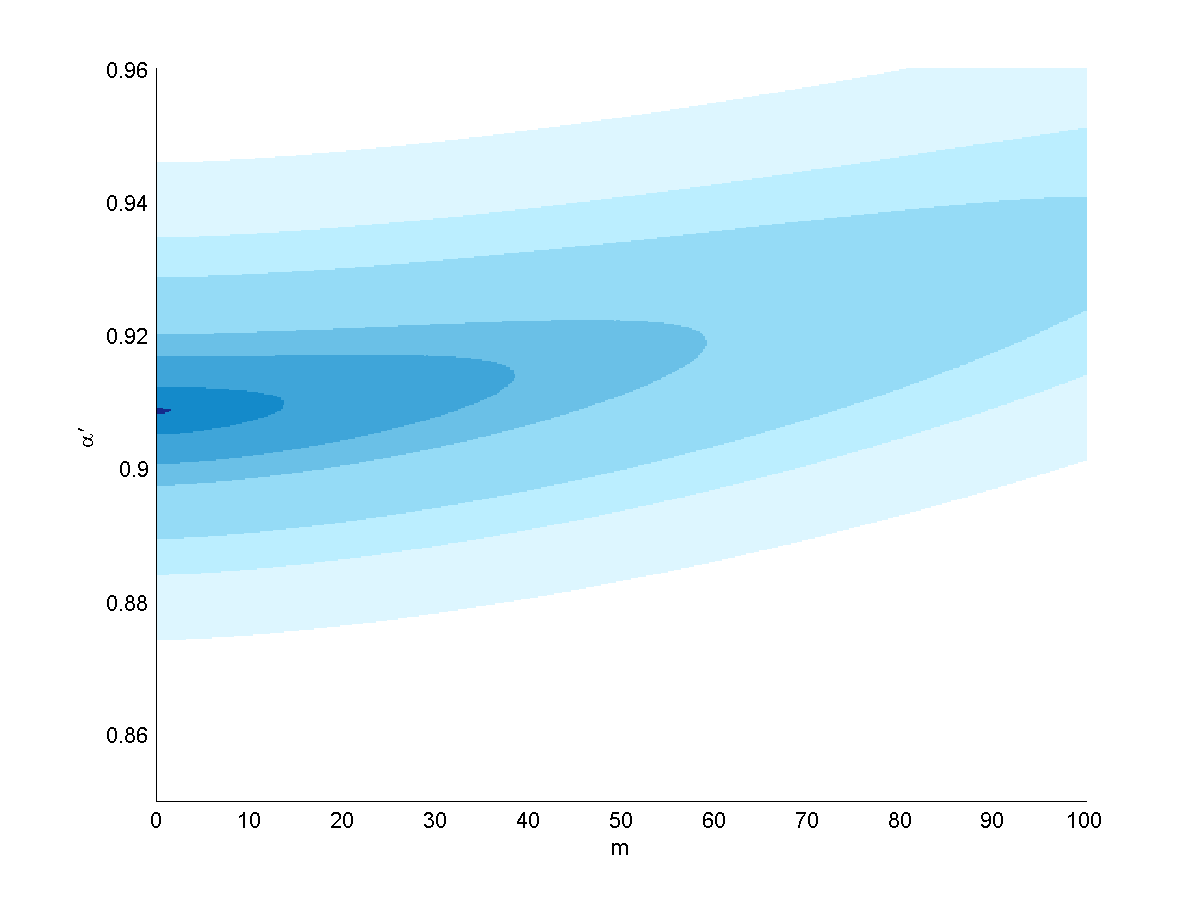}
						\caption{\label{fig:chi_3d_light} \(\chi^2\) as a function of \(\alp\) and \(m\) for the \((J,M^2)\) trajectory of the \(\rho\)  (left) and \(\omega\) (right) mesons. The intercept \(a\) is optimized to get a best fit for each point in the \((\alp,m)\) plane. \(\chi^2\) in these plots is normalized so that the value of the optimal linear fit \((m=0)\) is \(\chi^2 = 1\).}
\end{figure}
			We begin by looking at mesons consisting only of light quarks - \(u\) and \(d\). We assume for our analysis that the \(u\) and \(d\) quarks are equal in mass, as any difference between them would be too small to reveal itself in our fits.
			This sector is where we have the most data, but it is also where our fits are the least conclusive. The trajectories we have analyzed are those of the \(\pi/b\), \(\rho/a\), \(\eta/h\), and \(\omega/f\).
			
			Of the four \((J,M^2)\) trajectories, the two \(I = 1\) trajectories, of the \(\rho\) and the \(\pi\), show a weak dependence of \(\chi^2\) on \(m\). Endpoint masses anywhere between \(0\) and \(160\) MeV are nearly equal in terms of \(\chi^2\), and no clear optimum can be observed. For the two \(I = 0\) trajectories, of the \(\eta\) and \(\omega\), the linear fit is optimal. If we allow an increase of up to \(10\%\) in \(\chi^2\), we can add masses of only \(60\) MeV or less. Figure \ref{fig:chi_3d_light} presents the plots of \(\chi^2\) vs. \(\alp\) and \(m\) for the trajectories of the \(\omega\) and \(\rho\) and shows the difference in the allowed masses between them.
			
			The slope for these trajectories is between \(\alp = 0.81-0.86\) for the two trajectories starting with a pseudo-scalar (\(\eta\) and \(\pi\)), and \(\alp = 0.88-0.93\) for the trajectories beginning with a vector meson (\(\rho\) and \(\omega\)). The higher values for the slopes are obtained when we add masses, as increasing the mass generally requires an increase in \(\alp\) to retain a good fit to a given trajectory. This can also be seen in figure \ref{fig:chi_3d_light}, in the plot for the \(\rho\) trajectory fit.
			
		\paragraph{Strange and \texorpdfstring{$\ssb$}{s-sbar} mesons}
			We analyze three trajectories in the \((J,M^2)\) involving the strange quark. One is for mesons composed of one \(s\) quark and one light quark - the \(K^*\), the second is for \(\ssb\) mesons - the trajectory of the \(\phi\), and the last is for the charmed and strange \(D^*_s\), which is presented in the next subsection with the other charmed mesons.
					\begin{figure}[t!] \centering 
						\includegraphics[natwidth=1200bp, natheight=900bp, width=.48\textwidth]{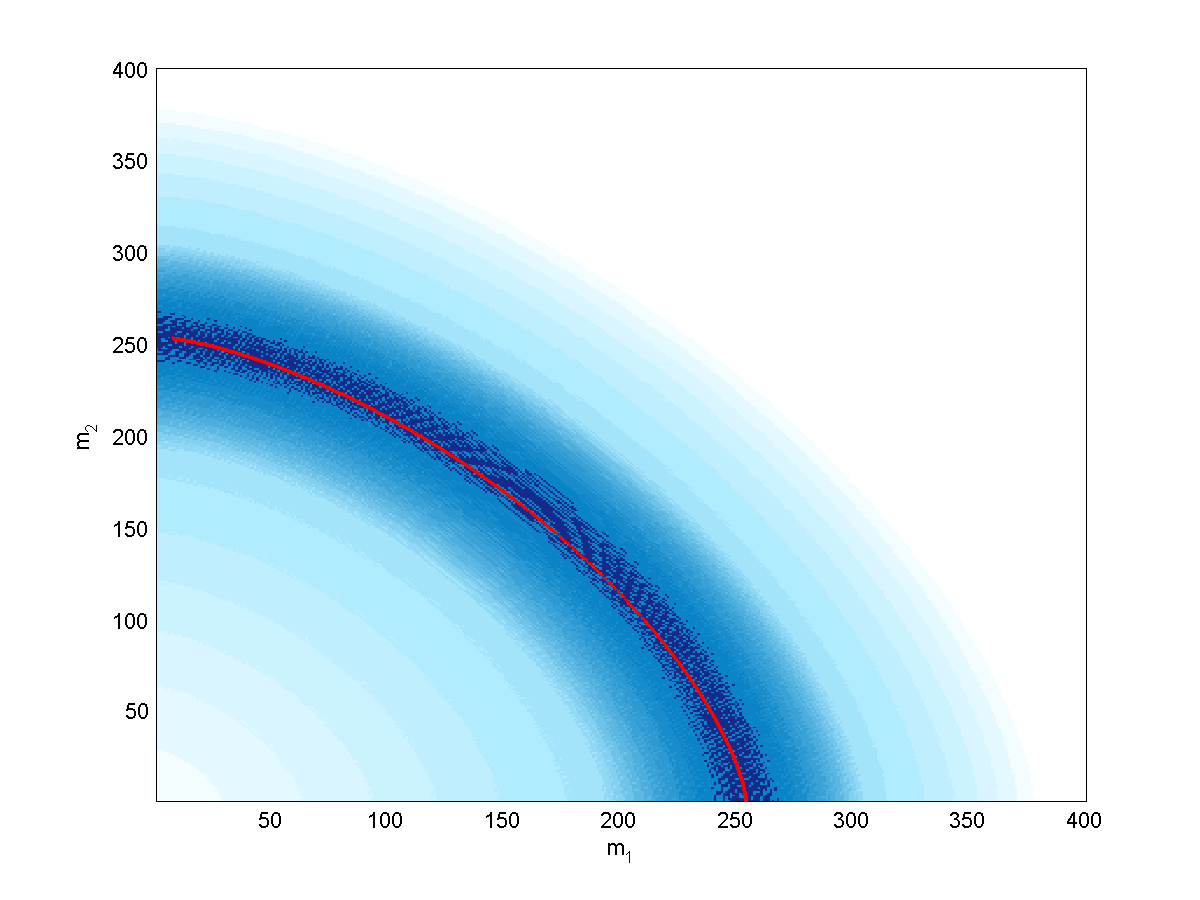}	 \hfill
						\includegraphics[natwidth=1200bp, natheight=900bp, width=.48\textwidth]{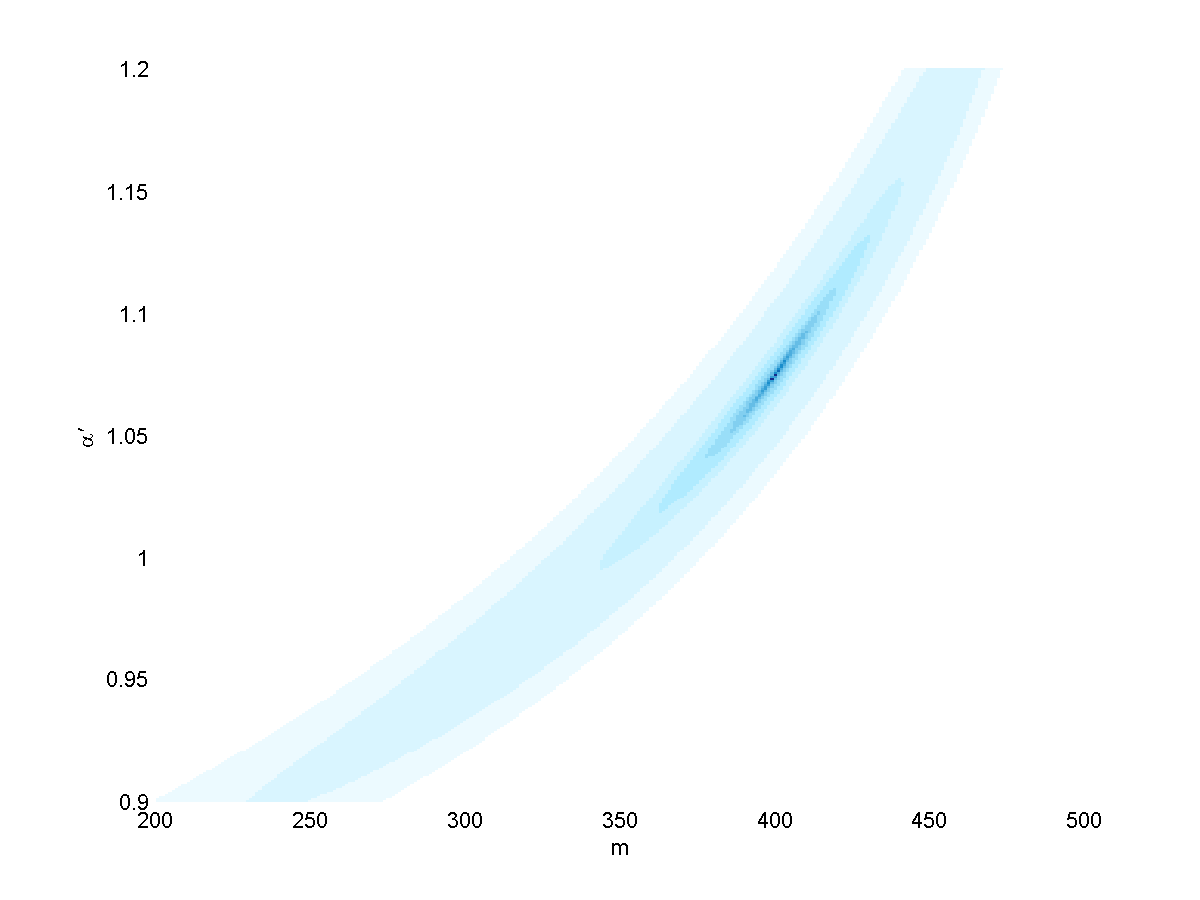}
						\caption{\label{fig:chi_j_strange} Left: \(\chi^2\) as a function of two masses for the \(K^*\) trajectory. \(a\) and \(\alp\) are optimized for each point. The red line is the curve \(m_1^{3/2}+m_2^{3/2} = 2\times(160)^{3/2}\) along which the minimum (approximately) resides. The minimum is \(\rchi{0.925}\) and the entire colored area has \(\chi^2_m/\chi^2_l < 1\). On the right is \(\chi^2\) as a function of \(\alp\) and \(m\) for the \((J,M^2)\) trajectory of the \(\phi\). The intercept \(a\) is optimized. The minimum is at \(\alp = 1.07, m = 400\) with \(\chi^2_m/\chi^2_l < 10^{-4}\) at the darkest spot. The lightest colored zone still has \(\chi^2_m/\chi^2_l < 1\), and the coloring is based on a logarithmic scale.}
				\end{figure}
			
			The \(K^*\) trajectory alone cannot be used to determine both the mass of the \(u/d\) quark and the mass of the \(s\). The first correction to the linear Regge trajectory in the low mass range is proportional to \(\alp\left(m_1^{3/2}+m_2^{3/2}\right)\sqrt{E}\). This is the result when eq. \eqref{eq:lowMass} is generalized to the case where there are two different (and small) masses. The plot on the left side of figure \ref{fig:chi_j_strange} shows \(\chi^2\) as a function of the two masses.
			
			The minimum for the \(K^*\) trajectory resides along the curve \(\mud^{3/2} + m_s^{3/2} = 2\times(160)^{3/2}\). If we take a value of around \(60\) MeV for the \(u/d\) quark, that means the preferred value for the \(m_s\) is around 220 MeV. The higher mass fits which are still better than the linear fit point to values for the \(s\) quark mass as high as 350 MeV, again when \(\mud\) is taken to be 60 MeV. The slope for the \(K^*\) fits goes from \(\alp = 0.85\) in the linear fit to \(0.89\) near the optimum to \(0.93\) for the higher mass fits.
			
			The trajectory of the \(\ssb\) mesons includes only three states, and as a result the optimum is much more pronounced than it was in previous trajectories. It is found at the point \(m_s = 400\), \(\alp = 1.07\). The value of \(\chi^2\) near that point approaches zero. The range in which the massive fits offer an improvement over the linear fit is much larger than that, as can be seen in the right side plot of figure \ref{fig:chi_j_strange}. Masses starting from around \(m_s = 250\) MeV still have \rchi{0.50} or less, and the slope then has a value close to that of the other fits, around \(0.9\) GeV\(^{-2}\).
		
		\paragraph{Charmed and \texorpdfstring{$\ccb$}{c-cbar} mesons}
		
		\begin{figure}[t!] \centering
						\includegraphics[natwidth=1200bp, natheight=900bp, width=.48\textwidth]{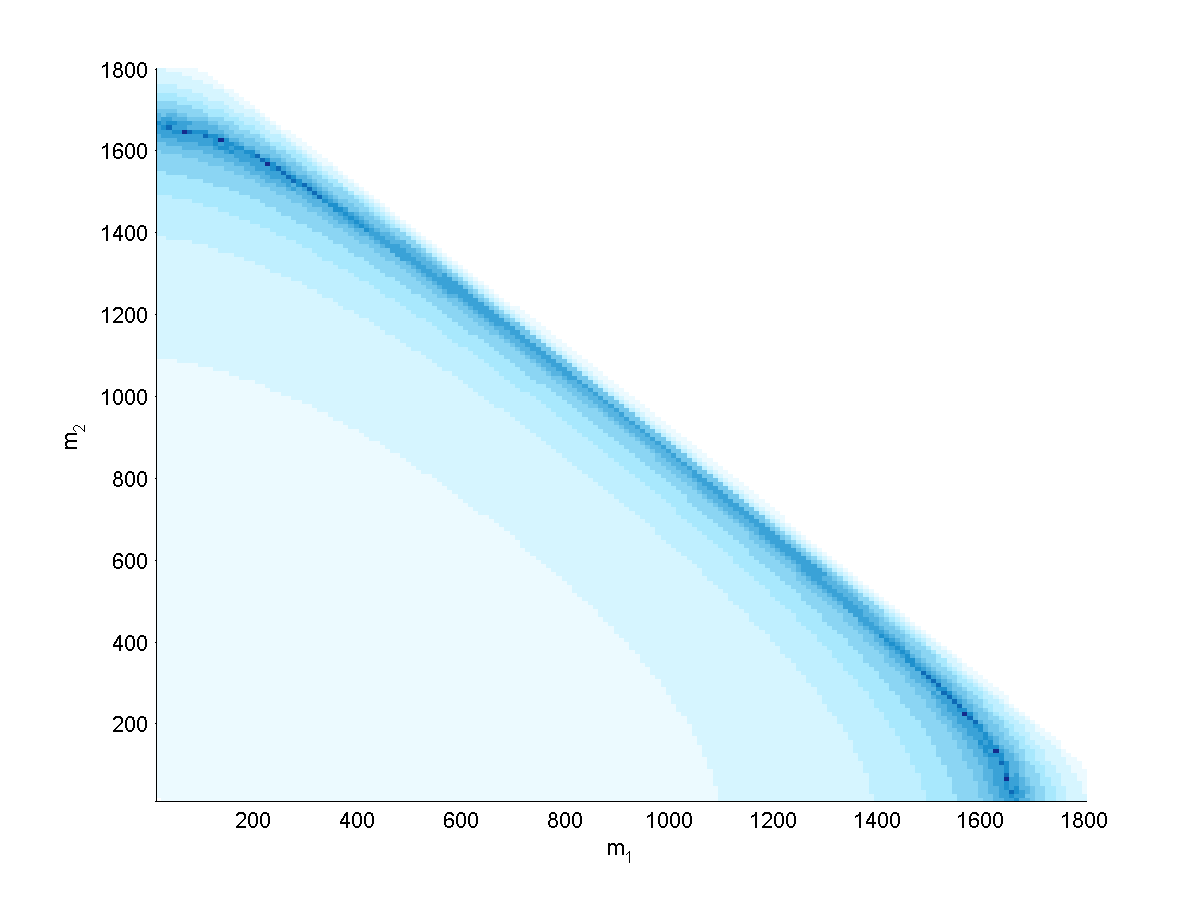}	 \hfill
						\includegraphics[natwidth=1200bp, natheight=900bp, width=.48\textwidth]{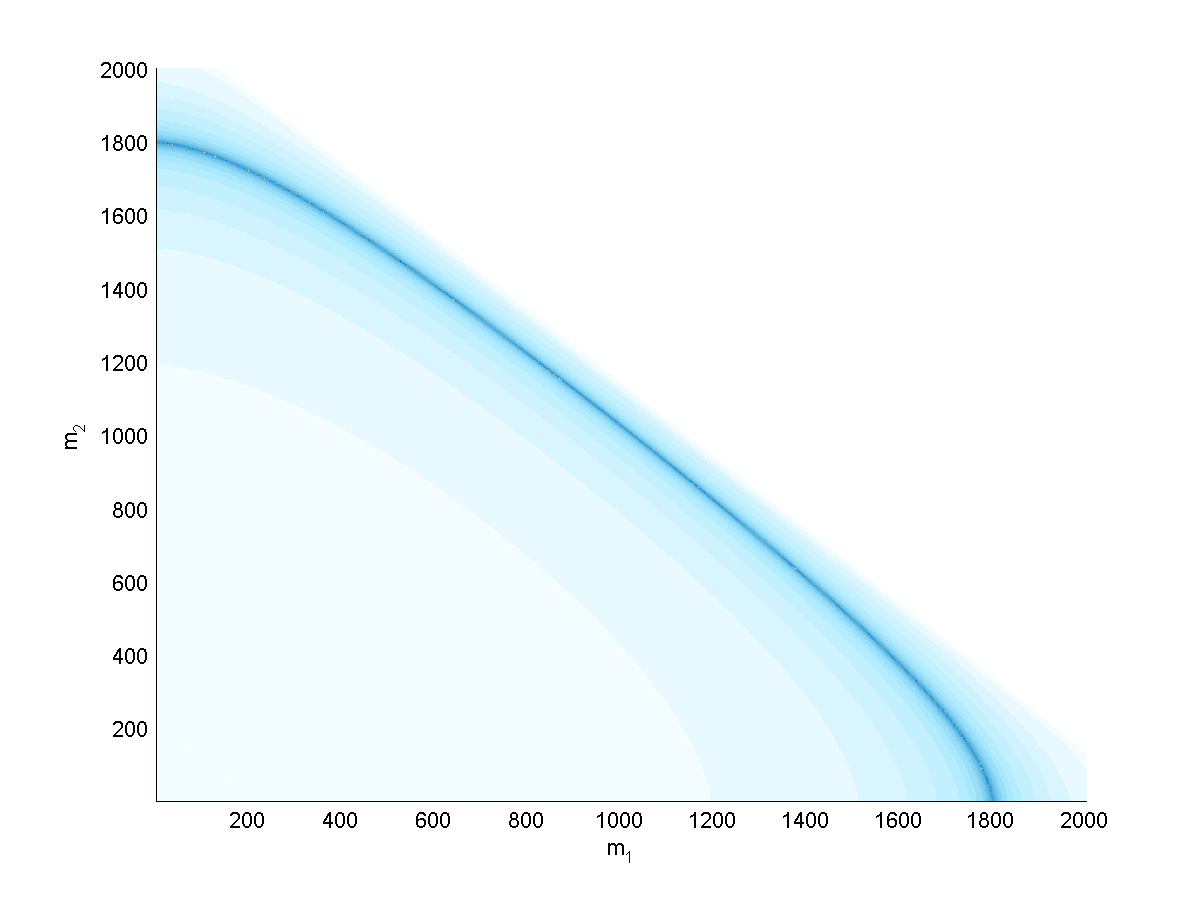} \\
						\caption{\label{fig:chi_j_heavy} Top: \(\chi^2\) as a function of two masses for the \(D\) (left) and \(D^*_s\) (right) trajectories. The coloring is based on a logarithmic scale, with the entire colored area having \(\chi^2_m/\chi^2_l < 1\). The minimum is \rchi{5\ten{-4}} for the \(D\), and \rchi{2\ten{-6}} for the \(D^*_s\). In both plots, \(a\) and \(\alp\) are optimized for each choice of the endpoint masses.}
				\end{figure}
				
			There are three trajectories we analyze involving a charm quark. The first is of the \(D\), comprised of a light quark and a \(c\) quark, the second is the \(D^*_s\) with a \(c\) and an \(s\), and the third is \(\ccb\) - the \(\Psi\). All trajectories have only three data points.
			
			For the \(D\) meson, the optimal fit has \(m_c = 1640\), \(\mud = 80\) and \(\alp = 1.07\). In this case, unlike the result for the \(K^*\) trajectory, there is a preference for an imbalanced choice of the masses, although with four fitting parameters and three data points we can't claim this with certainty. The fit for the \(D^*_s\) has a good fit consistent with the previous \(s\) and \(c\) fits at \(m_c = 1580\), \(m_s = 400\), and \(\alp = 1.09\). The plots of \(\chi^2\) vs. the two masses (\(m_c\) and \(\mud\)/\(m_s\)) can be seen in figure \ref{fig:chi_j_heavy}.
			
			In the same figure, we have \(\chi^2\) as a function of the single mass \(m_c\) for the \(\ccb\) \(\Psi\) trajectory. The minimum there is obtained at \(m_c = 1500\) MeV, where the slope is \(\alp = 0.98\) GeV\(^{-2}\).
			
			It is worth noting that while the linear fit results in values for \(\alp\) that are very far from the one obtained for the \(u\), \(d\), and \(s\) quark trajectories - \(0.42\), \(0.48\), and \(0.52\) for the \(\Psi\), \(D\), and \(D^*\) respectively - the massive fits point to a slope that is very similar to the one obtained for the previous trajectories.	This is also true, to a lesser extent, of the values of the intercept \(a\).
		
	\paragraph{\texorpdfstring{$\bbb$}{b-bbar} mesons}
		The last of the \((J,M^2)\) trajectories is that of the \(\bbb\) \(\Upsilon\) meson, again a trajectory with only three data points.
		The fits point to an optimal value of \(m_b = 4730\), exactly half the mass of the lowest particle in the trajectory. The slope is significantly lower than that obtained for other mesons, \(\alp = 0.64\) at the optimum.
		
\subsubsection{Meson trajectories in the \texorpdfstring{$(n,M^2)$}{(n,M2)} plane}
	\paragraph{Light quark mesons}
		In the light quark sector we fit the trajectories of the \(\pi\) and \(\pi_2\), the \(h_1\), the \(a_1\), and the \(\omega\) and \(\omega_3\).
		
		The \(h_1\) has a very good linear fit with \(\alp = 0.83\) GeV\(^{-2}\), that can be improved upon slightly by adding a mass of 100 MeV, with the whole range \(0-130\) MeV being nearly equal in \(\chi^2\).
		
		The \(a_1\) offers a similar picture, but with a higher \(\chi^2\) and a wider range of available masses. Masses between \(0\) and \(225\) are all nearly equivalent, with the slope rising with the added mass from \(0.78\) to \(0.80\) GeV\(^{-2}\).
		
		The \(\pi\) and \(\pi_2\) trajectories were fitted simultaneously, with a shared slope and mass between them and different intercepts. Again we have the range \(0\) to \(130\) MeV, \(\alp = 0.78-0.81\) GeV\(^{-2}\), with \(\mud = 100\) MeV being the optimum. The preference for the mass arises from non-linearities in the \(\pi\) trajectory, as the \(\pi_2\) when fitted alone results in the linear fit with \(\alp = 0.84\) GeV\(^{-2}\) being optimal.
		
		The \(\omega\) and \(\omega_3\) trajectories were also fitted simultaneously. Here again the higher spin trajectory alone resulted in an optimal linear fit, with \(\alp = 0.86\) GeV\(^{-2}\). The two fitted simultaneously are best fitted with a high mass, \(\mud = 340\), and high slope, \(\alp = 1.09\) GeV\(^{-2}\). Excluding the ground state \(\omega(782)\) from the fits eliminates the need for a mass and the linear fit with \(\alp = 0.97\) GeV\(^{-2}\) is then optimal. The mass of the ground state from the resulting fit is \(950\) MeV. This is odd, since we have no reason to expect the \(\omega(782)\) to have an abnormally low mass, especially since it fits in perfectly with its trajectory in the \((J,M^2)\) plane.
							
	\paragraph{\texorpdfstring{$\ssb$}{s-sbar} mesons}
		For the \(\ssb\) we have only one trajectory of three states, that of the \(\phi\). There are two ways to use these states. The first is to assign them the values \(n = 0,1,2\). Then, the linear fit with the slope \(\alp = 0.54\) GeV\(^{-2}\) is optimal.
		
		Since this result in inconsistent both in terms of the low value of the slope, and the absence of a mass for the strange quark, we tried a different assignment. We assumed the values \(n = 0,1,\) and \(3\) for the highest state and obtained the values \(\alp = 1.10, m_s = 515\) for the optimal fit. These are much closer to the values obtained in previous fits.
		
		The missing \(n = 2\) state is predicted to have a mass of around \(1960\) MeV. Interestingly, there is a state with all the appropriate quantum numbers at exactly that mass - the \(\omega(1960)\), and that state lies somewhat below the line formed by the linear fit to the radial trajectory of the \(\omega\). Even if the \(\omega(1960)\) is not the missing \(\ssb\) (or predominantly \(\ssb\)) state itself, this could indicate the presence of a \(\phi\) state near that mass.
		
	\paragraph{\texorpdfstring{$\ccb$}{c-cbar} mesons}
		Here we have the radial trajectory of the \(J/\Psi\), consisting of four states.
		
		The massive fits now point to the range \(1350-1475\) MeV for the \(c\) quark mass.	The biggest difference between the fits obtained here and the fits obtained before, in the \((J,M^2)\) plane is not in the mass, but in the slope, which now is in the range \(0.48-0.56\) GeV\(^{-2}\), around half the value obtained in the angular momentum trajectories involving a \(c\) quark - \(0.9-1.1\).
		
		It is also considerably lower than the slopes obtained in the \((n,M^2)\) trajectories of the light quark mesons, which would make it difficult to repeat the achievement of having a fit with a universal slope in the \((n,M^2)\) plane like the one we had in the \((J,M^2)\) plane.
	
	\paragraph{\texorpdfstring{$\bbb$}{b-bbar} mesons}
	There are two trajectories we use for the \(\bbb\) mesons.
	\begin{figure}[t!] \centering
						\includegraphics[natwidth=1200bp, natheight=900bp, width=.48\textwidth]{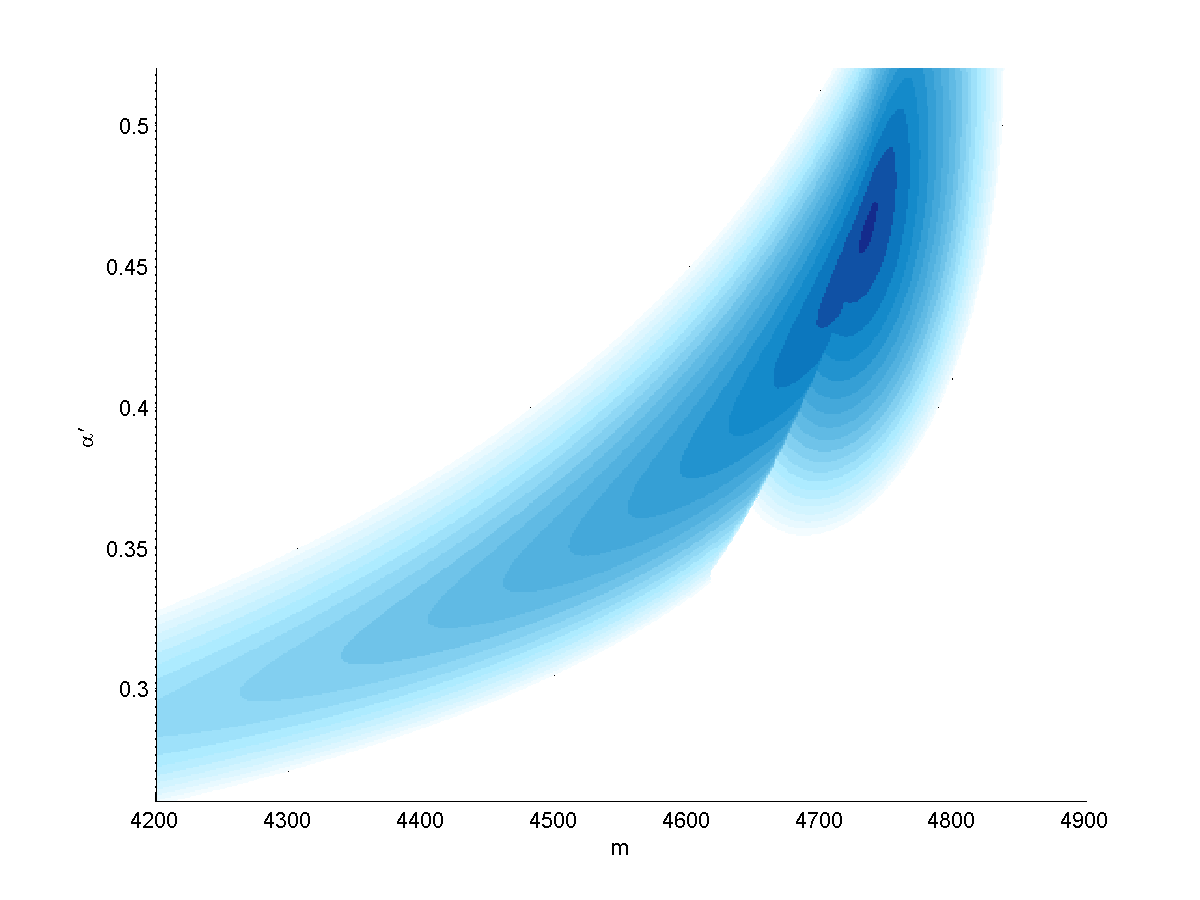}	 \hfill
						\includegraphics[natwidth=1200bp, natheight=900bp, width=.48\textwidth]{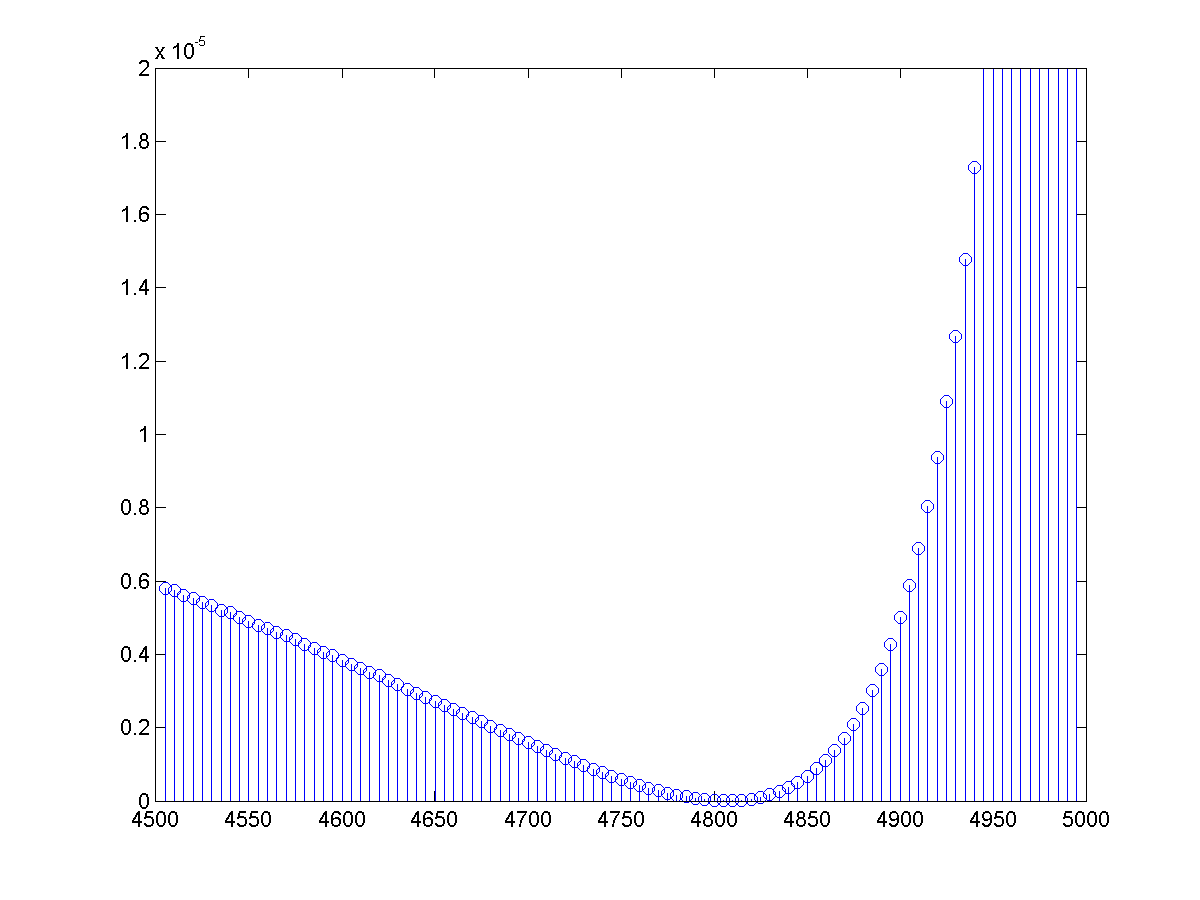}
						\caption{\label{fig:chi_n_bb} Left: \(\chi^2\) as a function of \(\alp\) and \(m_b\) for the \(\Upsilon\) radial trajectory. The discontinuity in the plot arises from the condition that the intercept \(a \leq 1\), otherwise the mass of the ground state is undefined. The two areas in the plot are then where \(a\) is still allowed to change (left) and where \(a\) is blocked from increasing further and is fixed at \(a = 1\) (oval shape on the right). Right: \(\chi^2\) as a function of \(m_b\) for the \(\chi_b\) trajectory.}
				\end{figure}
	
	The first is that of the \(\Upsilon\) meson, with six states in total, all with \(J^{PC} = 1^{--}\). For this trajectory we have an excellent fit with \(m_b = 4730\) MeV and the slope \(\alp = 0.46\) GeV\(^{-2}\). It is notable for having a relatively large number of states and still pointing clearly to a single value for the mass.
	
	The second \(\bbb\) trajectory is that of the \(\chi_b\) - \(J^{PC} = 1^{++}\). Here we have only three states and the best fit has a slightly higher mass for the \(b\) quark - \(m_b = 4800\) MeV - and a higher value for the slope \(\alp = 0.50\) GeV\(^{-2}\). Plots for \(\chi^2\) as a function of the \(b\) mass are shown in figure \ref{fig:chi_n_bb}.
\subsection{Review of baryon trajectory fits} \label{sec:baryons}
This section offers a discussion of the fit results for the baryon Regge trajectories. 

As the baryons are more complex objects than mesons, there are some issues we need to clarify before turning to the results of the fits.

We start by briefly discussing the results using the Y-shaped string model for the baryon, and the effects of including a massive baryonic vertex at the center of mass, via the replacement \(M \rightarrow M-m_{bv}\). The results prove to be against these options, so the rest of the section discusses the results when using the quark-diquark model for the baryons. This means we describe the baryons simply as a single string with two masses at its endpoints.

The slope tells us that there is a quark on one end of the string and a diquark on the other. The masses can tell us more on the structure of the baryon, for example it can say something about the composition of the diquark. However, as we explain later, the data used in the Regge trajectory fits is not enough to provide us that information.

One final note we make before the fit results is on the ``even-odd'' effect exhibited by the light baryons: the trajectories of light baryons are split into two parallel trajectories, one for states with even orbital angular momentum (or even parity), and another for the odd states.
	
We separate the results into three sections, one for the light quark baryons, the next for strange baryons, and the third for charmed baryons. In the light baryon section we also examine the radial trajectories of the \(N\) and \(\Delta\) baryons. The rest of the sections have trajectories only in the angular momentum plane \((J,M^2)\).

	\subsubsection{Y-shape and central mass}
	The Y-shaped string model is equivalent in terms of the Regge trajectories to a single string model with a higher effective string tension. In the picture we have before adding endpoint masses, we may assume (as a phenomenological model) linear trajectories for both the meson and baryon trajectories,
	with different slopes, \(\alp\!_m\) and \(\alp\!_b\) respectively. Now, the assumption that the baryons are Y-shaped strings while the mesons are straight single strings, and that there is a single universal string tension, would lead us to expect the relation
	\be \alp\!_b = \frac{2}{3}\alp\!_m \ee
	between the baryon and meson Regge slopes. When we fit the data we see no such relation. What we see in our results is that in fact, the meson and baryon slopes are very similar - \(\alp\!_b \approx \alp_m\) - supporting the same single string model for the baryons that was used for the mesons. This also excludes the triangle-shaped closed string baryon, which we have not analyzed in detail but predicts an effective slope \(\alp\!_b\) of between \(\frac{3}{8}\alp\!_m\) and \(\frac{1}{2}\alp\!_m\), depending on the type of solution \cite{Sharov:1998hi}.
	
Our addition of endpoint masses does not change this picture, as we would still need to see a similar relation between the baryon and meson slopes, with the baryon slope being consistently lower.
	
As for the baryonic vertex mass, the assumption that there is a central mass that contributes to the total mass of a state but not to the angular momentum was also found to be unsupported by the data. This does not rule out the presence of a mass due to a holographic baryonic vertex, but means it is either very small or located at the string endpoint, near the diquark, and not at its center.
	
	With these results in mind, we continue to present our fits using only the single string model with two masses at its endpoints, which is the same fitting model as the model used for the mesons.
	
	\subsubsection{Symmetric vs. imbalanced string} \label{sec:barsym}
	Now we turn to the different mass configurations in the single string model. As mentioned in the last subsection, there is no evidence to support any substantial mass located at the center of the string. To understand the structure of the baryon we would like to be able to tell how the mass is distributed between the two endpoints, but this is information we cannot gather from the Regge trajectory fits alone.
	In the low mass approximation for the single string, the leading order correction is proportional to \(\alp(m_1^{3/2}+m_2^{3/2})\sqrt{E}\). Therefore, for small masses we cannot distinguish from the Regge trajectory fits alone between different configurations with equal \(m_1^{3/2}+m_2^{3/2}\). There are higher order corrections, but our fits are not sensitive to them, and in practice, we see that fits with \(m_1^{3/2}+m_2^{3/2} = Const.\) are nearly equivalent even for masses of a few hundred MeV.
	
	When expanding \(J\) in \(E\) for two heavy masses, the combination in the leading term would be \(m_1 + m_2\). In the mid range that cannot be described accurately by either expansion there is a transition between the two different type of curves.In both cases the symmetric fit where \(m_1 = m_2 = m\) gives an indication of the total mass we can add to the endpoints for a good fit of a given trajectory. The best fitting masses are on a curve in the \((m_1,m_2)\) plane. The choice \(m_1 = m_2\) maximizes the total mass \(m_1 + m_2\), whether the masses are light or heavy.
	
	It should also be noted that for the trajectories we analyze we either have fits with low masses, where the \(m^{3/2}\) approximation is valid, or trajectories with only 3 data points where we would by default expect the optimum to be located on a curve in the \((m_1,m_2)\) plane, seeing how there are four fitting parameters in total.
	
	While the presentation in the following sections of the results is for the symmetric fit, this does not mean that we have found it is actually preferred by the data. The symmetric fit tells us whether there is a preference for non-zero endpoint masses or not, and allows us to obtain the values of the slope for a given total endpoints mass.
	
	\subsubsection{Splitting between trajectories of even and odd angular momentum}
	One of the interesting features of the baryon Regge trajectories is the splitting of the trajectories of even and odd orbital angular momentum states, which is seen in the trajectories of the light baryons, \(N\) and \(\Delta\). The states with even and odd orbital angular momentum do not lie on one single trajectory, but on two parallel linear trajectories, the odd \(L\) states being higher in mass and lying above the trajectory formed by the even states. The plot in figure \ref{fig:chi_b_light} shows this effect for the \((J,M^2)\) trajectory of the \(N\). The same effect is stronger for the \(\Delta\) trajectory.
	
	For the strange \(\Lambda\) baryon we do not see this effect. We assume that this effect is only manifest in the trajectories of the \(N\) and \(\Delta\), and is similarly suppressed for all subsequent trajectories of heavier baryons, which do not have enough data points to confirm this assumption directly.
		
		In our analysis, we fit the even and odd trajectories together, with the same endpoint masses and slope, and allow the intercept to carry the difference between the even and odd states.
	
	\subsubsection{Summary of results}
	\begin{table}[ht!] \centering
					\begin{tabular}{|c|c|cc|c|cc|} \hline
					Traj. & \(N\) & \multicolumn{2}{|c|}{\(m\)} & \alp & \multicolumn{2}{|c|}{\(a\)} \\ \hline
					
					\(N\) & \(7\) & \multicolumn{2}{|c|}{\(2m = 0-170\)} & \(0.944-0.959\) & \(a_e = (-0.32)-(-0.23)\) & \(a_o = (-0.75)-(-0.65)\) \\
					
					
					\(N\)\(^{[a]}\) & \(15\) & \multicolumn{2}{|c|}{\(2m = 0-425\)} & \(0.815-0.878\) & \(a_{1/2+} = (-0.22)-0.07\) & \(a_{3/2-} = (-0.36)-(-0.06)\) \\
					
					\(\Delta\) & \(7\) & \multicolumn{2}{|c|}{\(2m = 0-450\)} & \(0.898-0.969\) & \(a_e = 0.14-0.54\) & \(a_o = (-0.84)-(-0.42)\) \\
					
					\(\Delta\)\(^{[b]}\) & \(3\) & \multicolumn{2}{|c|}{\(2m = 0-175\)} & \(0.920-0.936\) & \multicolumn{2}{|c|}{\(a = 0.11-0.21\)} \\
					
					\(\Lambda\) & \(5\) & \multicolumn{2}{|c|}{\(2m = 0-125\)} & \(0.946-0.955\) & \multicolumn{2}{|c|}{\(a = (-0.68)-(-0.61)\)} \\
					
					\(\Sigma\) & \(3\) & \multicolumn{2}{|c|}{\(2m = 1190\)} & \(1.502\) & \multicolumn{2}{|c|}{\(a = (-0.15)\)} \\					 
					
					\(\Sigma\)\(^{[c]}\) & \(3\) & \multicolumn{2}{|c|}{\(2m = 1255\)} & \(1.459\) & \multicolumn{2}{|c|}{\(a = 1.37\)} \\
					
					\(\Xi\) & \(3\) & \multicolumn{2}{|c|}{\(2m = 1320\)} & \(1.455\) & \multicolumn{2}{|c|}{\(a = 0.50\)} \\
					
					\(\Lambda_c\) & \(3\) & \multicolumn{2}{|c|}{\(2m = 2010\)} & \(1.130\) & \multicolumn{2}{|c|}{\(a = 0.09\)} \\
					
					\hline \end{tabular}
					 \caption{\label{tab:summaryb} Summary table for the baryon fits. The ranges listed have \(\chi^2\) within 10\% of its optimal value.  \(N\) is the number of points in the trajectory. [a] is a fit to radial trajectories of the \(N\). The fifteen states used are four states with \(J^P = \jph{1}{+}\), three with \jph{3}{-}, and four pairs with other values of \(J^P\). [b] is the radial trajectory of the \(\Delta\) (\jph{3}{+}). [c] is a trajectory beginning with the state \(\Sigma(1385)\) \jph{3}{+}, as opposed to the \jph{1}{+} \(\Sigma\) ground state. The rest of the trajectories are all leading trajectories in the \((J,M^2)\) plane, and do not exclude any states.}
					\end{table}

		\begin{table}[ht!] \centering
					\begin{tabular}{|c|c|cc|cc|} \hline
					Traj. & \(N\) & \multicolumn{2}{|c|}{\(m\)} & \multicolumn{2}{|c|}{\(a\)} \\ \hline
					
					\(N\) & \(7\) & \multicolumn{2}{|c|}{\(2m = 0-180\)} & \(a_e = (-0.33)-(-0.22)\) & \(a_o = (-0.77)-(-0.65)\) \\
								
					\(\Delta\) & \(7\) & \multicolumn{2}{|c|}{\(2m = 300-530\)} & \(a_e = 0.31-0.66\) & \(a_o = (-0.71)-(-0.26)\) \\
					
					\(\Lambda\) & \(5\) & \multicolumn{2}{|c|}{\(2m = 0-10\)} & \multicolumn{2}{|c|}{\(a = (-0.68)-(-0.61)\)} \\
					
					\(\Sigma\) & \(3\) & \multicolumn{2}{|c|}{\(2m = 530-690\)} & \multicolumn{2}{|c|}{\(a = (-0.29)-(-0.04)\)} \\
					
					\(\Sigma\)* & \(3\) & \multicolumn{2}{|c|}{\(2m = 435-570\)} & \multicolumn{2}{|c|}{\(a = 0.15-0.38\)} \\
					
					\(\Xi\) & \(3\) & \multicolumn{2}{|c|}{\(2m = 750-930\)} & \multicolumn{2}{|c|}{\(a = (-0.22)-0.10\)} \\
					
					\(\Lambda_c\) & \(3\) & \multicolumn{2}{|c|}{\(2m = 1760\)} & \multicolumn{2}{|c|}{\(a = (-0.36)\)} \\
					
					\(\Xi_c\) & \(2\) & \multicolumn{2}{|c|}{\(2m = 2060\)} & \multicolumn{2}{|c|}{\(a = (-1.13)\)} \\
					
					\hline \end{tabular}
					 \caption{\label{tab:fixedslope} \((J,M^2)\) fits done with the slope fixed at \(\alp = 0.950\) GeV\(^{-2}\). Fits with \(m_1 = m_2\) generally maximize \(m_1 + m_2\). In this table we may also include a fit for two \(\Xi_c\) states. The ranges listed have \(\chi^2\) within 10\% of its optimal value. \(N\) is the number of points in the trajectory.}
					\end{table}

We present here the two summary tables: in table \ref{tab:summaryb} we list the results of the general fits, where we find the optimal slope and masses for each trajectory separately, while in table \ref{tab:fixedslope} are the results of fits done with a fixed slope, \(\alp = 0.95\) GeV\(^{-2}\).

As explained in previous sections, we assume the model of a single string connecting two endpoint masses (a quark and a diquark). We present our results in terms of the total mass of the endpoints, as we cannot determine the distribution of the mass between them from the Regge trajectory fits alone.

Our massive fits for the baryons are not always consistent. For the light quark trajectories, of the \(N\) and the \(\Delta\), we have seen there is no evidence for a string endpoint mass of the light quarks. These states are best fitted by linear trajectories with a slope similar to that of the light mesons - around 0.95 GeV\(^{-2}\) for the \(N\) and \(0.90\) GeV\(^{-2}\) for the \(\Delta\). The \((J,M^2)\) trajectories also exhibit a splitting between the trajectories of states with even and odd orbital angular momentum. In our fits we incorporate this difference into the intercept.

For the strange baryon trajectories there is a significant discrepancy between the \(\Lambda\), which is best (and very well) fitted by a linear, massless, trajectory, and the \(\Sigma\) baryons which are optimally fitted with a high total mass, of around 1200 MeV and an unusually high slope, 1.5 GeV\(^-2\). The fits when fixing the slope at the value obtained from the lighter baryon fits give a total mass in the more reasonable range (for the mass of an \(s\) quark) of \(500-600\) MeV. We have also looked into the doubly strange \(\Xi\) baryon, where we have a similar picture. We can fix the slope of the \(\Xi\) to \(\alp = 0.95\) GeV\(^{-2}\) and a total mass of around \(800\) MeV. This choice is not only more consistent with the slopes obtained from the lighter baryons, but also with the mass obtained for the \(s\) quark in the meson fits, \(m_s \approx 400\) MeV.

The last results are those of the charmed baryons, the heaviest baryons for which we have a trajectory.  The charmed \(\Lambda_c\) is again best fitted by a high slope, 1.2GeV\(^{-2}\) and a total mass of a little over 1800 MeV. Once more we can bring down the mass by fixing \(\alp\) at 0.95 GeV\(^{-2}\) and get \(2m = 1760\) MeV as the best fit. For the charmed-strange \(\Xi_c\) trajectory, including only two data points and therefore fitted only with the fixed slope of 0.95 GeV\(^{-2}\), we find a fit with \(2m = 2060\) MeV, a value consistent with the presence of both charmed and strange quarks.

	\subsubsection{Light quark baryons}
	
	\begin{figure}[t!] \centering
						\includegraphics[natwidth=1200bp, natheight=900bp, width=.44\textwidth]{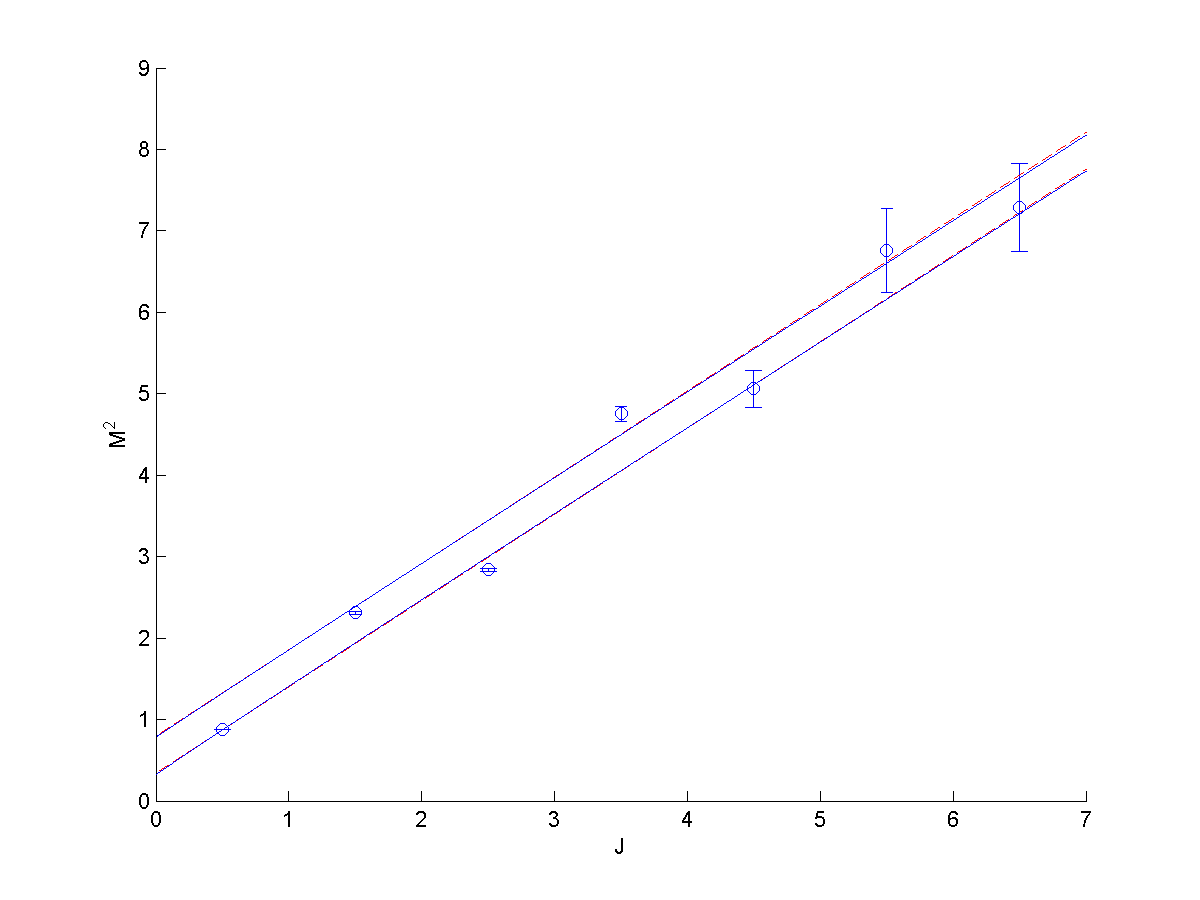}	 \hfill
					\includegraphics[natwidth=1200bp, natheight=900bp, width=.44\textwidth]{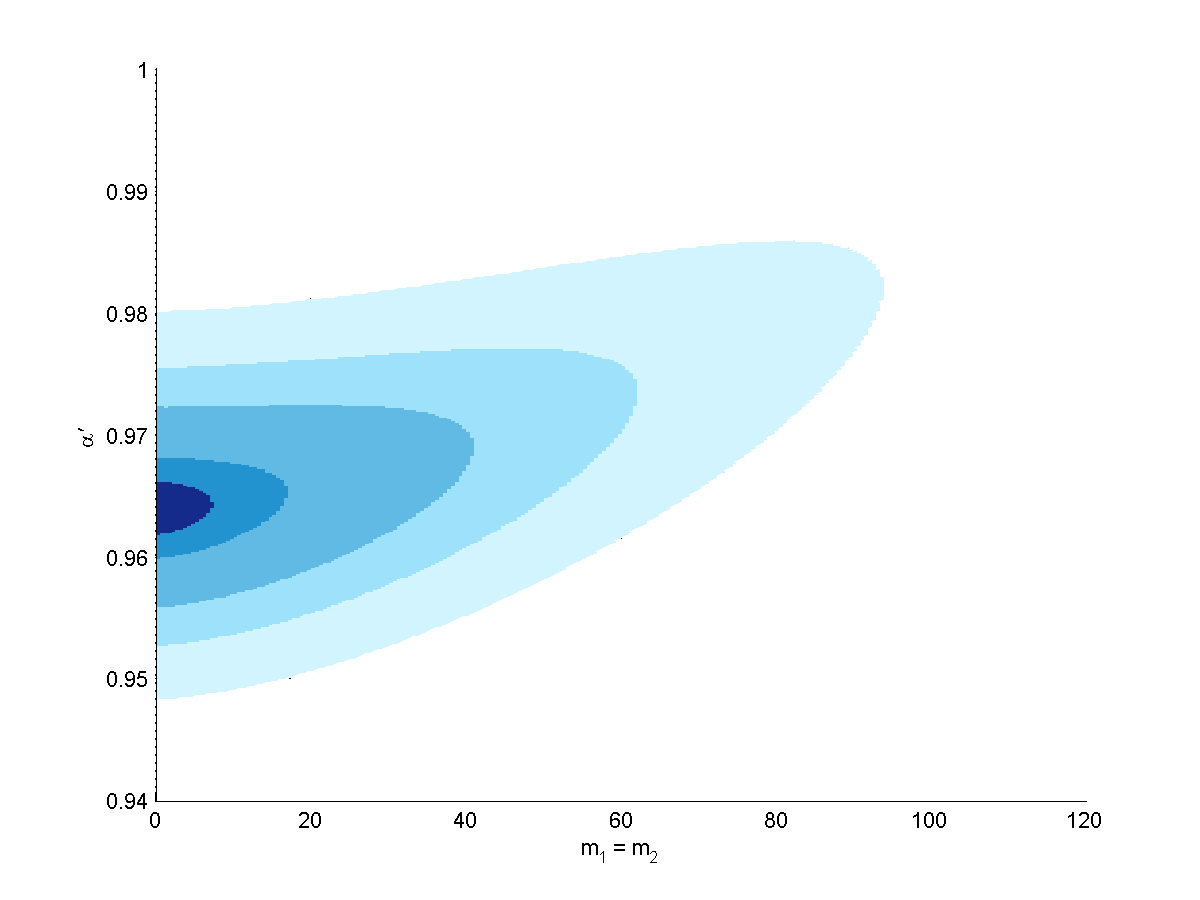}	 \hfill
\\
						\caption{\label{fig:chi_b_light} The light baryon trajectory fits. \textbf{Left:} The \(N\) trajectory and fits, showing the even/odd effect. \textbf{Right:} \(\chi^2\) vs. \((\alp,2m\)) for the \(N\). The \(\chi^2\) plots are for fits to the \((J,M^2)\) trajectories, and use only the even \(L\) states (results somewhat differ from those quoted for the fit to even and odd states together). The darkest areas in the \(\chi^2\) plots have \(\chi_m^2/\chi^2_l < 1\), lightest areas are \(\chi_m^2/\chi_l^2 < 1.1\).}
	\end{figure}
	
		In the light baryon sector, we look at the \(N\) and \(\Delta\) resonances.
		\paragraph{Trajectories in the \texorpdfstring{$(J,M^2)$}{(J,M2)} plane}
		One of the most interesting features of the baryon Regge trajectories is the splitting of the trajectories of even and odd \(L\) states. The states with even and odd orbital angular momentum do not lie on one single trajectory, but on two parallel linear trajectories, the odd \(L\) states being higher in mass and lying above the trajectory formed by the even states. The plot in figure \ref{fig:chi_b_light} shows this effect for the \((J,M^2)\) trajectory of the \(N\). In our analysis, we fit the even and odd trajectories together, with the same endpoint masses and slope, and allow the intercept to carry the difference between the even and odd states.
		
		In this way, we get that the \(N\) trajectory is best fitted with a slope of around \(0.95\) GeV\(^{-2}\), and that the linear fit is optimal. Only small masses, up to a total mass of \(2m = 170\) MeV, are allowed.\footnote{Masses in what we call the ``allowed'' range give a value of \(\chi^2\) that is within 10\(\%\) of its optimal value for that specific trajectory.} Trying a fit using only the four highest \(J\) states (two even and two odd), we achieve a weaker \(\chi^2\) dependence on the mass, and we can add a total mass of up to \(640\) MeV, with the slope being near \(1\) GeV\(^{-2}\) for the highest masses.
		
		The \(\Delta\) is also best fitted by the linear, massless, trajectory, but it allows for higher masses. The maximum for it is \(450\) MeV. The slope, for a given mass, is lower than that of the \(N\). It is between 0.9 GeV\(^{-2}\) for the linear fit, and 0.97 GeV\(^{-2}\) for the maximal massive fits.
		
		As for the even-odd effect, we quantify it by looking at the difference between the intercept obtained for the even \(L\) trajectory and the one obtained for the odd \(L\) trajectory. The magnitude of the even-odd splitting is higher for the \(\Delta\) than it is for the \(N\) baryons. For the \(\Delta\), the difference in the intercept is of almost one unit - \(a_e - a_o \approx 1\), while for the \(N\) it is less than half that: \(a_e - a_o \approx 0.45\). The difference in \(M^2\) is obtained by dividing by \(\alp\), so it is \(0.5\) GeV\(^{-2}\) for the \(N\) and \(1.1\) GeV\(^2\) for the \(\Delta\).
		
		\paragraph{Trajectories in the \texorpdfstring{$(n,M^2)$}{(n,M2)} plane}
		The radial trajectories we analyze are also best fitted with small, or even zero, endpoint masses.
		
		For the \(\Delta\) we have three states with \(J^P = \jph{3}{+}\).	The slope is between 0.92 and 0.94 GeV\(^{-2}\) and the maximal allowed total mass is less than 200 MeV.
		
		For the \(N\) we use a total of fifteen states: four with \(J^P = \jph{1}{+}\) (the neutron/proton and higher resonances), three with \jph{3}{-}, and four pairs with other \(J^P\) assignments. They are all fitted with the same slope and mass. The results show a lower slope here, from 0.82 GeV\(^{-2}\) for the linear fit to \(0.85\) GeV\(^{-2}\) for the highest mass fit, this time with \(2m = 425\) MeV.
		
\subsubsection{Strange baryons}
	In the strange section there are several trajectories we analyze.
	
\begin{figure}[t!] \centering
						\includegraphics[natwidth=1200bp, natheight=900bp, width=.44\textwidth]{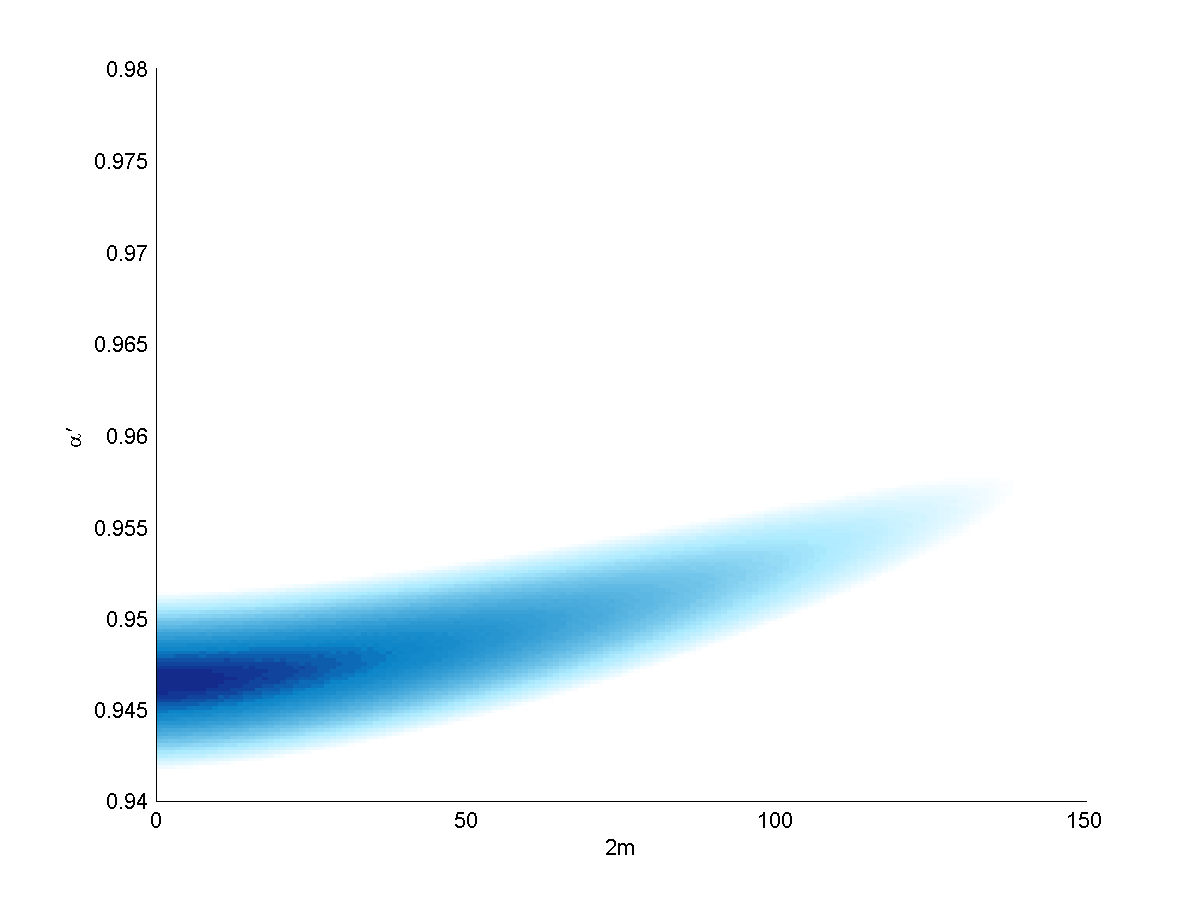}	
						\includegraphics[natwidth=1200bp, natheight=900bp, width=.44\textwidth]{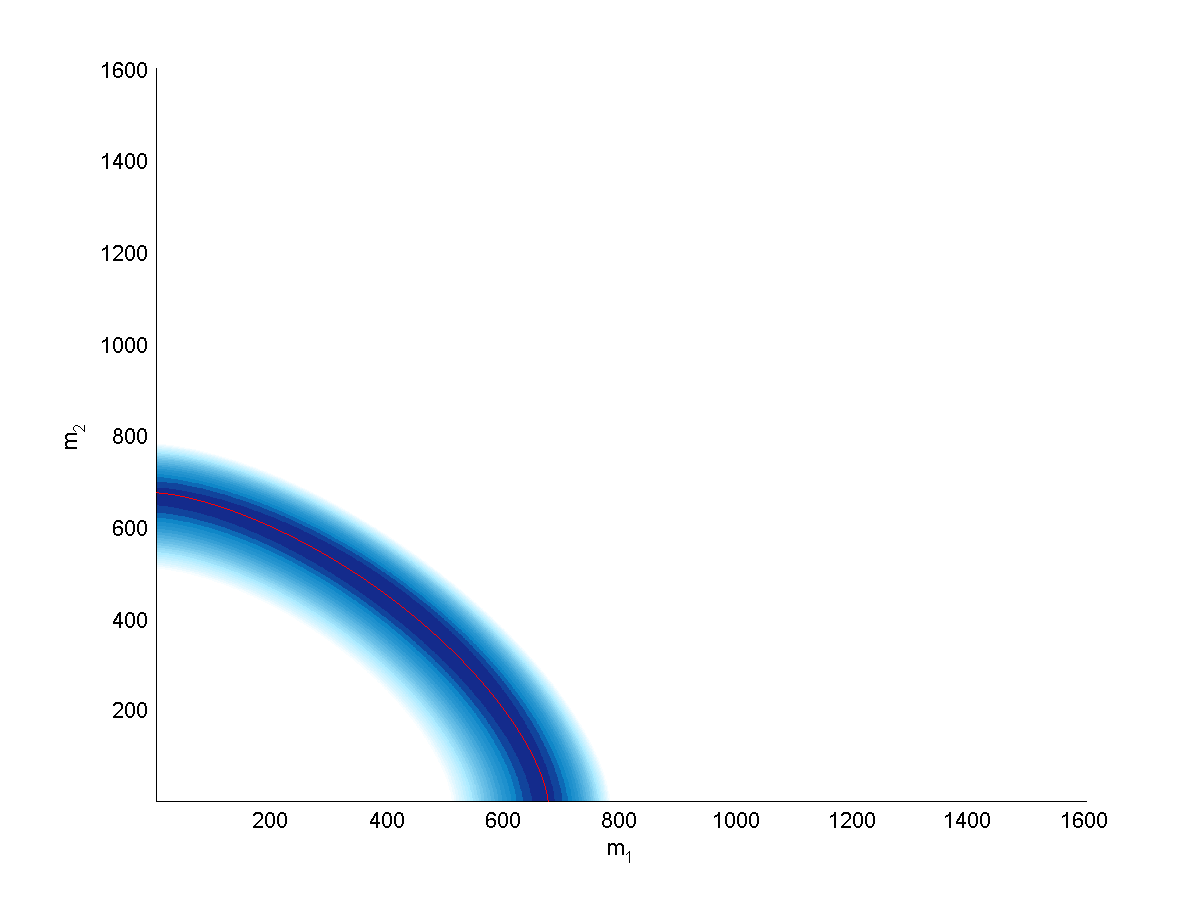}	
						\caption{\label{fig:chi_b_s} The strange baryon trajectory fits. \textbf{Left:} \(\chi^2\) vs. \((\alp,2m)\) for the \(\Lambda\) trajectory. \textbf{Right:} \(\chi^2\) vs. \((m_1,m_2\)) for the \(\Xi\) trajectory, for \(\alp = 0.950\). The red curve is \(m_1^{3/2}+m_2^{3/2} = 2\times(425)^{3/2}\).}
	\end{figure}

	The first is that of the \(\Lambda\). There are five states in this trajectory, enough for us to see that the even-odd effect is not present - or too weak to be noticeable. The linear fit, with \(\alp = 0.95\) GeV\(^{-2}\), is the optimal fit, and only small masses of the order of 60 MeV are allowed at each endpoint. Even if one of the masses is zero, the mass at the other end could not exceed 100 MeV. This is a puzzling result because the results of the meson fits, which will be compared in detail to the baryon fits in a later section, point toward a mass of \(200-400\) MeV for the \(s\) quark. We show the plot of \(\chi^2\) in figure \ref{fig:chi_b_s}.
	
	On the other hand, the other strange baryon trajectories we examine point to very high masses, with correspondingly high values of the slope.	These are the results of the two trajectories of the \(\Sigma\) baryon we examine. The first has three states with \(J^P = \) \jph{1}{+}, \jph{3}{-}, and \jph{5}{+}. The second has the parity reversed (for a given value of \(J\)): \(J^P\) = \jph{3}{+}, \jph{5}{-}, and \jph{7}{+}.
	
	Since there are only three states per trajectory we cannot determine from the data alone whether or not there is an even-odd splitting effect present here (and this is the case with all following trajectories). Assuming that there is no even-odd splitting, the best fits have \(2m \approx 1200\) MeV and a slope of about \(1.4-1.5\) GeV\(^{-2}\). Assuming splitting, we find in the case of the \(\Sigma\) that the linear fit connecting the two even states has \(\alp \approx 0.9\) GeV\(^{-2}\).
	
	A third option, is fixing the slope at a more reasonable low value - we chose \(\alp = 0.95\) GeV\(^{-2}\) - and redoing the massive fits (with the assumption that there is no splitting). The best fits then for the \(\Sigma\) are at around \(2m = 500\) MeV, with the mass being somewhat higher in the trajectory beginning with of the \jph{1}{+} state. This is certainly the choice that is most consistent with previous results, as we can distribute the total mass so there is a mass of \(m_s \approx 400\) MeV at one end and up to 100 MeV at the other. The cost in \(\chi^2\) is fairly high: for the first trajectory \(\chi^2\) is approximately \(10^{-4}\) for the higher slope and ten times larger for \(\alp = 0.95\), while for the second \(\chi^2\) is almost zero for the high slope fit\footnote{This is often the case with three point trajectories, where we may find a choice of the parameters for which the trajectory passes through all three data points. This makes the error in the measurement hard to quantify.} and about \(5\ten{-4}\) for the fixed slope fit. In any case, the fits with \(\alp = 0.95\) GeV\(^{-2}\) and with the added masses have a better \(\chi^2\) than the linear massless fits.
	
	There is one more possible trajectory we examine, of the doubly strange \(\Xi\) baryon. The best fit overall is again with \(\alp \approx 1.5\) GeV\(^{-2}\), and at a somewhat higher mass of \(2m = 1320\). Fixing the slope at \(0.95\) GeV\(^{-2}\) results in \(2m = 850\) being optimal. This is again the best choice in terms of consistency - the total mass is exactly in the range we would expect to see where there are two \(s\) quarks present. In \(\chi^2\), the fit with the high slope has \(\chi^2 \approx 10^{-4}\) while the latter has \(\chi^2 \approx 4\ten{-4}\). We plot the masses in the fixed slope fit in figure \ref{fig:chi_b_s}.
	
	\subsubsection{Charmed baryons}	
	In the charmed baryon section we have only one trajectory we can use, comprised of three states, that of the \(\Lambda_c\) baryon. The best fits are again at a relatively high slope, 1.1 GeV\(^{-2}\), with the mass \(2m = 2010\) MeV. This fit's \(\chi^2\) tends to zero. The fit with the slope fixed at 0.95 GeV\(^{-2}\) takes the mass down to \(2m = 1760\) MeV with \(\chi^2 = 3\ten{-5}\). The high slope fit is equivalent to a fit with \(m_1 = 1720\) MeV and \(m_2 = 90\), while a the fit with \(m_1 = 1400\) and \(m_2 = 90\) is roughly equal to the latter fixed slope fit.
	
	We can also do a fit using two \(\Xi_c\) states. These states are charmed and strange and are composed of \(dsc\) (\(\Xi_c^0\)) or \(usc\) (\(\Xi_c^+\)). Since we only have two states, we do only a fit with the fixed slope, \(\alp = 0.95\) GeV\(^{-2}\). The best massive fit then has \(2m = 2060\) MeV.

\subsubsection{Structure of the baryon in the quark-diquark model} \label{sec:structure}
For the light baryons our analysis of the spectrum cannot offer much new insight regarding the different baryons' structure,\footnote{\cite{Selem:2006nd} offers a discussion of the composition of the light baryons in a model of a quark and diquark joined by a flux tube. In the analysis of the spectrum done there, the light baryons are assigned different configurations of the diquark based on the energetics of the \(ud\) diquarks.} in particular because we have no way to distinguish between the two light quarks, given their small - possibly zero - masses, but also because we cannot in general make any comments regarding the mass distribution within the different baryons (both light and heavy). In spite of this, there is one interesting implication when interpreting our results in light of the underlying holographic models and the way they map the diquarks into flat space-time.

\begin{figure}[t!] \centering
					\includegraphics[width=.90\textwidth, natheight=1056bp,natwidth=1902bp]{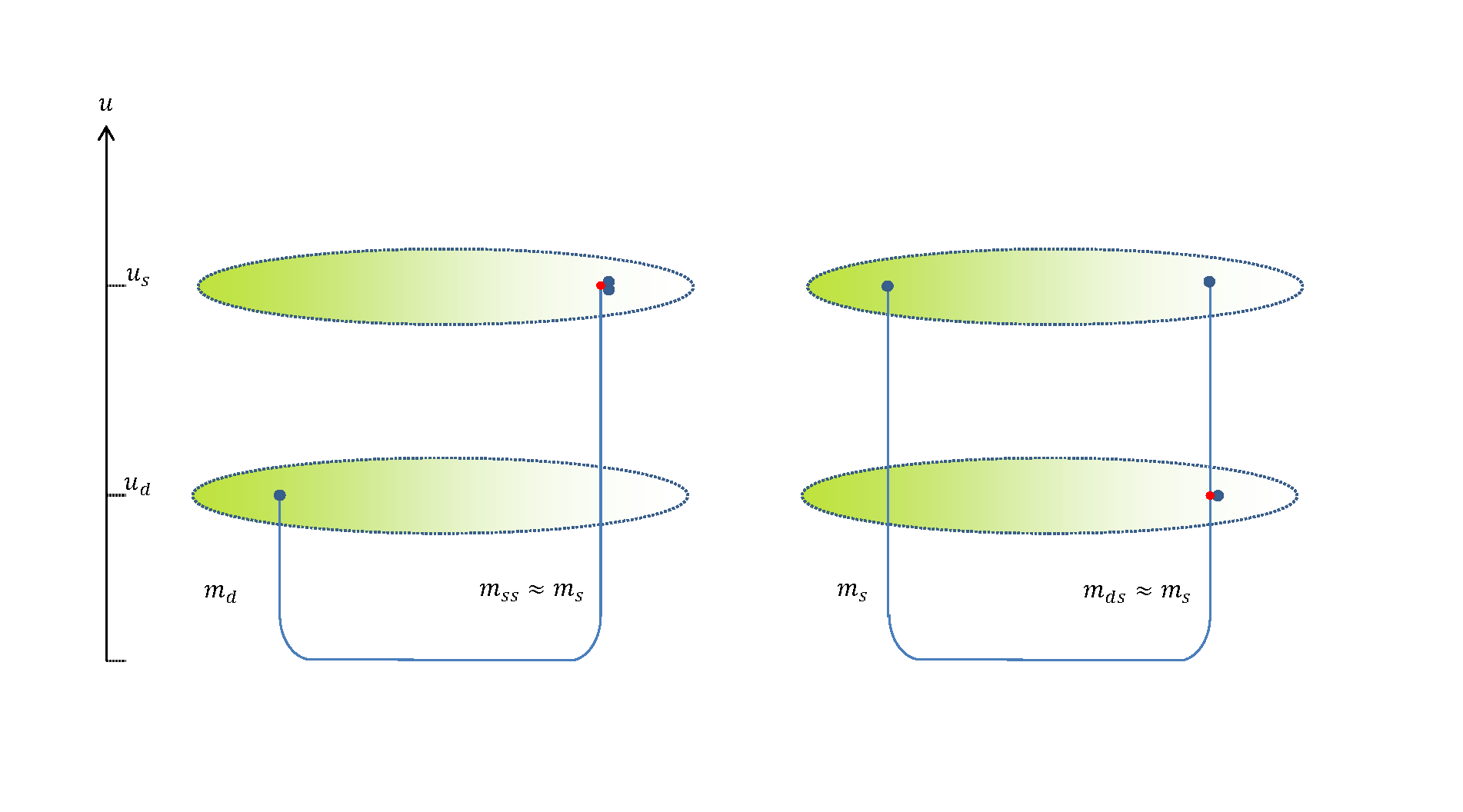}
					\caption{\label{fig:ssdbaryon}  The two holographic setups for the \(\Xi^-\) baryon, with different compositions of the diquarks. The vertical segments of the strings contribute the measured endpoint masses.}
\end{figure}

In our analysis of the meson spectrum, we argued that the mass parameter relevant to the analysis is the mass of the quark as a string endpoint, which generically was found to be between the usual QCD and constituent masses attributed to the respective quark. For the diquark the identification between string length and mass can have another implication, as illustrated in figure \ref{fig:ssdbaryon}. If the relevant mass parameter is the length of the vertical segment of the string connected to the flavor brane, and if the two quarks forming the diquark and the baryonic vertex to which they are both connected all lie close to each other on the flavor brane, then we would expect the mass of the diquark - in the holographic picture, as a string endpoint - to be approximately equal to the mass of a single quark:
\be m_{qq} \approx m_q \ee
This is because we have only one contribution to the mass from the string connecting the lone quark outside the diquark and the baryonic vertex. This is a prediction that can serve as a test of the holographic interpretation of the string endpoint masses. Since we do not have an accurate figure for the mass of the light \(u\) and \(d\) quarks, and since we lack data for charmed and heavier baryons, our best avenue for verifying this experimentally is by examining the doubly strange \(\Xi\) baryon.

The two options for the \(\Xi\) quark are one where the diquark is composed of an \(s\) and a light quark, and another where the diquark is composed of two \(s\) quarks. For the first option, our holographic interpretation would lead us to expect there to be two masses at the endpoints approximately equal to the \(s\) quark mass, leading to a total mass of \(2m \approx 2m_s\) at the endpoints. In the second option, the two \(s\) quarks in the diquark would contribute only once to the total mass we measure in the Regge trajectory fits, so the result for the total mass \(2m\) should be around, possibly a little higher than, the mass of a single \(s\) quark.

\begin{figure}[t!] \centering
					\includegraphics[width=.90\textwidth, natheight=1122bp,natwidth=1944bp]{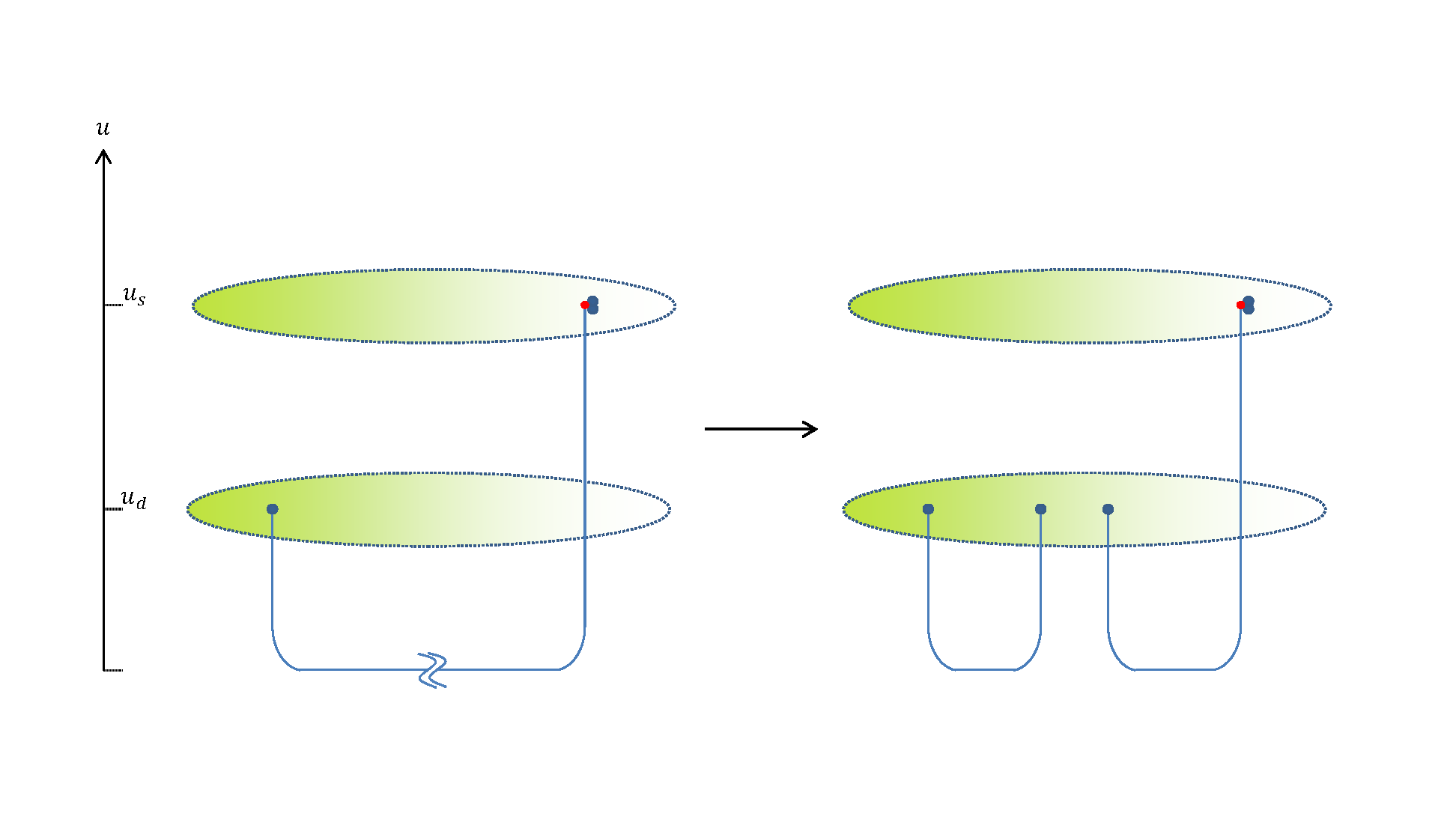}
					\caption{\label{fig:ssddecay}  A doubly strange baryon with an \(ss\) diquark decaying into a doubly strange baryon and a non-strange meson. First the string tears, and then the endpoints reconnect to the flavor brane, forming a quark-anti-quark pair.}
\end{figure}

The result from the fixed slope fit of the \(\Xi\) trajectory, \(2m = 750-930\) MeV, is consistent, from the holographic point of view, with a \(ds\) or \(us\) diquark, as opposed to \(ss\). This is because we expect the mass of the \(s\) to be somewhere near \(400\) MeV. Of course, from a purely flat space-time perspective, an \(ss\) diquark with a mass of roughly \(2m_s\) is not excluded.

If we look at the decay modes of the states used in the \(\Xi\) fits we might learn something about their structure \cite{Peeters:2005fq}. We look at a baryon's strong decays into a baryon and a meson, and our assumption is that in these decays the diquark and baryonic vertex go into the outgoing baryon while the third lone quark ends up in the meson. An illustration of this type of decay is in figure \ref{fig:ssddecay}.

The lightest doubly strange state does not have the phase space for strong decays. If we look at the two next states in the trajectory, the \(\Xi(1820)\) and the \(\Xi(2030)\), we see that they decay mainly into \(\Lambda K\) or \(\Sigma K\). The fact that they decay into a strange meson and strange baryon is against the \(ss\) diquark configuration. The leading modes of decay should leave the diquark intact, so if the diquark were \(ss\) the leading mode of decay would be \(\Xi\pi\) (as it is for some of the other observed doubly strange baryons, for which we do not have a trajectory).

For the baryons with a single strange quark, the \(\Lambda\) and the \(\Sigma\), we have seen an odd discrepancy between the obtained mass values. The mass in the \(\Lambda\) baryon was less than \(100\) MeV, while in the \(\Sigma\) we have seen masses of above \(400\) MeV. We cannot explain this discrepancy in terms of different configurations of the diquarks, as we expect the \(s\) to contribute to the mass whether it is in the diquark or not. The decay modes do not give a straightforward answer regarding the compositions of the \(\Lambda\) and \(\Sigma\), as the states decay both to \(NK\) and \(\Sigma \pi\)/\(\Lambda\pi\). We do not see a systematic preference for decays where the \(s\) remains in the baryon (implying it is near the baryonic vertex in the diquark) or vice versa in either of the trajectories.

For the charmed-strange \(\Xi_c\) we see a mass compatible compatible with \(m_1 + m_2 = m_s + m_c\). This implies to us that the possibility of a \(cs\) diquark is excluded, since we see both quarks' masses (from the holographic point of view we expect the diquark mass to be \(m_{cs} \approx m_c\)). Unfortunately we cannot test this based on the decay modes. If we look at the decays of the \(\Xi_c\) baryons, we find that the \(\Xi_c^0\)/\(\Xi_c^-\) does not have the phase space to decay strongly, and the next state we take in the trajectory, \(\Xi_c(2815)\), is also too light to provide information that would be useful to us. The \(\Xi_c(2815)\) cannot decay to a charmed meson and a strange baryon (which is the decay mode we will naively expect if the \(\Xi_c\) is a \(us\)/\(ds\) diquark joined to a \(c\) quark), simply because the lightest of these, \(D^\pm\)/\(D^0\) and \(\Lambda\) respectively, are still heavy enough so that the sum of their masses exceeds the mass of the \(\Xi_c(2815)\).

For the charmed \(\Lambda_c\) baryon, with one \(c\) and two light \(u/d\) quarks, we have no prediction based on the masses, because in any case we expect to see a total mass of approximately \(m_c = 1500\) MeV. The first state in the trajectory heavy enough to decay into a charmed meson and a baryon, the \(\Lambda_c(2800)^+\), was observed to decay both to \(pD^0\) and \(\Sigma_c \pi\), but there is no quantitative data to indicate which of these modes (if any) is dominant.
\subsection{Strings and the search for glueballs} \label{sec:glueballs}
\subsubsection{The glueball candidates: The \texorpdfstring{$f_0$}{f0} and \texorpdfstring{$f_2$}{f2} resonances}
The search for the glueballs is centered on the lightest states, the scalar ground state of \(J^{PC} = 0^{++}\), and the lightest tensor of \(2^{++}\). There is an abundance of isoscalar states with the quantum numbers \(J^{PC} = 0^{++}\) (the \(f_0\) resonances) or \(J^{PC} = 2^{++}\) (\(f_2\)). The Particle Data Group's (PDG) latest Review of Particle Physics \cite{PDG:2014}, which we we use as the source of experimental data throughout this section, lists 9 \(f_0\) states and 12 \(f_2\) states, with an additional 3 \(f_0\)'s and 5 \(f_2\)'s listed as unconfirmed ``further states''. These are all listed in table \ref{tab:glue_allf02}. In the following we make a naive attempt to organize the known \(f_0\) and \(f_2\) states into trajectories, first in the plane of orbital excitations \((J,M^2)\), then in the radial excitations plane \((n,M^2)\).

The states classified as ``further states'' are generally not used unless the prove to be necessary to complete the trajectories formed by the other states. The ``further states'' will be denoted with an asterisk below.\footnote{Note that the asterisk is not standard notation nor a part of the PDG given name of a state, we only use it to make clear the status of given states throughout the text.} For a more complete picture regarding the spectrum and specifically the interpretation of the different resonances as glueballs, the reader is referred to the relevant reviews \cite{Klempt:2007cp,Mathieu:2008me,Crede:2008vw,Ochs:2013gi} and citations therein.

\begin{table}[t!] \centering
		\begin{tabular}{|l|l||l|l|} \hline
		\textbf{State} & \textbf{Decay modes} & \textbf{State} & \textbf{Decay modes} \\ \hline\hline		
		\(f_0(500)/\sigma\) & \(\pi\pi\) dominant & \(f_2(1270)\) & \(\pi\pi\) \([85\%]\), \(4\pi\) \([10\%]\), \(KK\), \(\eta\eta\), \(\gamma\gamma\) \\ \hline
		
		\(f_0(980)\) & \(\pi\pi\) dominant, \(K\overline{K}\) seen & \(f_2(1430)\) & \(KK\), \(\pi\pi\) \\ \hline
		
		\(f_0(1370)\) & \(\pi\pi\), \(4\pi\), \(\eta\eta\), \(K\overline{K}\) & \(f^\prime_2(1525)\) & \(KK\) \([89\%]\), \(\eta\eta\) \([10\%]\), \(\gamma\gamma\) [seen] \\ \hline
		
		\(f_0(1500)\) & \(\pi\pi\) \([35\%]\), \(4\pi\) \([50\%]\), & \(f_2(1565)\) & \(\pi\pi\), \(\rho\rho\), \(4\pi\), \(\eta\eta\)\\
									& \(\eta\eta\)/\(\eta\eta\prime\) \([7\%]\), \(K\overline{K}\) \([9\%]\) & 
				 & \\ \hline
		
		\(f_0(1710)\) & \(K\overline{K}\), \(\eta\eta\), \(\pi\pi\) & \(f_2(1640)\) & \(\omega\omega\), \(4\pi\), \(KK\) \\ \hline
		
		\(f_0(2020)\) & \(\rho\pi\pi\), \(\pi\pi\), \(\rho\rho\), \(\omega\omega\), \(\eta\eta\) & \(f_2(1810)\) & \(\pi\pi\), \(\eta\eta\), \(4\pi\), \(KK\), \(\gamma\gamma\) [seen] \\ \hline
		
		\(f_0(2100)\) && \(f_2(1910)\) & \(\pi\pi\), \(KK\), \(\eta\eta\), \(\omega\omega\) \\ \hline
		
		\(f_0(2200)\) & & \(f_2(1950)\) & \(K^*K^*\), \(\pi\pi\), \(4\pi\), \(\eta\eta\), \(KK\), \(\gamma\gamma\), \(pp\) \\ \hline
		
		\(f_0(2330)\) & & \(f_2(2010)\) & \(KK\), \(\phi\phi\) \\ \hline
		
		*\(f_0\)(1200--1600) & & \(f_2(2150)\) & \(\pi\pi\), \(\eta\eta\), \(KK\), \(f_2(1270)\eta\), \(a_2\pi\), \(pp\) \\ \hline
		
		*\(f_0\)(1800) & & \(f_J(2220)\) & \(\pi\pi\), \(KK\), \(pp\), \(\eta\eta^\prime\) \\ \hline
		
		*\(f_0\)(2060) & &  \(f_2(2300)\) & \(\phi\phi\), \(KK\), \(\gamma\gamma\) [seen] \\ \hline
		
		& & \(f_2(2340)\) & \(\phi\phi\), \(\eta\eta\) \\ \cline{3-4}
		
		& & *\(f_2(1750)\) & \(KK\), \(\gamma\gamma\), \(\pi\pi\), \(\eta\eta\) \\ \cline{3-4}
		
		& & *\(f_2(2000)\) &  \\ \cline{3-4}
		
		& & *\(f_2(2140)\) & \\ \cline{3-4}
		
		& & *\(f_2(2240)\) & \\ \cline{3-4}
		
		& & *\(f_2(2295)\) & \\ \hline
		\end{tabular}
		\caption{\label{tab:glue_allf02} All the \(f_0\) and \(f_2\) states, with their observed decay modes, as listed by the PDG. The states marked by an asterisk are classified as ``further states'', i.e. in need of further experimental confirmation.}
	\end{table}

	\subsubsection{Assignment of the \texorpdfstring{$f_0$}{f0} into trajectories} \label{sec:glue_f0_fits}
	In a given assignment, we generally attempt to include all the \(f_0\) states listed in table \ref{tab:glue_allf02}, sorting them into meson and, if possible, glueball trajectories.
	
	We make an exception of the \(f_0(500)/\sigma\) resonance, which we do not use in any of the following sections. Its very large width (400--700 MeV) and low mass are enough to make it stand out among the other \(f_0\) states listed in the table. We find that it does not belong on a meson Regge trajectory, nor does it particularly suggest itself as a glueball candidate.\footnote{The authors of \cite{Nebreda:2011cp} state that the interpretation of the \(f_0(500)/\sigma\) as a glueball is ``strongly disfavored'', from what they consider a model independent viewpoint. We found no references that suggest the opposite.} Therefore, we simply ``ignore'' the \(f_0(500)\) in the following sections.

	\paragraph{Assignment of all states as mesons}
		Sorting the \(f_0\) states into trajectories with a meson-like slope leads to an assignment of the \(f_0\)'s into two groups of four:
	\[\mathrm{Light}:\qquad980, 1500, 2020, 2200, \]
	\[\ssb:\qquad1370, 1710, 2100, 2330. \]
While this simple assignment includes all the confirmed \(f_0\) states (except the \(f_0(500)\)) on two parallel trajectories, it remains unsatisfactory. If there are no glueballs we expect the states in the lower trajectory to be (predominantly) composed of light quarks, while the higher states should be \(\ssb\). This does not match what we know about the decay modes of the different states. For example, the \(f_0(1370)\) does not decay nearly as often to \(K\bar{K}\) as one would expect from an \(\ssb\) state. In fact, this assignment of the \(f_0\)'s into meson trajectories was proposed in some other works \cite{Anisovich:2000kxa,Anisovich:2002us,Masjuan:2012gc}, and the mismatch with the decay modes was already addressed in greater detail in \cite{Bugg:2012yt}. This suggests that the simple picture where all resonances belong on linear trajectories is wrong.

	\begin{figure}[tp!] \centering
	\includegraphics[width=0.48\textwidth]{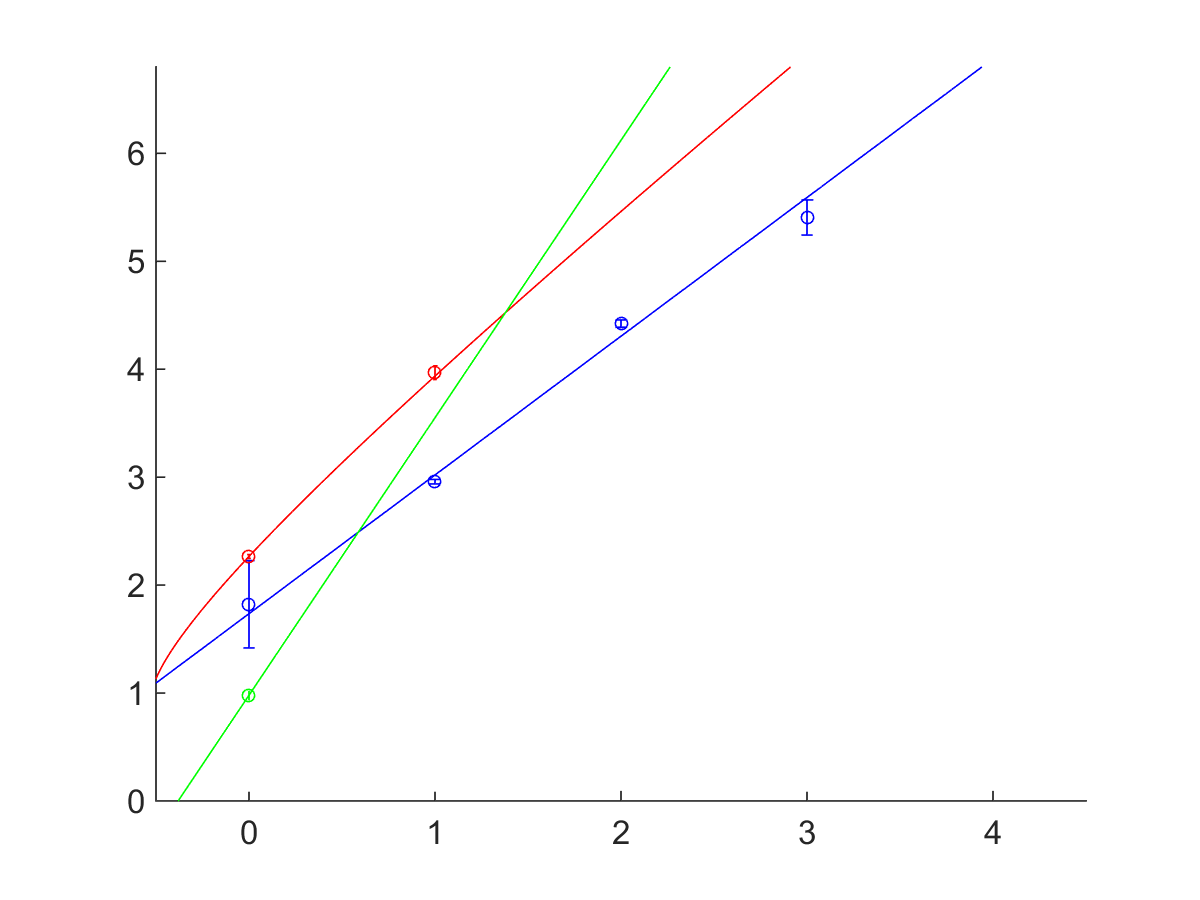} \,
	\includegraphics[width=0.48\textwidth]{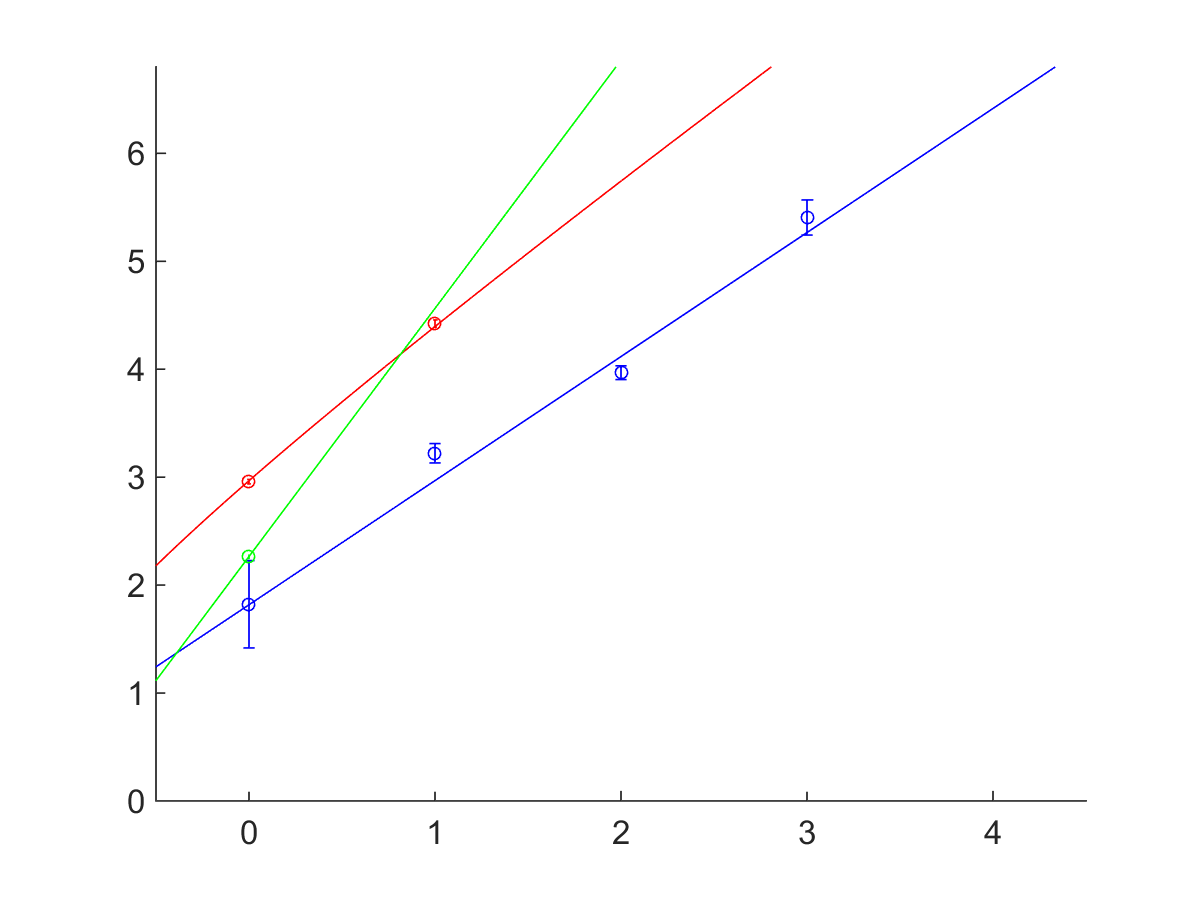} \\
				\caption{\label{tab:glue_fit980} Two examples of an assignment with glueball trajectories. In the \textbf{left} plot \(f_0(980)\) is the glueball ground state (green), \(f_0(1370)\) is the light quark meson (blue), and \(f_0(1505)\) is \(s\bar{s}\) (red). On the \textbf{right} we have \(f_0(1505)\) as glueball, \(f_0(1370)\) as light meson, and \(f_0(1710)\) as \(s\bar{s}\).}
			\end{figure}

	In the following assignments we pick and single out a state as the glueball ground state and try to build the meson trajectories without it.
	
	\paragraph{Assignment with the \(f_0(980)\) as glueball:}
	First is the the \(f_0(980)\). Assuming it is the glueball then the \(f_0(2330)\) is at the right mass to be its first excited (\(n = 2\)) partner. However, we find that the two meson trajectories given this assignment,
\[\mathrm{Light:}\qquad 1370, 1710, 2100,\]
\[\ssb\mathrm{:}\qquad 1500, 2020,\]
also predict a state very near the mass of the \(f_0(2330)\), and according to this assignment, there should be two more \(f_0\) states near the \(f_0(2330)\), for a total of three. The \(f_0(2200)\) has to be excluded.

We again have to put some states on trajectories that are not quite right for them: the \(f_0(1710)\) has a significant branching ratio for its decay into \(K\overline{K}\), while the \(f_0(1500)\), which is taken as the head of the \(\ssb\) trajectory, decays to \(K\overline{K}\) less than \(10\%\) of the time.

Note that the assignment above is the same as the one we would make if we excluded the \(f_0(980)\) on the grounds of it being an exotic (but non-glueball) state and assumed all the other states are mesons. The \(f_0(980)\) is commonly believed to be a multiquark state or a \(K\bar{K}\) bound state,\footnote{See the PDG's ``Note on scalar mesons below 2 GeV'' (in \cite{PDG:2014}) and references therein.} and in fact, we will find in following sections that even it is not a glueball, it is better to exclude it from the meson trajectories.
	
	\paragraph{Assignment with \texorpdfstring{$f_0$}{f0}(1370) as glueball}
	From here onwards the states singled out as glueballs are too high in mass for their excited states to be in the range of the measured \(f_0\) states listed in table \ref{tab:glue_allf02}, that is beneath 2.4 GeV.

Excluding the \(f_0(1370)\), we have:
\[\mathrm{Light:}\qquad [980], 1500, *1800, 2100, 2330\]
\[\ssb\mathrm{:}\qquad 1710, 2200.\]
The \(f_0(980)\) is put here in brackets to emphasize that it is optional. Including or excluding it can affect some of the fitting parameters but the trajectory is certainly not incomplete if we treat \(f_0(980)\) as a non-meson resonance and take \(f_0(1500)\) as the head of the trajectory.

The main issue here is that we have to use the state \(*f_0(1800)\) to fill in a hole in the meson trajectory, a state that is still considered unconfirmed by the PDG and whose nature is not entirely known. Its observers at BESIII \cite{Ablikim:2012ft} suggest it is an exotic state - a tetraquark, a hybrid, or itself a glueball. More experimental data is needed here.

Other than that we have \(f_0(2100)\) as a light meson and \(f_0(2200)\) as \(\ssb\). This is the option that is more consistent with the decays, as \(f_0(2200)\) is the one state of the two which is known to decay into \(K\overline{K}\) (we again refer to the comments in \cite{Bugg:2012yt} and references therein). However, in terms of the fit, we might do better to exchange them. It is possible that their proximity to each other affects their masses in such a way that our simple model does not predict, and this affects badly the goodness of our fit.

\paragraph{Assignment with \texorpdfstring{$f_0$}{f0}(1500) as glueball}
Taking the \(f_0(1500)\) to be the glueball, then the light meson trajectory will start with \(f_0(1370)\), giving:
	\[\mathrm{Light:}\qquad 1370, *1800, 2020, 2330,\]
	\[\ssb\mathrm{:}\qquad 1710, 2100.\]
With \(f_0(1500)\) identified as the glueball, this assignment includes all the states except \(f_0(2200)\). We could also use \(f_0(2200)\) as the \(\ssb\) state and leave out \(f_0(2100)\) instead.

There is no glaring inconsistency in this assignment with the decay modes, but we are again confronted with the state \(*f_0(1800)\), which we need to complete the light meson trajectory. The \(f_0(2020)\) is wider than other states in its trajectory. In particular, it is much wider than the following and last state in the trajectory, \(f_0(2330)\). We can assign the \(f_0(2330)\) to the \(\ssb\) trajectory instead, but there is no other argument for that state being \(\ssb\), considering it was observed only in its decays to \(\pi\pi\) and \(\eta\eta\). Perhaps the fact that \(f_0(1370)\) and \(f_0(2020)\) are both quite wide means that there should be two additional states, with masses comparable to those of \(*f_0(1800)\) and \(f_0(2330)\), that are also wide themselves, and those states will better complete this assignment.
			
\paragraph{Assignment with \texorpdfstring{$f_0$}{f0}(1710) as glueball}
Excluding the \(f_0(1710)\) from the meson trajectories we can make an assignment that includes all states except the \(f_0(500)\) and \(f_0(980)\):
\[\mathrm{Light:}\qquad 1370, *1800, 2100, 2330\]
\[\ssb\mathrm{:}\qquad 1500, 2020, 2200\]
\[\mathrm{Glue:}\qquad 1710\]
The disadvantage here is that we again have to use \(f_0(1500)\) as the head of the \(\ssb\) trajectory despite knowing that its main decay modes are to \(4\pi\) and \(\pi\pi\), as well as the fact the we - once again - need the \(*f_0(1800)\) resonance to fill in a hole for \(n = 1\) in the resulting light meson trajectory.

\paragraph{Conclusions from the \texorpdfstring{$f_0$}{f0} fits}
It is not hard to see that the \(f_0\) resonances listed in the PDG's Review of Particle Physics all fit in quite neatly on two parallel trajectories with a slope similar to that of other mesons. However, upon closer inspection, these trajectories - one for light quark mesons and one for \(\ssb\) - are not consistent with experimental data, as detailed above. For us the naive assignment is also inconsistent with what we have observed for the other \(\ssb\) trajectories when fitting the mesons, namely that the \(\ssb\) trajectories are not purely linear, and have to be corrected by adding a string endpoint mass for the \(s\) quark of at least 200 MeV.

The other novelty that we hoped to introduce, the half slope trajectories of the glueball, proved to be impractical - given the current experimental data which only goes up to less than \(2.4\) GeV for the relevant resonances.

	There is no one assignment that seems the correct one, although the two assignments singling out either \(f_0(1370)\) or \(f_0(1500)\) as the glueball ground states seem more consistent than the other possibilities. The best way to determine which is better is, as always, by finding more experimental data. We list our predictions for higher resonances based on these assignments in table \ref{tab:glue_predictions}.
	
	\begin{table}[ht!] \centering
	\begin{tabular}{|c|c|c|} \hline
Ground state 	&	 Next state mass 	&	 Next state width \\ \hline
\(f_0(980)\)	&	2385\plm70 	&	 405\plm175	\\ \hline
\(f_0(1370)\)	&	2555\plm110 	&	 1255\plm615	\\ \hline
\(f_0(1500)\)	&	2640\plm80 	&	 335\plm30	\\ \hline
\(f_0(1710)\)	&	2770\plm85 	&	 350\plm30	\\ \hline
\end{tabular} \caption{Predictions for the first excited glueball based on half slope Regge trajectories. The mass could correspond to either the \(0^{++}\) excited state with \(n = 2\), or to the first orbital excitation, a \(2^{++}\) state with \(n=0\). The width is calculated based on \(\Gamma \propto L^2 \propto M^2\) (the closed string has to tear twice, with the probability for each tear proportional to the string length). The errors take into account both the uncertainty in the experimental masses and widths, and the uncertainty in the slope.\label{tab:glue_predictions}}
\end{table}

\subsubsection{Assignment of the \texorpdfstring{$f_2$}{f2} into trajectories} \label{sec:glue_f2_fits}
We now turn to the \(f_2\) tensor resonances. We will first examine trajectories in the \((J,M^2)\) plane, then move on to the attempt to assign all the \(f_2\) states to trajectories in the \((n,M^2)\) plane.
\paragraph{Trajectories in the \texorpdfstring{$(J,M^2)$}{(J,M2)} plane}
The only way to get a linear trajectory connecting a \(0^{++}\) and a \(2^{++}\) state with the slope \(\alp\!_{gb} = \frac{1}{2}\alp_{meson}\) is to take the lightest \(f_0\) glueball candidate and the heaviest known \(f_2\). Then we have the pair \(f_0(980)\) and \(f_2(2340)\), and the straight line between them has a slope of 0.45 GeV\(^{-2}\). There is no \(J = 1\) resonance near the line stretched between them. However, this example mostly serves to demonstrate once again the difficulty of forming the glueball trajectories in practice.

It is a more sound strategy to look again for the meson trajectories, see what states are excepted from them, and check for overall consistency of the results. In forming the meson trajectories, we know that we can expect the \(\omega\) mesons with \(J^{PC} = 1^{--}\) to be part of the trajectories, in addition to some states at higher spin (\(3^{--}\), \(4^{++}\), \(\ldots\)), which will allow us to form trajectories with more points.

\begin{figure}[t!] \centering
	\includegraphics[width=0.49\textwidth]{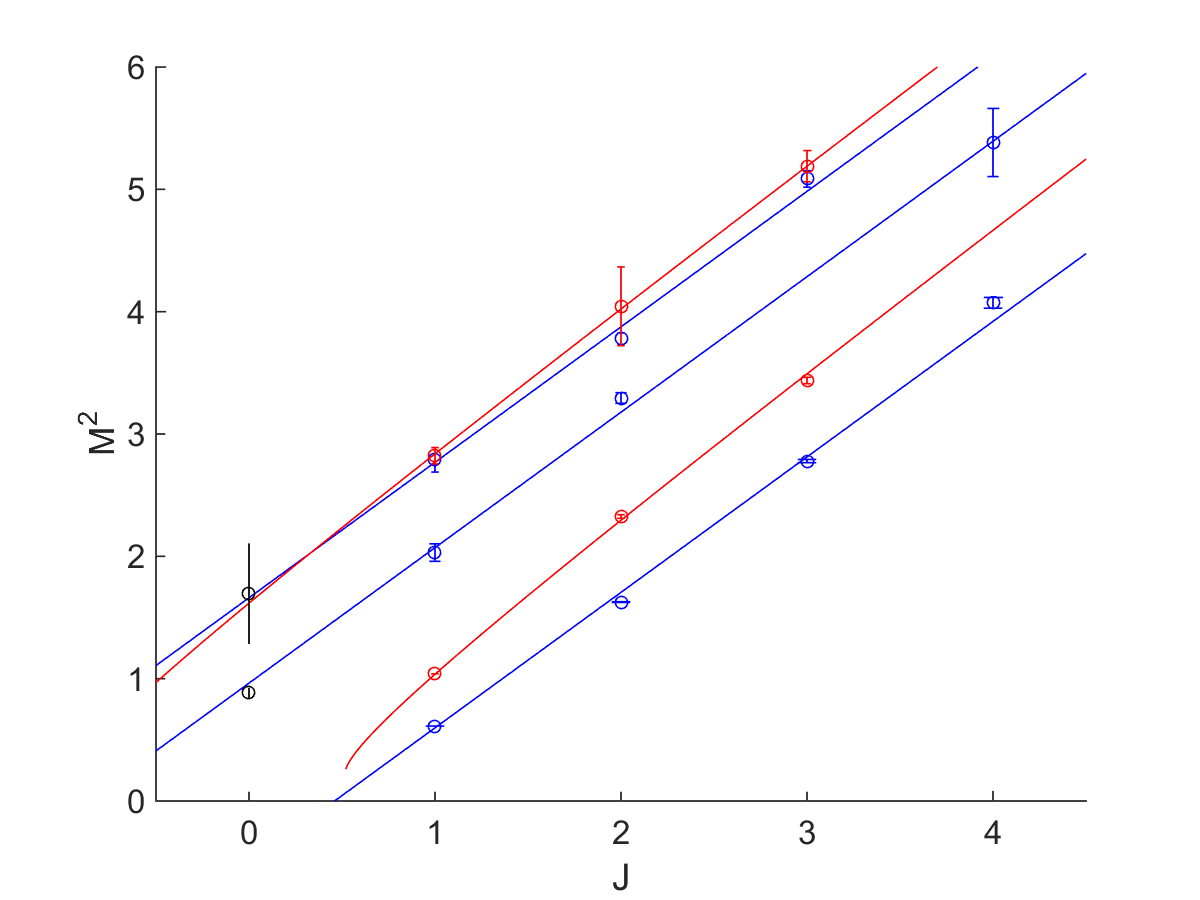}
	\includegraphics[width=0.49\textwidth]{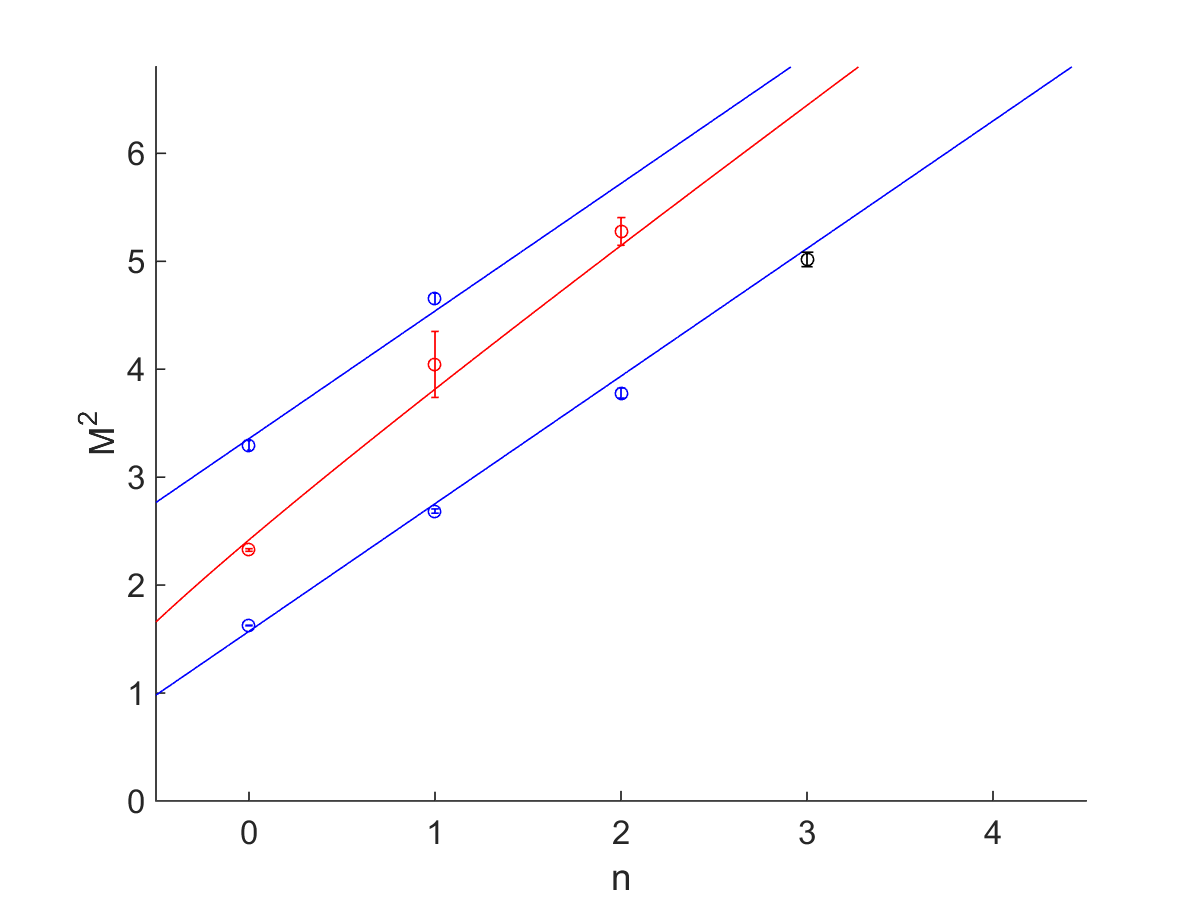}	
				\caption{\label{fig:glue_traj_f2} \textbf{Left:} The trajectory of the \(\omega\) (blue) and \(\phi\) (red) mesons in the \((J,M^2)\) plane and their daughter trajectories. The fits have the common slope \(\alp = 0.903\) GeV\(^{-2}\), and the \(\ssb\) trajectories are fitted using a mass of \(m_s = 250\) MeV for the \(s\) quark. The states forming the trajectories are as follows: With \(J^{PC} = 1^{--}\), \(\omega(782)\), \(\phi(1020)\), \(\omega(1420)\), \(\omega(1650)\), \(\phi(1680)\). With \(J^{PC} = 2^{++}\), \(f_2(1270)\), \(f_2^\prime(1525)\), \(f_2(1810)\), \(f_2(1950)\), and \(f_2(2010)\). With \(J^{PC} = 3^{--}\), \(\omega_3(1670)\), \(\phi_3(1850)\), \(*\omega_3(2255)\), and \(\omega_3(2285)\). And with \(J^{PC} = 4^{++}\), \(f_4(2050)\) and \(f_4(2300)\). We also plot at \(J^{PC} = 0^{++}\) the \(f_0(980)\) and \(f_0(1370)\) which are found to lie near the trajectories fitted, but were not included themselves in the fits, as they are not theoretically expected to belong to them. \textbf{Right}: Some radial trajectories of the \(f_2\), with blue lines for light mesons and red for \(\ssb\). The fits have the common slope \(\alp = 0.846\) GeV\(^{-2}\), and the \(\ssb\) trajectories are fitted using a mass of \(m_s = 400\) MeV for the \(s\) quark. The states forming the trajectories are as follows: The first light meson trajectory with \(f_2(1270)\), \(f_2(1640)\), and \(f_2(1950)\), and followed by the unconfirmed state \(*f_2(2240)\) which was not used in the fit. The \(\ssb\) trajectory with \(f_2^\prime(1525)\), \(f_2(2010)\), and \(f_2(2300)\). And the second light meson trajectory with \(f_2(1810)\) and \(f_2(2150)\).}
			\end{figure}

We can several meson trajectories in the \((J,M^2)\) plane of at least three states, illustrated in figure \ref{fig:glue_traj_f2}. The \(f_2\) states classified in this assignment as mesons are \(f_2(1270)\), \(f_2^\prime(1525)\), \(f_2(1810)\), \(f_2(1950)\), and \(f_2(2010)\). These can perhaps be partnered to existing \(f_0\) states as members of triplets of states with \(J = 0, 1, 2\) and \(PC = ++\) split by spin-orbit interactions. We do not know the exact magnitude of the splitting. There are some \(f_0\) states close (within 20--100 MeV) to the \(f_2\) states mentioned above, and the PDG lists some \(f_1\) (\(1^{++}\)) resonances that may be useful, but we do not find any such trio of states with similar properties and masses that could be said to belong to such a spin-orbit triplet. Therefore, we limit our conclusions from these Regge trajectories to the \(f_2\) which we found we could directly place on them.

\paragraph{\texorpdfstring{Trajectories in the $(n,M^2)$}{Trajectories in the (n,M2)} plane} \label{sec:glue_f2_radial}
Sorting the \(f_2\) resonances into trajectories, the situation is somewhat simpler than with the \(f_0\) scalars, as here we have two states that belong on meson trajectories in the \((J,M^2)\) plane, as we found in previous sections. In particular, the \(f_2(1270)\) belongs to the trajectory of the \(\omega\) meson, and the \(f^\prime_2(1525)\) is an \(s\bar{s}\) and sits on the \(\phi\) trajectory. Their decay modes and other properties are also well known and there is no real doubt about their nature.

The linear trajectory beginning with the \(f_2(1270)\) meson includes the states \(f_2(1640)\)and \(f_2(1950)\). We can include one of the further states \(*f_2(2240)\) as the fourth point in the trajectory. We can also use the \(f_J(2220)\) in place of the \(*f_2(2240)\), but it seems an unnatural choice because of the widths of the states involved (the \(f_J(2220)\) is much narrower than the others with \(\Gamma = 23\pm8 MeV\)).

The projected trajectory of the \(f^\prime_2(1525)\), using the same slope as the \(f_2(1270)\) trajectory and adding mass corrections for the \(s\) quark, includes the \(f_2(2010)\) and the \(f_2(2300)\).

This leaves out the states \(f_2(1430)\), \(f_2(1565)\), \(f_2(1810)\), \(f_2(1910)\), \(f_J(2220)\), and \(f_2(2340)\), as well as the five resonances classified as further states.

The next state we look at is \(f_2(1810)\), classified as a light meson in the \((J,M^2)\) fits of the previous section. It can be used as the head of another light meson trajectory, and the state that would follow it then is \(f_2(2150)\). The next state could be \(f_2(2340)\), except that it has been observed to decay to \(\phi\phi\), making it very unlikely to be a light quark meson.

The state \(f_2(1430)\) is intriguing, in part because of the very small width reported by most (but not all) experiments cited in the PDG, but also because it is located in mass between the two lightest mesons of \(J^{PC} = 2^{++}\), that is between \(f_2(1270)\) (light) and \(f_2^\prime(1525)\) (\(\ssb\)). If we had to assign the \(f_2(1430)\) to a Regge trajectory, then it is best placed preceding the \(f_2(1810)\) and \(f_2(2150)\) in the linear meson trajectory discussed in the last paragraph.

The \(f_J(2200)\), previously known as \(\xi(2230)\), is also a narrow state. It has been considered a candidate for the tensor glueball \cite{Bai:1996wm,Crede:2008vw}. It can be assigned to a linear meson trajectory, as already discussed, but it is clear already from its narrow width that it is not the best choice.

The \(f_2(1565)\) is also left out, but it could be paired with \(f_2(1910)\) to form another linear meson trajectory. To continue we need another state with a mass of around 2200 MeV.

To summarize, we may organize the \(f_2\) resonances by picking first the resonances for the trajectories of the two known mesons,
\[\mathrm{Light:}\qquad 1270, 1640, 1950\]
\[\ssb\mathrm{:}\qquad 1525, 2010, 2300 \]
then find the trajectories starting with the lightest states not yet included. This gives us another meson trajectory using the states
\[\mathrm{Light:}\qquad 1810, 2150\]

The trajectories formed by these eight states are drawn in figure \ref{fig:glue_traj_f2}.

To summarize, there are some simplifications in assigning the \(f_2\) to radial trajectories compared to assigning the \(f_0\) resonances. The reason is that we can look at both orbital and radial trajectories and it is easier to classify some states as mesons. The radial trajectories are consistent with the orbital trajectories: states classified as mesons in the latter are also classified as mesons in the former, and with the same quark contents.

The most interesting states after that remain the \(f_2(1430)\) and \(f_J(2220)\). While the latter has been considered a candidate for the glueball and has been the object of some research (see papers citing \cite{Bai:1996wm}), the former is rarely addressed, despite its curious placement in the spectrum between the lightest \(2^{++}\) light and \(\ssb\) mesons. It seems a worthwhile experimental question to clarify its status - and its quantum numbers, as the most recent observation \cite{Vladimirsky:2001ek} can not confirm whether it is a \(0^{++}\) or \(2^{++}\) state, a fact which led to at least one suggestion \cite{Vijande:2004he} that the \(f_2(1430)\) could be itself the scalar glueball.

\subsubsection{Assignments with non-linear trajectories for the glueball} \label{sec:glue_holo_fits}
In this section we would like check the applicability of a glueball trajectory of the form
\be J = \alp_{gb}E^2-2\alp_{gb}m_0E + a \,, \label{eq:glue_holotraj} \ee
which is the general form we expect from a semi-classical calculation of the corrections to the trajectory in a curved background, and as put forward in section \ref{sec:closed_string_curved}. The novelty here is a term linear in the mass \(E\), which makes the Regge trajectory \(\alpha(t)\) non-linear in \(t = E^2\). The constant \(m_0\) can be either negative or positive, depending on the specific holographic background, and a priori we have to examine both possibilities. It was also noted in section \label{sec:closed_string_curved} that there may be a correction to the slope, but we assume it is small compared to the uncertainty in the phenomenological value of the Regge slope, and we use
\be \alp_{gb} = \frac{1}{2}\alp \ee
throughout this section. We also substitute \(J \rightarrow J + n\) as usual to apply the formula to radial trajectories.

With the \(m_0\) term we can write
\be \frac{\partial J}{\partial E^2} = \frac{\alp}{2}\left(1-\frac{m_0}{E}\right) \,. \label{eq:glue_eff_holo_slope} \ee
We can look at this as an effective slope, and it is the easiest way to see that when \(m_0\) is negative, the effective slope is higher than that of the linear trajectory, and vice versa.

\paragraph{Fits using the holographic formula}
Using the simple linear formula we could not, in most cases, find glueball trajectories among the observed \(f_0\) and \(f_2\) states. This is because the first excited state is expected to be too high in mass and outside the range of the states measured in experiment.

Adding an appropriate \(m_0\) term can modify this behavior enough for us to find some pairs of states on what we would then call glueball trajectories, and by appropriate we mean a negative value that will make the effective slope of eq. \ref{eq:glue_eff_holo_slope} higher. The problem is then that we have only pairs of states, with two fitting parameters: \(m_0\) and \(a\) (and \(\alp\) which is fixed by the meson trajectory fits). We form these pairs by picking a state left out from the meson trajectories proposed in sections \ref{sec:glue_f0_fits} and \ref{sec:glue_f2_fits} and assigning it as the excited partner of the appropriate glueball candidate.

There is a solution for \(m_0\) and \(a\) for any pair of states which we can take, and the question then becomes whether there is a reason to prefer some values of the two parameters over others. We list some other values obtained for \(m_0\) and \(a\) in table \ref{tab:glue_holo_fits}.

\begin{table}[t!] \centering
	\begin{tabular}{|c|c|c|c|c|} \hline
	Ground state & Excited state & \(\alp\) [GeV\(^{-2}\)] & $m_0$ [GeV] & \(a\) \\ \hline
	
	\(f_0(980)\) & \(f_0(2200)\) & 0.79 & -0.52 & -0.78 \\
	
	\(f_0(1370)\) & \(f_0(2020)\) & 0.87 & -2.00 & -3.21 \\
	
	\(f_0(1500)\) & \(f_0(2200)\) & 0.87 & -1.51 & -2.97 \\
	
	\(f_2(1430)\) & \(f_J(2220)\) & 0.81 & -1.33 & -2.42 \\ \hline
	
	\end{tabular} \caption{\label{tab:glue_holo_fits} Values obtained for the parameters \(m_0\) and \(a\) for some of the possible pairs of states on glueball trajectories. The states selected as the excited state of the glueball are those not included in the meson trajectories of the assignments of sections \ref{sec:glue_f0_fits} and \ref{sec:glue_f2_fits}, and the slopes are selected based on the results of the meson fits presented in the same sections.}
\end{table}

\paragraph{Using the holographic formula with a constrained intercept}
\cite{Bigazzi:2004ze} implies that a universal form of the first semi-classical correction of the Regge trajectory of the rotating folded string is
\be J + n = \frac{1}{2}\alp(E-m_0)^2 \,, \ee
up to further (model dependent) modifications of the slope, which in the cases calculated are small. In other words, the intercept obtained then from the semi-classical calculation is
\be a = \frac{1}{2}\alp m_0^2 \,. \ee

The intercept is always positive in this scenario. If we want to include the ground state with \(J = n = 0\) the only way to do it is to take a positive \(m_0\), specifically we should take \(m_0 = M_{gs}\), where \(M_{gs}\) is the mass of the ground state. There is no problem with the resulting expression theoretically, but it is not very useful in analyzing the observed spectrum. The trouble is that when using this expression the energy rises much too fast with \(J\) and we end up very quickly with masses outside the range of the glueball candidates. If we take, for instance, \(f_0(980)\) as the ground state then the first excited state is expected to have a mass of around 2500 MeV, and the heavier candidates naturally predict even heavier masses for the excited states.

Another way to use eq. \ref{eq:glue_holotraj} is to begin the trajectory with a \(J = 2\) state. Then \(m_0\) can be either positive or negative.  We can then proceed as usual: we pick the head of a trajectory and see if there are any matches for its predicted excited states. We can see, for example, that we can again pair \(f_2(1430)\) with \(f_J(2220)\). Constraining \(\alp\) to be \(0.90\) GeV\(^{-2}\), the best fit has \(m_0 = -0.72\) GeV, and the masses calculated are 1390 and 2260 MeV for the experimental values of \(1453\pm4\) and \(2231\pm4\) MeV.

\subsubsection{Glueball Regge trajectories in lattice QCD} \label{sec:glue_lattice}
The glueball spectrum has been studied extensively in lattice QCD. Some works have compared results with different stringy models, e.g. \cite{Athenodorou:2010cs,Bochicchio:2013aha,Bochicchio:2013sra,Caselle:2015tza}. However, the question whether or not the glueballs form linear Regge trajectories is not often addressed, due to the difficulty involved in computing highly excited states. When linear Regge trajectories are discussed, it is often when trying to identify the glueball with the pomeron and searching for states along the given pomeron trajectory,
\be \alpha(t) = \alp_p t + 1 + \epsilon \, \ee
 where the slope and the intercept are known from experiment to be \(\alp_p = 0.25\) GeV\(^{-2}\) and \(1 + \epsilon \approx 1.08\) \cite{Donnachie:1984xq}.

The most extensive study of glueball Regge trajectories is that of Meyer and Teper \cite{Meyer:2004jc,Meyer:2004gx}, where a relatively large number of higher mass states is computed, including both high spin states and some highly excited states at low spin. We present here fits to some trajectories with more than two states, based on the results in \cite{Meyer:2004gx}.

Results in lattice computations are for the dimensionless ratio between the mass of a state and the square root of the string tension: \(M/\sqrt{T}\). To get the masses \(M\) in MeV one has to fix the scale by setting the value of \(T\). This introduces an additional uncertainty in the obtained values. For the purpose of identifying Regge trajectories we can work directly with dimensionless quantities, avoiding this extra error. For the following, our fitting model will be
\be \frac{M^2}{T} = \frac{2\pi}{\eta} (N + a) \ee
In this notation the ratio \(\eta\), which is the primary fitting parameter (in addition to the intercept \(a\)), is expected to be 1 for open strings and \(1/2\) for closed strings. It is referred to below as the ``relative slope''. \(N\) will be either the spin \(J\) or the radial excitation number \(n\).

\begin{figure}[tp!] \centering
	\includegraphics[width=0.49\textwidth]{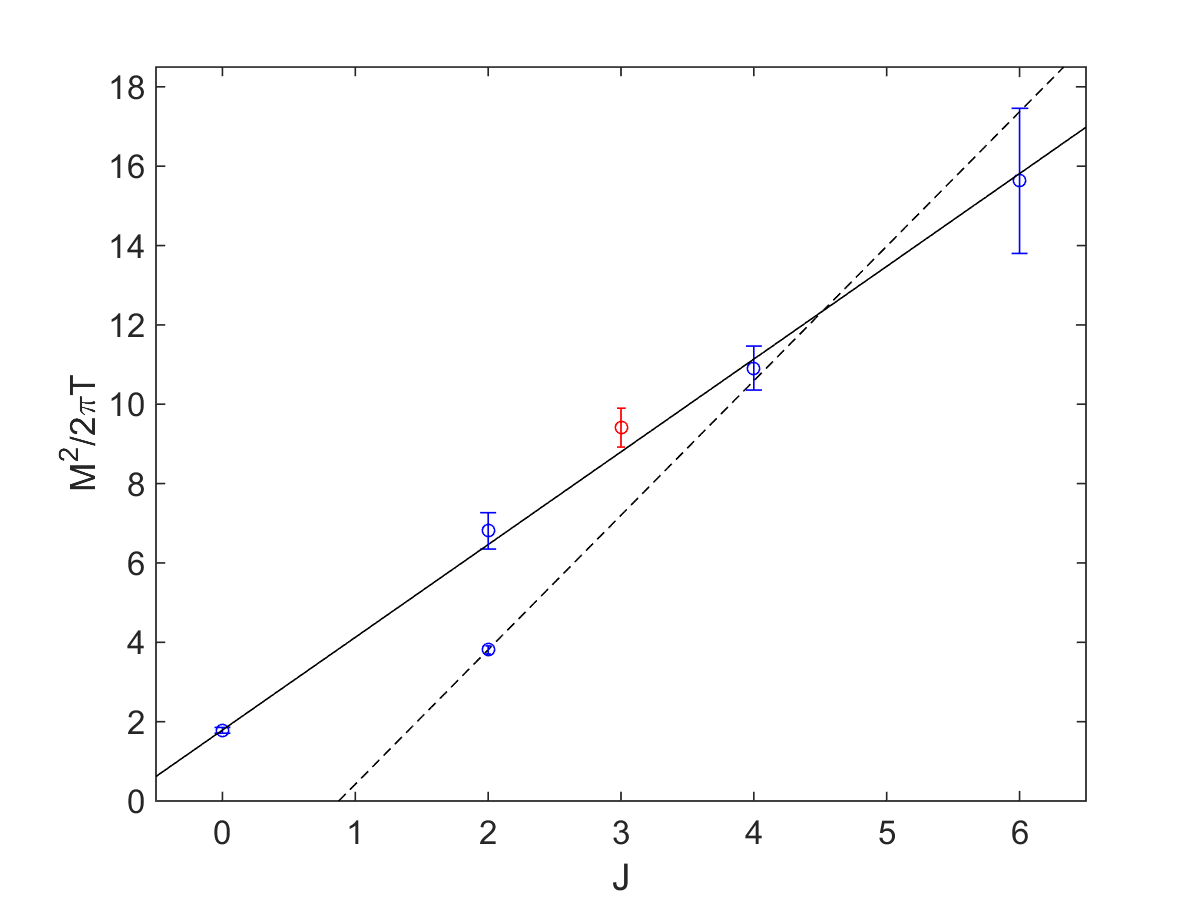}
	\includegraphics[width=0.49\textwidth]{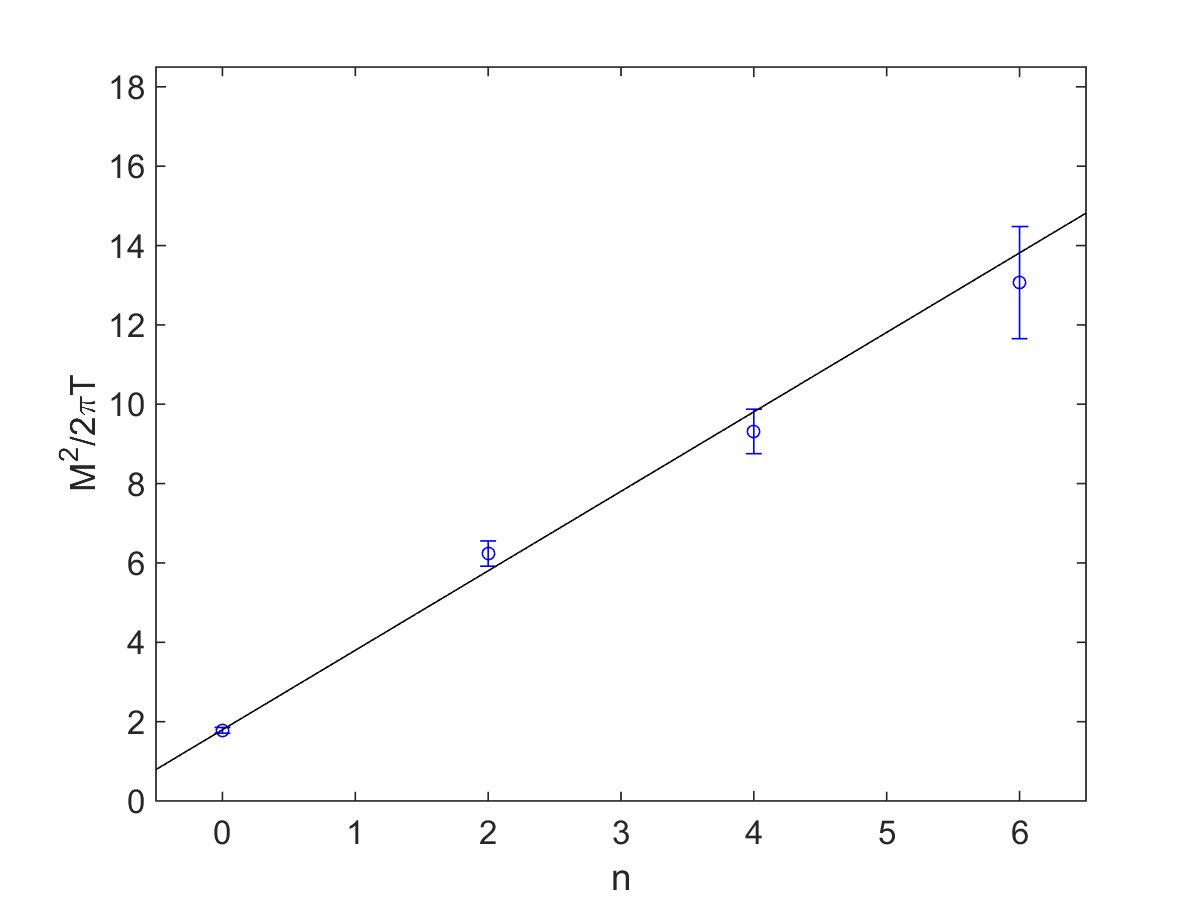}
	\caption{\label{fig:glue_lat_Meyer} The trajectories of the \(PC = ++\) glueball states found in lattice calculations in \cite{Meyer:2004gx}. \textbf{Left:} Trajectories in the \((J,M^2)\) plane. The full line is the fit to a proposed trajectory using four states with \(J = 0, 2, 4, 6\), where the relative slope is \(0.43\) and the lightest tensor is excluded (\(\chi^2 = 0.37\)). The dotted line is the leading trajectory proposed in the analysis in \cite{Meyer:2004gx}, with a pomeron-like slope. It includes the \(J = 2, 4,\) and \(6\) states. (\(\chi^2 = 1.24\)). In this second option the scalar is excluded. Also plotted is the \(3^{++}\) state, which was not used in the fit. \textbf{Right:} trajectory of four states with \(J^{PC} = 0^{++}\). The relative slope is exactly \(0.50\) (\(\chi^2 = 1.48\)).}
\end{figure}

\paragraph{Trajectories in the \texorpdfstring{$(J,M^2)$}{(J,M2)} plane:} As mentioned above, \cite{Meyer:2004gx} has the most high spin states. The analysis there observes that the first \(2^{++}\)and \(4^{++}\) states can be connected by a line with the relative slope
\be \eta = 0.28\pm0.02, \ee
which, when taking a typical value of the string tension \(\sqrt{T} = 430\) MeV (\(\alp = 0.84\) GeV\(^{-2}\)), gives a slope virtually identical to that expected for the pomeron, \(0.25\) GeV\(^{-2}\). This trajectory can be continued with the calculated \(6^{++}\) state. A fit to the three state trajectory gives the result
\be \eta = 0.29\pm0.15.\ee
This trajectory leaves out the \(0^{++}\) ground state. In \cite{Meyer:2004gx} the lowest \(0^{++}\) is paired with the second, excited, \(2^{++}\) state, giving a trajectory with
\be \eta = 0.40\pm0.04.\ee
A possibility not explored in \cite{Meyer:2004gx} is that of continuing this trajectory, of the first \(0^{++}\) and the excited \(2^{++}\), and with the \(4^{++}\) and \(6^{++}\) states following. Then we have the result
\be q = 0.43\pm0.03\ee
This second option not only includes more points, it is also a better fit in terms of \(\chi^2\) per degrees of freedom (0.37 instead of 1.24). The lowest \(2^{++}\) state is then left out of a trajectory. There is also a \(J = 3\) state in the \(PC = ++\) sector that lies very close to the trajectory of the \(0^{++}\) ground state. In the closed string model the \(J = 3\) state is not expected to belong to the trajectory, so that state is also left out of the fit. The trajectories of the \(PC = ++\) states are in the left side of figure \ref{fig:glue_lat_Meyer}.

\paragraph{Trajectories in the \texorpdfstring{$(n,M^2)$}{(n,M2)} plane:}
In trajectories in the \((n,M^2)\) plane we assume \(n\) takes only even values, i.e. \(n = 0,2,4,\ldots\), as it does for the closed string. The results when taking \(n = 0,1,2,\ldots\) will be half those listed.

We again take the results from \cite{Meyer:2004gx}, as it offers calculations of several excited states with the same \(J^{PC}\). Most notably we see there four states listed with \(J^{PC} = 0^{++}\). We observe that those points are well fitted by a trajectory with the slope
\be q = 0.50\plm0.07, \ee
where \(\chi^2 = 1.48\) for the fit. It is interesting to compare this with the trajectory that can be drawn from the \(0^{++}\) ground state in the \((J,M^2)\) plane. The \((n,M^2)\) trajectory with \(n = 0,2,4,6\) is very similar to the trajectory beginning with the same state and continuing to \(J = 2, 4,\) and \(6\). This is what we see also for mesons and baryons in experiment: two analogous trajectories with similar slopes in the different planes.

Other than the trajectory of the four \(0^{++}\) states (plotted in figure \ref{fig:glue_lat_Meyer}), we list the slopes calculated for pairs of states who share other quantum numbers. We also include there some results based on different lattice calculations. This is in table \ref{tab:glue_lat_n}.
\begin{table} \centering
	\begin{tabular}{|c|c|c|c|c|c|} \hline
		\(J^{PC}\) & \(0^{++}\) & \(2^{++}\) & \(4^{++}\) & \(0^{-+}\) & \(2^{-+}\) \\ \hline\hline
		Meyer \cite{Meyer:2004gx} &
				0.50\plm0.07 & 0.67\plm0.10 & 0.30\plm0.06 & 0.39\plm0.07 & 0.56\plm0.13 \\ \hline

		M\&P \cite{Morningstar:1999rf} &
				0.51\plm0.12 & -						& -						 & 0.32\plm0.02 & 0.38\plm0.03 \\ \hline
	\end{tabular}
	\caption{\label{tab:glue_lat_n} Relative slopes \(q\) of trajectories in the \((n,M^2)\) plane. The first result (Meyer/\(0^{++}\)) is that of a fit to the four point trajectory drawn in \ref{fig:glue_lat_Meyer}. The other results are obtained when calculating the slopes between pairs of states, where the lowest state is assumed to have \(n = 0\), and the next calculated state is taken to be the first excited state with \(n = 2\).}
\end{table}
\section{ Summary and future directions}

The model of holography inspired stringy hadrons (HISH)  is still in its infant stage. The story is two folded. One side is the development of a fully consistent quantum theory from which one should be able to extract the spectra, decay  and scattering processes and on the other side is the phenomenology of comparing the theoretical predictions to the actual data and to feedback from the data to the model. In the first part of the review we  described the string configurations of certain  holographic models that correspond to hadrons and we sketched the derivation of the HISH model. In the second part we
confronted the PDG hadronic spectrum with the model. A significant part of the spectra does fit nicely the model but there are also exceptional cases.

Not surprisingly there  are still quite a lot of  open questions and correspondingly  directions for future investigation. Here we describe only few of them:
\begin{itemize}
\item
Our model does not incorporate spin degrees of freedom. It is well known that the spin and the spin-orbit interaction play an important role in the spectra of mesons. Thus the simple rotating string models that we are using have to be improved by introducing spin degrees of freedom to the endpoints. One way to achieve it is by replacing the spinless relativistic particle with one that carries spin or in the holographic framework associating spin to the vertical segments of the holographic string.
\item
Our model assumes chargeless massive endpoint particles. The endpoint of a string on a flavor brane carries a charge associated with the symmetry group of the flavor branes. Thus it is natural to add an interaction, for instance Abelian interaction, between the two string endpoints. It is easy to check that this change will introduce a modification of the intercept.

\item
As was discussed in the introduction, the models we are using are not the outcome of a full quantization of the system. We have been either using a WKB approximation for the spectra in the $(n,M^2)$ plane or using an ansatz of $J\rightarrow J+n-a$ for passing from the classical to the quantum model. In \cite{Hellerman:2013kba} the quantization of the rotating string without massive endpoints was determined. The quantum Regge trajectories associated with strings with massive endpoints require determining the contributions to the intercept to order $J^0$ from both the `` Casimir" term and the Polchinski-Strominger term\cite{Polchinski:1991ax}. Once a determination of the intercept as a function of $\frac{m^2}{T}$ is made, an improved fit and a re-examination of the deviations from a universal model should be made.
\item
We have looked in the present work into only one feature of meson physics - the Regge trajectories of the spectra. One additional property that can be explored is the width of the decay of a meson into two mesons. The stringy holographic width was computed in \cite{Peeters:2005fq}. A detailed comparison with decay width of mesons can provide an additional way to extract string endpoint masses that can be compared to the one deduced from the spectra.

\item
Eventually we have in mind to perform ``precise comparisons" using holographic rotating string models instead of the model of rotating string with massive endpoints in flat space-time.
\item
As was emphasized in this note regarding the search for glueballs, one urgent issue is to gain additional data about flavorless hadrons. This calls for a further investigation of experiments that yield this kind of resonances and for proposing future experiments of potential glueball production, in particular in the range above 2.4 GeV. This can follow the predictions of the masses and width of the resonances as were listed in \cite{Sonnenschein:2015zaa}.
\item
Related to the exploration of experimental data is the investigation of efficient mechanisms of creating glueballs. This issue was not addressed in this paper. Among possible glueball formation one finds radiative $J/\psi$ decays, pomeron pomeron collisions in hadron-hadron central production and in $p$-$\bar p$ annihilation. Naturally, we would like to understand possible glueball formation in LHC experiments. It is known that we can find in the latter processes of gluon-gluon scattering and hence it may serve as a device for glueball creation.
\item
As was mentioned in section $\S 5.2$, the quantization of folded closed strings in D non-critical dimensions has not yet been deciphered. In \cite{Hellerman:2013kba} the expression derived for the intercept is singular in the case where is only one rotation plane - as it naturally is in $D=4$. We mentioned a potential avenue to resolve this issue by introducing massive particles on the folds, quantize the system as that of a string with massive endpoints \cite{ASY}, and then take the limit of zero mass.
\item
We have mentioned that the rotating closed strings are in fact rotating folded closed strings. However, we did not make any attempt in this note to explore the role of the folds. In fact it seems that very few research has been devoted to the understanding of folded strings \cite{Ganor:1994rm}. It would be interesting to use the rotating closed string as a venue to the more general exploration of strings with folds which may be related to certain systems in nature.
\item
A mystery related to the closed string description of glueballs is the relation between the pomeron and the glueball. Supposedly  both the glueball and the pomeron are described by a closed string. As we have emphasized in this note the slope of the closed string is half that of the open string and hence we advocated the search of trajectories with that slope. However, it was found from fitting the differential cross section of $p$-$p$ collisions that the slope of the pomeron is $\alp_{pomeron}\approx 0.25$ GeV$^{-2}$. That is, a slope which is closer to a quarter of that of the meson open string rather than half. Thus the stringy structure of the pomeron and its exact relation to the glueball is still an open question. The stringy holographic description of the pomeron physics was discussed in particular in \cite{Shuryak:2013sra}.
\item
The closed string description of the glueball faces a very obvious question. In QCD one can form a glueball as a bound state of two, three, or in fact any number of gluons. The stringy picture seems to describe the composite of two gluons, and it is not clear how to realize those glueballs constructed in QCD from more than two gluons.
\item
Another prediction from holography is the presence of a baryonic vertex. Our fits exclude the presence of a baryonic vertex mass at the center of mass of the rotating baryon, but in the quark-diquark model which we prefer it is expected to be found at one of the string endpoints, with the diquark. If there is a baryonic vertex at an endpoint, there is no evidence to suggest it contributes greatly to the endpoint mass.
\item
We could also attempt enhancements of the model. Adding spin degrees of freedom to the endpoints would give us a much better chance of constructing a universal model that would describe the entire baryon spectrum. In \cite{Selem:2006nd} the distinction between spin zero and spin one diquarks played an important part in analyzing the spectrum, while we have only discussed the flavor structure of the diquark and ascribed its mass to the holographic string alone (with a possible small addition from the baryonic vertex). Spin interactions could also help explain the even-odd splitting observed in the light baryons - our simple classical model will not explain a phenomenon that distinguishes between symmetric and anti-symmetric states without some additional interaction.
\item
We should also strive to gain a better understanding of the intercept. While all previous sections discuss results for the slope and endpoint masses alone, the intercept, the results for which are listed in the summary tables of section $\S 6.3.4$, is an interesting parameter from a theoretical point of view, and understanding its behavior is an important goal in constructing a truly universal model of the baryon. Added interactions should contribute, in leading order, a correction to the intercept, which should also be affected by the endpoint masses, and understanding the intercept's behavior may also help us distinguish between different configurations of the baryon without requiring additional information from experiment.
\end{itemize}

{\bf Note added} While the HISH models have been developed, an alternative way to describe the hadronic spectrum  based on light-front holography was proposed (\cite{Brodsky:2015oia} and references therein). Though the two approaches are inspired by holography, they are different mainly since the latter is essentially a QFT description whereas the HISH is a stringy one . 

{\bf Acknowledgments}

I would like to thank all my collaborators in the various projects reviewed in this review paper: O. Aharony, A. Brandhuber, B. Burrington, O. Bergman, R. Casero, A. Dymarsky, E. Federovsky, G. Harpaz, N. Itzhaki V. Kaplunovsky, Y. Kinar, U. Kol  M. Kruczenski,  S. Kuperstein,  D. Melnikov, O. Mintakevich, Pando Zayas,  A. Paredes,  K. Peeters, T. Sakai, E. Schreiber, S. Seki,  D. Vaman, D. Weissman, S. Yankielowicz, M.  Zamaklar. I would also like to thank D. Weissman for his help in preparing  this review and O. Aharony for his remarks on manuscript. This  work  was  supported  in  part  by  a  centre  of  excellence  supported  by  the
Israel  Science  Foundation  (grant  number  1989/14),  and  by  the  US-Israel  bi-national  fund
(BSF)  grant  number  2012383  and  the  Germany  Israel  bi-national  fund  GIF  grant  number
I-244-303.7-2013
\providecommand{\href}[2]{#2}\begingroup\raggedright
\endgroup

\begin{thebibliography}{10}

	\bibitem{Collins:book} P.~Collins, {\it {An Introduction to Regge Theory and High Energy Physics}},
  {\em Cambridge Univeristy Press} (1977) 456.

\bibitem{Sonnenschein:1999if}
  J.~Sonnenschein,
  ``What does the string / gauge correspondence teach us about Wilson loops?,''
  hep-th/0003032.
	\bibitem{Sonnenschein:2014jwa} J.~Sonnenschein and D.~Weissman, {\it {Rotating strings confronting PDG
  mesons}},  {\em JHEP} {\bf 1408} (2014) 013,
  [\href{http://xxx.lanl.gov/abs/1402.5603}{{\tt arXiv:1402.5603}}].

	\bibitem{Sonnenschein:2014bia} J.~Sonnenschein and D.~Weissman, {\it {A rotating string model versus baryon
  spectra}},  {\em JHEP} {\bf 1502} (2015) 147,
  [\href{http://xxx.lanl.gov/abs/1408.0763}{{\tt arXiv:1408.0763}}].


	\bibitem{Sonnenschein:2015zaa} J.~Sonnenschein and D.~Weissman,
  {\it{``Glueballs as rotating folded closed strings}}
  JHEP {\bf 1512}, 011 (2015)
  doi:10.1007/JHEP12(2015)011
  [arXiv:1507.01604 [hep-ph]].

	\bibitem{Baker:2002km} M.~Baker and R.~Steinke, {\it {Semiclassical quantization of effective string
  theory and Regge trajectories}},  {\em Phys.Rev.} {\bf D65} (2002) 094042,
  [\href{http://xxx.lanl.gov/abs/hep-th/0201169}{{\tt hep-th/0201169}}].

	\bibitem{Schreiber:2004ie} E.~Schreiber, {\it {Excited mesons and quantization of string endpoints}},
  \href{http://xxx.lanl.gov/abs/hep-th/0403226}{{\tt hep-th/0403226}}.

	\bibitem{Hellerman:2013kba} S.~Hellerman and I.~Swanson, {\it {String Theory of the Regge Intercept}},
  \href{http://xxx.lanl.gov/abs/1312.0999}{{\tt arXiv:1312.0999}}.

	\bibitem{Zahn:2013yma} J.~Zahn, {\it {The excitation spectrum of rotating strings with masses at the
  ends}},  {\em JHEP} {\bf 1312} (2013) 047,
  [\href{http://xxx.lanl.gov/abs/1310.0253}{{\tt arXiv:1310.0253}}].

	\bibitem{Aharony:2013ipa} O.~Aharony and Z.~Komargodski, {\it {The Effective Theory of Long Strings}},
  {\em JHEP} {\bf 1305} (2013) 118,
  [\href{http://xxx.lanl.gov/abs/1302.6257}{{\tt arXiv:1302.6257}}].


	\bibitem{Erdmenger:2007cm} J.~Erdmenger, N.~Evans, I.~Kirsch and E.~Threlfall,
  ``Mesons in Gauge/Gravity Duals - A Review,''
  Eur.\ Phys.\ J.\ A {\bf 35} (2008) 81
  doi:10.1140/epja/i2007-10540-1
  [arXiv:0711.4467 [hep-th]].

	\bibitem{Polchinski:2000uf} J.~Polchinski and M.~J.~Strassler,
  ``The String dual of a confining four-dimensional gauge theory,''
  hep-th/0003136.


	\bibitem{Aharony:2002up} O.~Aharony,
  ``The NonAdS / nonCFT correspondence, or three different paths to QCD,''
  hep-th/0212193.


	\bibitem{Maldacena:2000yy} J.~M.~Maldacena and C.~Nunez,
  ``Towards the large N limit of pure N=1 superYang-Mills,''
  Phys.\ Rev.\ Lett.\  {\bf 86}, 588 (2001)
  doi:10.1103/PhysRevLett.86.588
  [hep-th/0008001].


	\bibitem{Witten:1998zw} E.~Witten, {\it {Anti-de Sitter space, thermal phase transition, and
  confinement in gauge theories}},  {\em Adv.Theor.Math.Phys.} {\bf 2} (1998)
  505--532, [\href{http://xxx.lanl.gov/abs/hep-th/9803131}{{\tt
  hep-th/9803131}}].


	\bibitem{Klebanov:2000hb} I.~R.~Klebanov and M.~J.~Strassler,
  ``Supergravity and a confining gauge theory: Duality cascades and chi SB resolution of naked singularities,''
  JHEP {\bf 0008}, 052 (2000)
  doi:10.1088/1126-6708/2000/08/052
  [hep-th/0007191].


	\bibitem{Itzhaki:1998dd} N.~Itzhaki, J.~M.~Maldacena, J.~Sonnenschein and S.~Yankielowicz,
  ``Supergravity and the large N limit of theories with sixteen supercharges,''
  Phys.\ Rev.\ D {\bf 58}, 046004 (1998)
  doi:10.1103/PhysRevD.58.046004
  [hep-th/9802042].

	\bibitem{Kuperstein:2004yf} S.~Kuperstein and J.~Sonnenschein, {\it {Non-critical, near extremal AdS(6)
  background as a holographic laboratory of four dimensional YM theory}},  {\em
  JHEP} {\bf 0411} (2004) 026,
  [\href{http://xxx.lanl.gov/abs/hep-th/0411009}{{\tt hep-th/0411009}}].


	\bibitem{Polchinski:2002jw} J.~Polchinski and M.~J.~Strassler,
  ``Deep inelastic scattering and gauge / string duality,''
  JHEP {\bf 0305}, 012 (2003)
  doi:10.1088/1126-6708/2003/05/012
  [hep-th/0209211].

	\bibitem{oai:arXiv.org:hep-ph/0602229} A.~Karch, E.~Katz, D.~T. Son, and M.~A. Stephanov, {\it {Linear confinement and
  AdS/QCD}},  {\em Phys.Rev.} {\bf D74} (2006) 015005,
  [\href{http://xxx.lanl.gov/abs/hep-ph/0602229}{{\tt hep-ph/0602229}}].

	\bibitem{Gursoy:2007cb} U.~Gursoy and E.~Kiritsis,
  ``Exploring improved holographic theories for QCD: Part I,''
  JHEP {\bf 0802}, 032 (2008)
  doi:10.1088/1126-6708/2008/02/032
  [arXiv:0707.1324 [hep-th]].




	\bibitem{Karch:2002sh} A.~Karch and E.~Katz,
  ``Adding flavor to AdS / CFT,''
  JHEP {\bf 0206}, 043 (2002)
  doi:10.1088/1126-6708/2002/06/043
  [hep-th/0205236].

	\bibitem{Kruczenski:2003be} M.~Kruczenski, D.~Mateos, R.~C.~Myers and D.~J.~Winters,
  ``Meson spectroscopy in AdS / CFT with flavor,''
  JHEP {\bf 0307} (2003) 049
  doi:10.1088/1126-6708/2003/07/049
  [hep-th/0304032].




	\bibitem{Burrington:2004id} B.~A.~Burrington, J.~T.~Liu, L.~A.~Pando Zayas and D.~Vaman,
  ``Holographic duals of flavored N=1 super Yang-mills: Beyond the probe approximation,''
  JHEP {\bf 0502}, 022 (2005)
  doi:10.1088/1126-6708/2005/02/022
  [hep-th/0406207].

	\bibitem{Sakai:2003wu} T.~Sakai and J.~Sonnenschein,
  ``Probing flavored mesons of confining gauge theories by supergravity,''
  JHEP {\bf 0309}, 047 (2003)
  doi:10.1088/1126-6708/2003/09/047
  [hep-th/0305049].


	\bibitem{Dymarsky:2009cm} A.~Dymarsky, S.~Kuperstein, and J.~Sonnenschein, {\it {Chiral Symmetry Breaking
  with non-SUSY D7-branes in ISD backgrounds}},  {\em JHEP} {\bf 0908} (2009)
  005, [\href{http://xxx.lanl.gov/abs/0904.0988}{{\tt arXiv:0904.0988}}].

	\bibitem{Jarvinen:2011qe} M.~Jarvinen and E.~Kiritsis,
  ``Holographic Models for QCD in the Veneziano Limit,''
  JHEP {\bf 1203}, 002 (2012)
  doi:10.1007/JHEP03(2012)002
  [arXiv:1112.1261 [hep-ph]].

	\bibitem{SakSug} T.~Sakai and S.~Sugimoto, {\it {Low energy hadron physics in holographic QCD}},
   {\em Prog.Theor.Phys.} {\bf 113} (2005) 843--882,
  [\href{http://xxx.lanl.gov/abs/hep-th/0412141}{{\tt hep-th/0412141}}].

	\bibitem{Brandhuber:1998er} A.~Brandhuber, N.~Itzhaki, J.~Sonnenschein and S.~Yankielowicz,
  {\it{Wilson loops, confinement, and phase transitions in large N gauge theories from supergravity,}}
  JHEP {\bf 9806}, 001 (1998)
  doi:10.1088/1126-6708/1998/06/001
  [hep-th/9803263].

	\bibitem{Kinar:1998vq} Y.~Kinar, E.~Schreiber, and J.~Sonnenschein, {\it {Q anti-Q potential from
  strings in curved space-time: Classical results}},  {\em Nucl.Phys.} {\bf
  B566} (2000) 103--125, [\href{http://xxx.lanl.gov/abs/hep-th/9811192}{{\tt
  hep-th/9811192}}].


	\bibitem{Aharony:2006da} O.~Aharony, J.~Sonnenschein and S.~Yankielowicz,
  Annals Phys.\  {\bf 322}, 1420 (2007)
  doi:10.1016/j.aop.2006.11.002
  [hep-th/0604161].
	

	\bibitem{Kaplunovsky:2010eh} V.~Kaplunovsky and J.~Sonnenschein, {\it {Searching for an Attractive Force in
  Holographic Nuclear Physics}},  {\em JHEP} {\bf 1105} (2011) 058,
  [\href{http://xxx.lanl.gov/abs/1003.2621}{{\tt arXiv:1003.2621}}].

	\bibitem{Bergman:2007pm} O.~Bergman, S.~Seki and J.~Sonnenschein,
  JHEP {\bf 0712}, 037 (2007)
  doi:10.1088/1126-6708/2007/12/037
  [arXiv:0708.2839 [hep-th]].
	

	\bibitem{Casero:2007ae} R.~Casero, E.~Kiritsis and A.~Paredes,
  ``Chiral symmetry breaking as open string tachyon condensation,''
  Nucl.\ Phys.\ B {\bf 787}, 98 (2007)
  doi:10.1016/j.nuclphysb.2007.07.009
  [hep-th/0702155 [HEP-TH]].


	\bibitem{Dhar:2007bz} A.~Dhar and P.~Nag,
  ``Sakai-Sugimoto model, Tachyon Condensation and Chiral symmetry Breaking,''
  JHEP {\bf 0801}, 055 (2008)
  doi:10.1088/1126-6708/2008/01/055
  [arXiv:0708.3233 [hep-th]].

	\bibitem{Aharony:2008an} O.~Aharony and D.~Kutasov,
  ``Holographic Duals of Long Open Strings,''
  Phys.\ Rev.\ D {\bf 78}, 026005 (2008)
  doi:10.1103/PhysRevD.78.026005
  [arXiv:0803.3547 [hep-th]].

	\bibitem{Burrington:2007qd} B.~A.~Burrington, V.~S.~Kaplunovsky and J.~Sonnenschein,
  ``Localized Backreacted Flavor Branes in Holographic QCD,''
  JHEP {\bf 0802}, 001 (2008)
  doi:10.1088/1126-6708/2008/02/001
  [arXiv:0708.1234 [hep-th]].
	

	\bibitem{Maldacena:1998im} J.~M.~Maldacena,
  ``Wilson loops in large N field theories,''
  Phys.\ Rev.\ Lett.\  {\bf 80}, 4859 (1998)
  doi:10.1103/PhysRevLett.80.4859
  [hep-th/9803002].


	\bibitem{Kol:2010fq} U.~Kol and J.~Sonnenschein,
  JHEP {\bf 1105}, 111 (2011)
  doi:10.1007/JHEP05(2011)111
  [arXiv:1012.5974 [hep-th]].


	\bibitem{Witten:1998xy} E.~Witten, {\it {Baryons and branes in anti-de Sitter space}},  {\em JHEP} {\bf
  9807} (1998) 006, [\href{http://xxx.lanl.gov/abs/hep-th/9805112}{{\tt
  hep-th/9805112}}].

	\bibitem{Dymarsky:2010ci} A.~Dymarsky, D.~Melnikov, and J.~Sonnenschein, {\it {Attractive Holographic
  Baryons}},  {\em JHEP} {\bf 1106} (2011) 145,
  [\href{http://xxx.lanl.gov/abs/1012.1616}{{\tt arXiv:1012.1616}}].

	\bibitem{Seki:2008mu} S.~Seki and J.~Sonnenschein, {\it {Comments on Baryons in Holographic QCD}},
  {\em JHEP} {\bf 0901} (2009) 053,
  [\href{http://xxx.lanl.gov/abs/0810.1633}{{\tt arXiv:0810.1633}}].

	\bibitem{Bali:2000un} G.~S. Bali, {\it {Casimir scaling of SU(3) static potentials}},  {\em
  Phys.Rev.} {\bf D62} (2000) 114503,
  [\href{http://xxx.lanl.gov/abs/hep-lat/0006022}{{\tt hep-lat/0006022}}].

	\bibitem{Isgur:1984bm} N.~Isgur and J.~E. Paton, {\it {A Flux Tube Model for Hadrons in QCD}},  {\em
  Phys.Rev.} {\bf D31} (1985) 2910.

	\bibitem{Meyer:2004jc} H.~B. Meyer and M.~J. Teper, {\it {Glueball Regge trajectories and the pomeron:
  A Lattice study}},  {\em Phys.Lett.} {\bf B605} (2005) 344--354,
  [\href{http://xxx.lanl.gov/abs/hep-ph/0409183}{{\tt hep-ph/0409183}}].

	\bibitem{Meyer:2004gx} H.~B. Meyer, {\it {Glueball regge trajectories}},
  \href{http://xxx.lanl.gov/abs/hep-lat/0508002}{{\tt hep-lat/0508002}}.

	\bibitem{Burden:1982zb} C.~Burden and L.~Tassie, {\it {Rotating Strings, Glueballs and Exotic Mesons}},
   {\em Austral.J.Phys.} {\bf 35} (1982) 223--233.

	\bibitem{PandoZayas:2003yb} L.~A. Pando~Zayas, J.~Sonnenschein, and D.~Vaman, {\it {Regge trajectories
  revisited in the gauge / string correspondence}},  {\em Nucl.Phys.} {\bf
  B682} (2004) 3--44, [\href{http://xxx.lanl.gov/abs/hep-th/0311190}{{\tt
  hep-th/0311190}}].

	\bibitem{Kruczenski:2004me} M.~Kruczenski, L.~A. Pando~Zayas, J.~Sonnenschein, and D.~Vaman, {\it {Regge
  trajectories for mesons in the holographic dual of large-N(c) QCD}},  {\em
  JHEP} {\bf 0506} (2005) 046,
  [\href{http://xxx.lanl.gov/abs/hep-th/0410035}{{\tt hep-th/0410035}}].

	\bibitem{'tHooft:2004he} G.~'t~Hooft, {\it {Minimal strings for baryons}},
  \href{http://xxx.lanl.gov/abs/hep-th/0408148}{{\tt hep-th/0408148}}.

	\bibitem{Sharov:2000pg} G.~Sharov, {\it {Quasirotational motions and stability problem in dynamics of
  string hadron models}},  {\em Phys.Rev.} {\bf D62} (2000) 094015,
  [\href{http://xxx.lanl.gov/abs/hep-ph/0004003}{{\tt hep-ph/0004003}}].

	\bibitem{ThesisFederovsky} E.~Federovsky, {\it {Stringy baryons and scattering amplitudes}},  Master's
  thesis, Tel Aviv University, August, 2010.

	\bibitem{ThesisHarpaz} G.~Harpaz, {\it {Simulating stringy baryons}},  Master's thesis, Tel Aviv
  University, June, 2008.

	\bibitem{Sharov:2013tga} G.~Sharov, {\it {String Models, Stability and Regge Trajectories for Hadron
  States}},  \href{http://xxx.lanl.gov/abs/1305.3985}{{\tt arXiv:1305.3985}}.

	\bibitem{Selem:2006nd} A.~Selem and F.~Wilczek, {\it {Hadron systematics and emergent diquarks}},
  \href{http://xxx.lanl.gov/abs/hep-ph/0602128}{{\tt hep-ph/0602128}}.

	\bibitem{Dubovsky:2015zey} S.~Dubovsky and V.~Gorbenko,
  ``Towards a Theory of the QCD String,''
  arXiv:1511.01908 [hep-th].


	\bibitem{Arvis:1983fp} J.~Arvis, {\it {The Exact $q \bar{q}$ Potential in Nambu String Theory}},  {\em
  Phys.Lett.} {\bf B127} (1983) 106.

	\bibitem{Aharony:2009gg} O.~Aharony and E.~Karzbrun, {\it {On the effective action of confining
  strings}},  {\em JHEP} {\bf 0906} (2009) 012,
  [\href{http://xxx.lanl.gov/abs/0903.1927}{{\tt arXiv:0903.1927}}].

	\bibitem{Bigazzi:2004ze} F.~Bigazzi, A.~Cotrone, L.~Martucci, and L.~Pando~Zayas, {\it {Wilson loop,
  Regge trajectory and hadron masses in a Yang-Mills theory from semiclassical
  strings}},  {\em Phys.Rev.} {\bf D71} (2005) 066002,
  [\href{http://xxx.lanl.gov/abs/hep-th/0409205}{{\tt hep-th/0409205}}].

	\bibitem{Frolov:2002av} S.~Frolov and A.~A.~Tseytlin,
  ``Semiclassical quantization of rotating superstring in AdS(5) x S**5,''
  JHEP {\bf 0206}, 007 (2002)
  doi:10.1088/1126-6708/2002/06/007
  [hep-th/0204226].

	\bibitem{Polchinski:1991ax} J.~Polchinski and A.~Strominger, {\it {Effective string theory}},  {\em
  Phys.Rev.Lett.} {\bf 67} (1991) 1681--1684.

	\bibitem{Hellerman:2014cba} S.~Hellerman, S.~Maeda, J.~Maltz and I.~Swanson,
  ``Effective String Theory Simplified,''
  JHEP {\bf 1409}, 183 (2014)
  doi:10.1007/JHEP09(2014)183
  [arXiv:1405.6197 [hep-th]].

	\bibitem{Lambiase:1995st} G.~Lambiase and V.~V.~Nesterenko,
  ``Quark mass correction to the string potential,''
  Phys.\ Rev.\ D {\bf 54}, 6387 (1996)
  doi:10.1103/PhysRevD.54.6387
  [hep-th/9510221].


	\bibitem{ASY} O.~Aharony, J.~Sonnenschein, and S.~Yankielowicz, {\it {On the quantization of
  rotating open strings with massive endpoints [In preparation]}}, .

	\bibitem{oai:arXiv.org:hep-th/9911123} Y.~Kinar, E.~Schreiber, J.~Sonnenschein, and N.~Weiss, {\it {Quantum
  fluctuations of Wilson loops from string models}},  {\em Nucl.Phys.} {\bf
  B583} (2000) 76--104, [\href{http://xxx.lanl.gov/abs/hep-th/9911123}{{\tt
  hep-th/9911123}}].



	\bibitem{Sharov:1998hi} G.~Sharov, {\it {String baryon model 'triangle': Hypocycloidal solutions}},
  {\em Phys.Rev.} {\bf D58} (1998) 114009,
  [\href{http://xxx.lanl.gov/abs/hep-th/9808099}{{\tt hep-th/9808099}}].

	\bibitem{Peeters:2005fq} K.~Peeters, J.~Sonnenschein, and M.~Zamaklar, {\it {Holographic decays of
  large-spin mesons}},  {\em JHEP} {\bf 0602} (2006) 009,
  [\href{http://xxx.lanl.gov/abs/hep-th/0511044}{{\tt hep-th/0511044}}].

	\bibitem{PDG:2014} {\bf Particle Data Group} Collaboration, K.~Olive et~al., {\it {Review of
  Particle Physics}},  {\em Chin.Phys.} {\bf C38} (2014) 090001.

	\bibitem{Klempt:2007cp} E.~Klempt and A.~Zaitsev, {\it {Glueballs, Hybrids, Multiquarks. Experimental
  facts versus QCD inspired concepts}},  {\em Phys.Rept.} {\bf 454} (2007)
  1--202, [\href{http://xxx.lanl.gov/abs/0708.4016}{{\tt arXiv:0708.4016}}].

	\bibitem{Mathieu:2008me} V.~Mathieu, N.~Kochelev, and V.~Vento, {\it {The Physics of Glueballs}},  {\em
  Int.J.Mod.Phys.} {\bf E18} (2009) 1--49,
  [\href{http://xxx.lanl.gov/abs/0810.4453}{{\tt arXiv:0810.4453}}].

	\bibitem{Crede:2008vw} V.~Crede and C.~Meyer, {\it {The Experimental Status of Glueballs}},  {\em
  Prog.Part.Nucl.Phys.} {\bf 63} (2009) 74--116,
  [\href{http://xxx.lanl.gov/abs/0812.0600}{{\tt arXiv:0812.0600}}].

	\bibitem{Ochs:2013gi} W.~Ochs, {\it {The Status of Glueballs}},  {\em J.Phys.} {\bf G40} (2013)
  043001, [\href{http://xxx.lanl.gov/abs/1301.5183}{{\tt arXiv:1301.5183}}].

	\bibitem{Nebreda:2011cp} J.~Nebreda, J.~Pelaez, and G.~Rios, {\it {Enhanced non-quark-antiquark and
  non-glueball Nc behavior of light scalar mesons}},  {\em Phys.Rev.} {\bf D84}
  (2011) 074003, [\href{http://xxx.lanl.gov/abs/1107.4200}{{\tt
  arXiv:1107.4200}}].

	\bibitem{Anisovich:2000kxa} A.~Anisovich, V.~Anisovich, and A.~Sarantsev, {\it {Systematics of q anti-q
  states in the (n, M**2) and (J, M**2) planes}},  {\em Phys.Rev.} {\bf D62}
  (2000) 051502, [\href{http://xxx.lanl.gov/abs/hep-ph/0003113}{{\tt
  hep-ph/0003113}}].

	\bibitem{Anisovich:2002us} V.~Anisovich, {\it {Systematics of quark anti-quark states and scalar exotic
  mesons}},  {\em Phys.Usp.} {\bf 47} (2004) 45--67,
  [\href{http://xxx.lanl.gov/abs/hep-ph/0208123}{{\tt hep-ph/0208123}}].

	\bibitem{Masjuan:2012gc} P.~Masjuan, E.~Ruiz~Arriola, and W.~Broniowski, {\it {Systematics of radial and
  angular-momentum Regge trajectories of light non-strange
  $q\overline{q}$-states}},  {\em Phys.Rev.} {\bf D85} (2012) 094006,
  [\href{http://xxx.lanl.gov/abs/1203.4782}{{\tt arXiv:1203.4782}}].

	\bibitem{Bugg:2012yt} D.~Bugg, {\it {Comment on "Systematics of radial and angular-momentum Regge
  trajectories of light nonstrange $q\overline{q}$-states"}},  {\em
  Phys.Rev.} {\bf D87} (2013), no.~11 118501,
  [\href{http://xxx.lanl.gov/abs/1209.3481}{{\tt arXiv:1209.3481}}].

	\bibitem{Ablikim:2012ft} {\bf BESIII} Collaboration, M.~Ablikim et~al., {\it {Study of the
  near-threshold ?? mass enhancement in doubly OZI-suppressed J/?????
  decays}},  {\em Phys.Rev.} {\bf D87} (2013), no.~3 032008,
  [\href{http://xxx.lanl.gov/abs/1211.5668}{{\tt arXiv:1211.5668}}].

	\bibitem{Bai:1996wm} {\bf BES Collaboration} Collaboration, J.~Bai et~al., {\it {Studies of xi
  (2230) in J / psi radiative decays}},  {\em Phys.Rev.Lett.} {\bf 76} (1996)
  3502--3505.

	\bibitem{Vladimirsky:2001ek} V.~Vladimirsky, V.~Grigorev, O.~Erofeeva, Y.~Katinov, V.~Lisin, et~al., {\it
  {Resonance maximum in the system of two K(S) mesons at 1450-MeV}},  {\em
  Phys.Atom.Nucl.} {\bf 64} (2001) 1895--1897.

	\bibitem{Vijande:2004he} J.~Vijande, F.~Fernandez, and A.~Valcarce, {\it {Constituent quark model study
  of the meson spectra}},  {\em J.Phys.} {\bf G31} (2005) 481,
  [\href{http://xxx.lanl.gov/abs/hep-ph/0411299}{{\tt hep-ph/0411299}}].

	\bibitem{Athenodorou:2010cs} A.~Athenodorou, B.~Bringoltz, and M.~Teper, {\it {Closed flux tubes and their
  string description in D=3+1 SU(N) gauge theories}},  {\em JHEP} {\bf 02}
  (2011) 030, [\href{http://xxx.lanl.gov/abs/1007.4720}{{\tt
  arXiv:1007.4720}}].

	\bibitem{Bochicchio:2013aha} M.~Bochicchio, {\it {Yang-Mills mass gap, Floer homology, glueball spectrum,
  and conformal window in large-N QCD}},
  \href{http://xxx.lanl.gov/abs/1312.1350}{{\tt arXiv:1312.1350}}.

	\bibitem{Bochicchio:2013sra} M.~Bochicchio, {\it {Glueball and meson spectrum in large-N massless QCD}},
  \href{http://xxx.lanl.gov/abs/1308.2925}{{\tt arXiv:1308.2925}}.

	\bibitem{Caselle:2015tza} M.~Caselle, A.~Nada, and M.~Panero, {\it {Hagedorn spectrum and thermodynamics
  of SU(2) and SU(3) Yang-Mills theories}},
  \href{http://xxx.lanl.gov/abs/1505.0110}{{\tt arXiv:1505.0110}}.

	\bibitem{Donnachie:1984xq} A.~Donnachie and P.~Landshoff, {\it {Elastic Scattering and Diffraction
  Dissociation}},  {\em Nucl.Phys.} {\bf B244} (1984) 322.

	\bibitem{Morningstar:1999rf} C.~J. Morningstar and M.~J. Peardon, {\it {The Glueball spectrum from an
  anisotropic lattice study}},  {\em Phys.Rev.} {\bf D60} (1999) 034509,
  [\href{http://xxx.lanl.gov/abs/hep-lat/9901004}{{\tt hep-lat/9901004}}].

	\bibitem{Ganor:1994rm} O.~Ganor, J.~Sonnenschein, and S.~Yankielowicz, {\it {Folds in 2-D string
  theories}},  {\em Nucl.Phys.} {\bf B427} (1994) 203--244,
  [\href{http://xxx.lanl.gov/abs/hep-th/9404149}{{\tt hep-th/9404149}}].


	\bibitem{Shuryak:2013sra} E.~Shuryak and I.~Zahed,
  ``New regimes of the stringy (holographic) Pomeron and high-multiplicity $pp$ and $pA$ collisions,''
  Phys.\ Rev.\ D {\bf 89}, no. 9, 094001 (2014)
  doi:10.1103/PhysRevD.89.094001
  [arXiv:1311.0836 [hep-ph]].

	\bibitem{Casero:2005se} R.~Casero, A.~Paredes, and J.~Sonnenschein, {\it {Fundamental matter, meson
  spectroscopy and non-critical string/gauge duality}},  {\em JHEP} {\bf 0601}
  (2006) 127, [\href{http://xxx.lanl.gov/abs/hep-th/0510110}{{\tt
  hep-th/0510110}}].

	\bibitem{oai:arXiv.org:0806.0152} O.~Mintakevich and J.~Sonnenschein, {\it {On the spectra of scalar mesons from
  HQCD models}},  {\em JHEP} {\bf 0808} (2008) 082,
  [\href{http://xxx.lanl.gov/abs/0806.0152}{{\tt arXiv:0806.0152}}].

	\bibitem{Peeters:2006iu} K.~Peeters, J.~Sonnenschein, and M.~Zamaklar, {\it {Holographic melting and
  related properties of mesons in a quark gluon plasma}},  {\em Phys.Rev.} {\bf
  D74} (2006) 106008, [\href{http://xxx.lanl.gov/abs/hep-th/0606195}{{\tt
  hep-th/0606195}}].

	\bibitem{Imoto:2010ef} T.~Imoto, T.~Sakai, and S.~Sugimoto, {\it {Mesons as Open Strings in a
  Holographic Dual of QCD}},  {\em Prog.Theor.Phys.} {\bf 124} (2010) 263--284,
  [\href{http://xxx.lanl.gov/abs/1005.0655}{{\tt arXiv:1005.0655}}].

	\bibitem{cobitalks} J.~Sonnenschein, {\it Holographic stingy hadrons},  {\em Talks delivered in
  September IPMU, October NBI and November NS/WAS 2013}.

	\bibitem{Chodos:1973gt} A.~Chodos and C.~B. Thorn, {\it {Making the Massless String Massive}},  {\em
  Nucl.Phys.} {\bf B72} (1974) 509.

	\bibitem{PDG:2012} {\bf Particle Data Group} Collaboration, J.~Beringer et~al., {\it {Review of
  Particle Physics (RPP)}},  {\em Phys.Rev.} {\bf D86} (2012) 010001.

	\bibitem{Eichten:2007qx} E.~Eichten, S.~Godfrey, H.~Mahlke, and J.~L. Rosner, {\it {Quarkonia and their
  transitions}},  {\em Rev.Mod.Phys.} {\bf 80} (2008) 1161--1193,
  [\href{http://xxx.lanl.gov/abs/hep-ph/0701208}{{\tt hep-ph/0701208}}].

	\bibitem{Brambilla:2010cs} N.~Brambilla, S.~Eidelman, B.~Heltsley, R.~Vogt, G.~Bodwin, et~al., {\it {Heavy
  quarkonium: progress, puzzles, and opportunities}},  {\em Eur.Phys.J.} {\bf
  C71} (2011) 1534, [\href{http://xxx.lanl.gov/abs/1010.5827}{{\tt
  arXiv:1010.5827}}].

	\bibitem{Gershtein:2006ng} S.~Gershtein, A.~Likhoded, and A.~a. Luchinsky, {\it {Systematics of heavy
  quarkonia from Regge trajectories on (n,M**2) and (M**2,J) planes}},  {\em
  Phys.Rev.} {\bf D74} (2006) 016002,
  [\href{http://xxx.lanl.gov/abs/hep-ph/0602048}{{\tt hep-ph/0602048}}].

	\bibitem{Ebert:2009ub} D.~Ebert, R.~Faustov, and V.~Galkin, {\it {Mass spectra and Regge trajectories
  of light mesons in the relativistic quark model}},  {\em Phys.Rev.} {\bf D79}
  (2009) 114029, [\href{http://xxx.lanl.gov/abs/0903.5183}{{\tt
  arXiv:0903.5183}}].

	\bibitem{Ebert:2009ua} D.~Ebert, R.~Faustov, and V.~Galkin, {\it {Heavy-light meson spectroscopy and
  Regge trajectories in the relativistic quark model}},  {\em Eur.Phys.J.} {\bf
  C66} (2010) 197--206, [\href{http://xxx.lanl.gov/abs/0910.5612}{{\tt
  arXiv:0910.5612}}].

	\bibitem{Ebert:2011jc} D.~Ebert, R.~Faustov, and V.~Galkin, {\it {Spectroscopy and Regge trajectories
  of heavy quarkonia and $B_c$ mesons}},  {\em Eur.Phys.J.} {\bf C71} (2011)
  1825, [\href{http://xxx.lanl.gov/abs/1111.0454}{{\tt arXiv:1111.0454}}].

	\bibitem{Aaij:2013sza} {\bf LHCb} Collaboration, R.~Aaij et~al., {\it {Study of $D_J$ meson decays to
  $D^+\pi^-$, $D^0 \pi^+$ and $D^{*+}\pi^-$ final states in pp collision}},
  {\em JHEP} {\bf 1309} (2013) 145,
  [\href{http://xxx.lanl.gov/abs/1307.4556}{{\tt arXiv:1307.4556}}].

	\bibitem{Wang:2013tka} Z.-G. Wang, {\it {Analysis of strong decays of the charmed mesons $D_J(2580)$,
  $D_J^*(2650)$, $D_J(2740)$, $D^*_J(2760)$, $D_J(3000)$, $D_J^*(3000)$}},
  {\em Phys.Rev.} {\bf D88} (2013) 114003,
  [\href{http://xxx.lanl.gov/abs/1308.0533}{{\tt arXiv:1308.0533}}].



	\bibitem{Inopin:1999nf} A.~Inopin and G.~Sharov, {\it {Hadronic Regge trajectories: Problems and
  approaches}},  {\em Phys.Rev.} {\bf D63} (2001) 054023,
  [\href{http://xxx.lanl.gov/abs/hep-ph/9905499}{{\tt hep-ph/9905499}}].

	\bibitem{Tang:2000tb} A.~Tang and J.~W. Norbury, {\it {Properties of Regge trajectories}},  {\em
  Phys.Rev.} {\bf D62} (2000) 016006,
  [\href{http://xxx.lanl.gov/abs/hep-ph/0004078}{{\tt hep-ph/0004078}}].

	\bibitem{Klempt:2009pi} E.~Klempt and J.-M. Richard, {\it {Baryon spectroscopy}},  {\em Rev.Mod.Phys.}
  {\bf 82} (2010) 1095--1153, [\href{http://xxx.lanl.gov/abs/0901.2055}{{\tt
  arXiv:0901.2055}}].

	\bibitem{Hata:2007mb} H.~Hata, T.~Sakai, S.~Sugimoto, and S.~Yamato, {\it {Baryons from instantons in
  holographic QCD}},  {\em Prog.Theor.Phys.} {\bf 117} (2007) 1157,
  [\href{http://xxx.lanl.gov/abs/hep-th/0701280}{{\tt hep-th/0701280}}].

	\bibitem{Kaplunovsky:2012gb} V.~Kaplunovsky, D.~Melnikov, and J.~Sonnenschein, {\it {Baryonic Popcorn}},
  {\em JHEP} {\bf 1211} (2012) 047,
  [\href{http://xxx.lanl.gov/abs/1201.1331}{{\tt arXiv:1201.1331}}].

	\bibitem{Kaplunovsky:2013iza} V.~Kaplunovsky and J.~Sonnenschein, {\it {Dimension Changing Phase Transitions
  in Instanton Crystals}},  {\em JHEP} {\bf 1404} (2014) 022,
  [\href{http://xxx.lanl.gov/abs/1304.7540}{{\tt arXiv:1304.7540}}].



	\bibitem{Brandhuber:1998xy} A.~Brandhuber, N.~Itzhaki, J.~Sonnenschein, and S.~Yankielowicz, {\it {Baryons
  from supergravity}},  {\em JHEP} {\bf 9807} (1998) 020,
  [\href{http://xxx.lanl.gov/abs/hep-th/9806158}{{\tt hep-th/9806158}}].

	\bibitem{Callan:1999zf} J.~Callan, Curtis~G., A.~Guijosa, K.~G. Savvidy, and O.~Tafjord, {\it {Baryons
  and flux tubes in confining gauge theories from brane actions}},  {\em
  Nucl.Phys.} {\bf B555} (1999) 183--200,
  [\href{http://xxx.lanl.gov/abs/hep-th/9902197}{{\tt hep-th/9902197}}].

	\bibitem{Friedmann:2014qpa} T.~Friedmann, {\it {QCD vs. the Centrifugal Barrier: a New QCD Effect}},  {\em
  EPJ Web Conf.} {\bf 70} (2014) 00027.

	\bibitem{Decays} J.~Sonnenschein and D.~Weissman, {\it {On decays of stringy hadrons [work in
  progress]}}, .


	\bibitem{Csaki:1998qr} C.~Csaki, H.~Ooguri, Y.~Oz, and J.~Terning, {\it {Glueball mass spectrum from
  supergravity}},  {\em JHEP} {\bf 9901} (1999) 017,
  [\href{http://xxx.lanl.gov/abs/hep-th/9806021}{{\tt hep-th/9806021}}].

	\bibitem{Brower:2000rp} R.~C. Brower, S.~D. Mathur, and C.-I. Tan, {\it {Glueball spectrum for QCD from
  AdS supergravity duality}},  {\em Nucl. Phys.} {\bf B587} (2000) 249--276,
  [\href{http://xxx.lanl.gov/abs/hep-th/0003115}{{\tt hep-th/0003115}}].

	\bibitem{Elander:2013jqa} D.~Elander, A.~F. Faedo, C.~Hoyos, D.~Mateos, and M.~Piai, {\it {Multiscale
  confining dynamics from holographic RG flows}},  {\em JHEP} {\bf 05} (2014)
  003, [\href{http://xxx.lanl.gov/abs/1312.7160}{{\tt arXiv:1312.7160}}].

	\bibitem{Bhanot:1980fx} G.~Bhanot and C.~Rebbi, {\it {SU(2) String Tension, Glueball Mass and
  Interquark Potential by Monte Carlo Computations}},  {\em Nucl.Phys.} {\bf
  B180} (1981) 469.

	\bibitem{Niemi:2003hb} A.~J. Niemi, {\it {Are glueballs knotted closed strings?}},
  \href{http://xxx.lanl.gov/abs/hep-th/0312133}{{\tt hep-th/0312133}}.

	\bibitem{Sharov:2007ag} G.~Sharov, {\it {Closed String with Masses in Models of Baryons and
  Glueballs}},  \href{http://xxx.lanl.gov/abs/0712.4052}{{\tt
  arXiv:0712.4052}}.

	\bibitem{Solovev:2000nb} L.~Solovev, {\it {Glueballs in the string quark model}},  {\em
  Theor.Math.Phys.} {\bf 126} (2001) 203--211,
  [\href{http://xxx.lanl.gov/abs/hep-ph/0006010}{{\tt hep-ph/0006010}}].

	\bibitem{Talalov:2001cp} S.~Talalov, {\it {The Glueball Regge trajectory from the string inspired
  theory}},  \href{http://xxx.lanl.gov/abs/hep-ph/0101028}{{\tt
  hep-ph/0101028}}.

	\bibitem{LlanesEstrada:2000jw} F.~J. Llanes-Estrada, S.~R. Cotanch, P.~J. de~A.~Bicudo, J.~E.~F. Ribeiro, and
  A.~P. Szczepaniak, {\it {QCD glueball Regge trajectories and the Pomeron}},
  {\em Nucl.Phys.} {\bf A710} (2002) 45--54,
  [\href{http://xxx.lanl.gov/abs/hep-ph/0008212}{{\tt hep-ph/0008212}}].

	\bibitem{Szczepaniak:2003mr} A.~P. Szczepaniak and E.~S. Swanson, {\it {The Low lying glueball spectrum}},
  {\em Phys.Lett.} {\bf B577} (2003) 61--66,
  [\href{http://xxx.lanl.gov/abs/hep-ph/0308268}{{\tt hep-ph/0308268}}].

	\bibitem{Pons:2004dk} J.~Pons, J.~Russo, and P.~Talavera, {\it {Semiclassical string spectrum in a
  string model dual to large N QCD}},  {\em Nucl.Phys.} {\bf B700} (2004)
  71--88, [\href{http://xxx.lanl.gov/abs/hep-th/0406266}{{\tt
  hep-th/0406266}}].

	\bibitem{Abreu:2005uw} E.~Abreu and P.~Bicudo, {\it {Glueball and hybrid mass and decay with string
  tension below Casimir scaling}},  {\em J.Phys.} {\bf G34} (2007) 195207,
  [\href{http://xxx.lanl.gov/abs/hep-ph/0508281}{{\tt hep-ph/0508281}}].

	\bibitem{Brau:2004xw} F.~Brau and C.~Semay, {\it {Semirelativistic potential model for glueball
  states}},  {\em Phys.Rev.} {\bf D70} (2004) 014017,
  [\href{http://xxx.lanl.gov/abs/hep-ph/0412173}{{\tt hep-ph/0412173}}].

	\bibitem{Mathieu:2005wc} V.~Mathieu, C.~Semay, and F.~Brau, {\it {Casimir scaling, glueballs and hybrid
  gluelumps}},  {\em Eur.Phys.J.} {\bf A27} (2006) 225--230,
  [\href{http://xxx.lanl.gov/abs/hep-ph/0511210}{{\tt hep-ph/0511210}}].

	\bibitem{Simonov:2006re} Y.~Simonov, {\it {Glueballs, gluerings and gluestars in the d=2+1 SU(N) gauge
  theory}},  {\em Phys.Atom.Nucl.} {\bf 70} (2007) 44--52,
  [\href{http://xxx.lanl.gov/abs/hep-ph/0603148}{{\tt hep-ph/0603148}}].

	\bibitem{Mathieu:2006bp} V.~Mathieu, C.~Semay, and B.~Silvestre-Brac, {\it {Semirelativistic potential
  model for low-lying three-gluon glueballs}},  {\em Phys.Rev.} {\bf D74}
  (2006) 054002, [\href{http://xxx.lanl.gov/abs/hep-ph/0605205}{{\tt
  hep-ph/0605205}}].

	\bibitem{BoschiFilho:2002vd} H.~Boschi-Filho and N.~R. Braga, {\it {Gauge / string duality and scalar
  glueball mass ratios}},  {\em JHEP} {\bf 0305} (2003) 009,
  [\href{http://xxx.lanl.gov/abs/hep-th/0212207}{{\tt hep-th/0212207}}].

	\bibitem{Polchinski:Vol1} J.~Polchinski, {\it {String theory. Vol. 1: An introduction to the bosonic
  string}}, .

	\bibitem{Armoni:2006ri} A.~Armoni and B.~Lucini, {\it {Universality of k-string tensions from
  holography and the lattice}},  {\em JHEP} {\bf 06} (2006) 036,
  [\href{http://xxx.lanl.gov/abs/hep-th/0604055}{{\tt hep-th/0604055}}].

	\bibitem{Cotrone:2005fr} A.~L. Cotrone, L.~Martucci, and W.~Troost, {\it {String splitting and strong
  coupling meson decay}},  {\em Phys. Rev. Lett.} {\bf 96} (2006) 141601,
  [\href{http://xxx.lanl.gov/abs/hep-th/0511045}{{\tt hep-th/0511045}}].

	\bibitem{Bigazzi:2006jt} F.~Bigazzi and A.~L. Cotrone, {\it {New predictions on meson decays from string
  splitting}},  {\em JHEP} {\bf 11} (2006) 066,
  [\href{http://xxx.lanl.gov/abs/hep-th/0606059}{{\tt hep-th/0606059}}].

	\bibitem{Hashimoto:2007ze} K.~Hashimoto, C.-I. Tan, and S.~Terashima, {\it {Glueball decay in holographic
  QCD}},  {\em Phys. Rev.} {\bf D77} (2008) 086001,
  [\href{http://xxx.lanl.gov/abs/0709.2208}{{\tt arXiv:0709.2208}}].

	\bibitem{Brunner:2015oqa} F.~Br{\"u}nner, D.~Parganlija, and A.~Rebhan, {\it {Glueball Decay Rates in the
  Witten-Sakai-Sugimoto Model}},  {\em Phys. Rev.} {\bf D91} (2015), no.~10
  106002, [\href{http://xxx.lanl.gov/abs/1501.0790}{{\tt arXiv:1501.0790}}].

	\bibitem{Brunner:2015yha} F.~Br{\"u}nner and A.~Rebhan, {\it {Nonchiral enhancement of scalar glueball
  decay in the Witten-Sakai-Sugimoto model}},
  \href{http://xxx.lanl.gov/abs/1504.0581}{{\tt arXiv:1504.0581}}.

	\bibitem{Benslama:2002pa} {\bf CLEO} Collaboration, K.~Benslama et~al., {\it {Anti-search for the
  glueball candidate f(J)(2220) in two -photon interactions}},  {\em Phys.Rev.}
  {\bf D66} (2002) 077101, [\href{http://xxx.lanl.gov/abs/hep-ex/0204019}{{\tt
  hep-ex/0204019}}].

	\bibitem{Gregory:2012hu} E.~Gregory, A.~Irving, B.~Lucini, C.~McNeile, A.~Rago, et~al., {\it {Towards
  the glueball spectrum from unquenched lattice QCD}},  {\em JHEP} {\bf 1210}
  (2012) 170, [\href{http://xxx.lanl.gov/abs/1208.1858}{{\tt
  arXiv:1208.1858}}].

	\bibitem{Albanese:1987ds} {\bf APE} Collaboration, M.~Albanese et~al., {\it {Glueball Masses and String
  Tension in Lattice QCD}},  {\em Phys.Lett.} {\bf B192} (1987) 163--169.

	\bibitem{Chen:2005mg} Y.~Chen, A.~Alexandru, S.~Dong, T.~Draper, I.~Horvath, et~al., {\it {Glueball
  spectrum and matrix elements on anisotropic lattices}},  {\em Phys.Rev.} {\bf
  D73} (2006) 014516, [\href{http://xxx.lanl.gov/abs/hep-lat/0510074}{{\tt
  hep-lat/0510074}}].

	\bibitem{Bali:1993fb} {\bf UKQCD Collaboration} Collaboration, G.~Bali et~al., {\it {A Comprehensive
  lattice study of SU(3) glueballs}},  {\em Phys.Lett.} {\bf B309} (1993)
  378--384, [\href{http://xxx.lanl.gov/abs/hep-lat/9304012}{{\tt
  hep-lat/9304012}}].

	\bibitem{Lucini:2014paa} B.~Lucini, {\it {Glueballs from the Lattice}},  {\em PoS} {\bf QCD-TNT-III}
  (2013) 023, [\href{http://xxx.lanl.gov/abs/1401.1494}{{\tt
  arXiv:1401.1494}}].

	\bibitem{Lucini:2004my} B.~Lucini, M.~Teper, and U.~Wenger, {\it {Glueballs and k-strings in SU(N)
  gauge theories: Calculations with improved operators}},  {\em JHEP} {\bf
  0406} (2004) 012, [\href{http://xxx.lanl.gov/abs/hep-lat/0404008}{{\tt
  hep-lat/0404008}}].

\cite{Brodsky:2015oia}
\bibitem{Brodsky:2015oia}
  S.~J.~Brodsky, A.~Deur, G.~F.~de Téramond and H.~G.~Dosch,
  ``Light-Front Holography and Superconformal Quantum Mechanics: A New Approach to Hadron Structure and Color Confinement,''
  Int.\ J.\ Mod.\ Phys.\ Conf.\ Ser.\  {\bf 39} (2015) 1560081
  doi:10.1142/S2010194515600812
  [arXiv:1510.01011 [hep-ph]].
\end{thebibliography}
\end{document}